\documentclass[11pt,twoside]{article}
\usepackage[makeroom]{cancel}
\usepackage{latexsym}
\usepackage{longtable}
\usepackage{epsfig}
\usepackage{graphicx,bbm,psfrag}
\usepackage{amssymb,amsmath,amsthm,booktabs,mathtools}
\usepackage{bm}
\usepackage{color}
\usepackage{dutchcal}
\usepackage{hyperref}
\hypersetup{
    colorlinks,
    citecolor=red,
    linkcolor=blue
}

\newtheorem{theorem}{Theorem}[section]
\newtheorem{rema}[theorem]{Remark}
\newtheorem{defi}[theorem]{Definition}
\newtheorem{lemma}[theorem]{Lemma}
\newtheorem{corol}[theorem]{Corollary}
\newtheorem{assum}{Property}

\newcommand{\bc}{\begin{center}}
\newcommand{\ec}{\end{center}}
\def\ba#1{\begin{array}{#1}\displaystyle}
\newcommand{\ea}{\end{array}}

\newcommand{\beq}{\begin{equation}}
\newcommand{\eeq}{\end{equation}}
\newcommand{\beqa}{\begin{eqnarray}}
\newcommand{\eeqa}{\end{eqnarray}}
\newcommand{\no}{\nonumber}
\newcommand{\n}{\nonumber\\}
\newcommand{\bi}{\begin{itemize}}
\newcommand{\ei}{\end{itemize}}

\def\lt#1{\left#1}
\def\rt#1{\right#1}
\def\t#1{\tilde{#1}}
\def\h#1{\hat{#1}}

\def\frc#1#2{\frac{#1}{#2}}

\newcommand{\p}{\partial}

\newcommand{\bra}{\langle}
\newcommand{\ket}{\rangle}
\newcommand{\Z}{{\mathbb{Z}}}
\newcommand{\N}{{\mathbb{N}}}
\newcommand{\R}{{\mathbb{R}}}
\newcommand{\C}{{\mathbb{C}}}
\newcommand{\Tr}{{\rm Tr}}

\newcommand{\ep}{\epsilon}

\newcommand{\ri}{{\rm i}}
\newcommand{\re}{{\rm e}}
\newcommand{\dd}{{\rm d}}
\newcommand{\1}{{\bf 1}}

\DeclareMathOperator{\supp}{{\rm supp}}
\DeclareMathOperator*{\balim}{{\widetilde{\rm lim}}}
\newcommand{\li}{\Lambda^{\rm eul}}
\DeclareMathOperator{\im}{im}
\DeclareMathOperator{\ran}{ran}
\DeclareMathOperator{\spa}{span}

\newcommand{\lo}{{\cal V}}
\newcommand{\lon}{{\cal N}}
\newcommand{\hi}{{\cal H}}

\newcommand{\hicons}{{\cal Q}}

\newcommand{\dbra}{\langle}
\newcommand{\dket}{\rangle}

\def\la#1{\mathcal{#1}}
\def\hla#1{\check{\mathcal{#1}}}

\newcommand{\halmos}{\rule{1ex}{1.4ex}}
\newcommand{\eproof}{\hspace*{\fill}\mbox{$\halmos$}}
\def\proofof#1{\noindent {\em Proof of #1.\ }}

\begin{document}

\begin{titlepage}

\begin{center}
{\Large {\bf Hydrodynamic projections and the emergence \\[0.2cm]  of linearised Euler equations in one-dimensional isolated systems}}

\vspace{1cm}

{\large Benjamin Doyon$^*$}
\vspace{0.2cm}

{\small\em
Department of Mathematics, King's College London, Strand, London WC2R 2LS, U.K.}

\end{center}

\vspace{1cm}

\noindent One of the most profound questions of mathematical physics is that of establishing from first principles the hydrodynamic equations in large, isolated, strongly interacting many-body systems. This involves understanding relaxation at long times under reversible dynamics, determining the space of emergent collective degrees of freedom (the ballistic waves), showing that projection occurs onto them, and establishing their dynamics (the hydrodynamic equations). We make progress in these directions, focussing for simplicity on one-dimensional systems. Under a model-independent definition of the complete space of extensive conserved charges, we show that hydrodynamic projection occurs in Euler-scale two-point correlation functions. A fundamental ingredient is a property of relaxation: we establish ergodicity of correlation functions along almost every direction in space-time. We further show that to every extensive conserved charge with a local density is associated a local current and a continuity equation; and that Euler-scale two-point correlation functions of local conserved densities satisfy a hydrodynamic equation. The results are established rigorously within a general framework based on Hilbert spaces of observables. These spaces occur naturally in the $C^*$ algebra description of statistical mechanics by the Gelfand-Naimark-Segal construction. Using Araki's exponential clustering and the Lieb-Robinson bound, we show that the results hold, for instance, in every nonzero-temperature Gibbs state of short-range quantum spin chains. Many techniques we introduce are generalisable to higher dimensions. This provides a precise and universal theory for the emergence of ballistic waves at the Euler scale and how they propagate within homogeneous, stationary states.

\vfill

$^*$email: {\tt benjamin.doyon@kcl.ac.uk}\hfill \today

\end{titlepage}


\tableofcontents

\section{Introduction}

The passage from short-scale, microscopic motion to large-scale, emergent collective behaviours is at the heart of some of the deepest questions in modern theoretical physics. The problem may be framed as determining, from the intricate microscopic dynamics of a myriad constituents in interaction, the emergent degrees of freedom that are relevant for observations at large scales of space and time, and their own dynamics.

Take the example of travelling surface-water waves. A local disturbance on a steady water surface -- say a finger touching it -- produces a complicated rearrangement of water molecules at microscopic distances. But the strongest effect on any local probe that is far enough away -- say a nearby floating leaf -- occurs when the surface wave, propagating out of the local disturbance, hits it. The surface wave is an emergent behaviour, with its own, new dynamics. In this case, it is obtained by linear response from the Euler equations with boundary conditions at the surface. Similarly, in a large class of many-body systems, strong correlations are expected to occur along trajectories associated with the propagation of ballistic, or slowly decaying modes, such as surface water waves or sound waves, and hydrodynamics is their emergent theory \cite{Spohn-book}.

Despite the simplicity of the above example, a full mathematical understanding of how ballistic correlations emerge from the microscopic dynamics of particles ruled by the fundamental, deterministic equations of physics is still missing -- this is in contrast to significant progress for stochastic systems, see \cite{demasi2006mathematical,kipnis2013scalint}. Probing behaviours at long times is a monumental task. As far as the author is aware, the {\em ergodicity} property of correlation functions -- that long-time averages do not fluctuate -- remains without a rigorous proof. Fully relating Euler-scale correlations to a universally defined {\em space of ballistic modes} has also been beyond reach. Further, there are only a few rigorous proofs of {\em hydrodynamic equations}: certain particle systems with conservative noise \cite{olla1993hydro,even2014hydro,komorowski2016ballistic,marchesani2018hydro}, and special examples in strongly interacting systems whose dynamics is Hamiltonian or more generally reversible and deterministic, including the completely integrable hard rod gas \cite{Boldrighini1983}, classical and quantum disordered anhamornic chains \cite{bernardin2019hydro,hannani2020hydro}, and the Rule 54 cellular automaton \cite{klobas2019time}. These are some of the most important challenges of mathematical physics 

The goal of this paper is to make progress on these problems. We focus on one-dimensional systems, both for their relevance to recent research (see the reviews \cite{1742-5468-2016-6-064002,vidmar2016generalized,1742-5468-2016-6-064005,GogolinEquilibration2016,spohn2018leshouches,DoyonLecture2020}), and in order to illustrate the techniques in the simplest possible setting. We concentrate on correlation functions at large wavelengths and large times, and the linearised Euler equations they satisfy. All results are established within a general framework encoding basic properties of many-body systems. They apply to all short-range quantum spin chains with Hamiltonian dynamics, are not restricted to these.

Crucially, we make no a priori assumption as to the type of ballistic modes emerging. Intuitively, it is well understood that ballistic modes are related to conserved charges admitted by the model. The conventional assumption that mass, momentum and energy are the only relevant quantities, leads to the standard Euler equations (with natural relativistic generalisations). However, it is now well established that this assumption is broken in integrable models, where an infinity of conserved charges must be taken into account, such as in classical soliton gases \cite{El-2003,El-Kamchatnov-2005} and many-body quantum systems \cite{PhysRevLett.98.050405,PhysRevLett.115.157201,PhysRevX.6.041065,PhysRevLett.117.207201}, as confirmed by experiments \cite{kinoshita,AmerongenYang08,Langen207,Schemmerghd,malvania2020ghd}. This highlights the importance of characterising the space of ballistic modes, in a manner that does not depend on the specific properties of the dynamics.

\subsection{Linearised Euler equation and Boltzmann-Gibbs principle}

The problem of establishing hydrodynamics is fruitfully divided as such: (1) obtaining the universal theory of emergent ballistic waves in a model-independent fashion, and (2) specialising the space of ballistic waves and their dynamics to given sub-families of models, such as integrable and non-integrable models. In this paper, we address the first point.

A crucial step is to prove the universal form of the {\em linearised Euler equation}.  Consider a statistical model on the one-dimensional lattice\footnote{This setup is natural for describing quantum and classical chains, but can also be applied to gases and field theories lying on the line $\R$, by grouping observables into adjacent cells parametrised by $\Z$.} $\Z$: a $C^*$-algebra of ``local" observables $\mathfrak U$, a state (positive linear functional) $\omega$ giving their statistical averages, and space-time-translation $*$-isomorphisms under which $\omega$ is invariant; $\la a(x,t)$ is the translate of $\la a\in \mathfrak U$ by distance $x\in\Z$ and time $t\in\R$  (see e.g.~\cite{IsraelConvexity,BratelliRobinson12,Na07,SimsReviewLR}). Under what conditions and for what (say countable) set of observables $\la \{q_i\}\subset\mathfrak U$ -- representing ballistic waves -- do the Fourier transforms of two-point correlation functions $C_{ij}(k,t) =  \sum_{x\in\Z} \re^{\ri kx} \big[\omega\big(\la q_i(x,t)\la q_j(0,0)\big) - \omega(\la q_i)\omega(\la q_j)\big]$ satisfy the linear differential equation
\beq\label{eulerheuristic}
	\frc{\p}{\p t} C_{ij}(k,t) = \ri k \sum_{l} \mathsf A_i^{~l}
	C_{lj}(k,t)\ ?
\eeq
The flux Jacobian $\mathsf A_i^{~l}$ \cite{Spohn-book,toth2003onsager,SciPostPhys.3.6.039,DoyonLecture2020} is usually defined by assuming that for every $\la q_i$ there exists a current $\la j_i$ such that a continuity equation holds,
\beq\label{contintro}
	\frc{\p \la q_i(x,t)}{\p t} + \la j_i(x,t) - \la j_i(x-1,t) = 0
\eeq
(that is, $\la q_i$'s are densities of extensive conserved charges), and then $\mathsf A_i^{~l} = \p \omega(\la j_i)/\p\omega(\la q_l)$, where variations of $\omega$ are taken within a manifold of (generalised) Gibbs states.  In interacting models, Eq.~\eqref{eulerheuristic} is expected to hold at large scales, $k\to0$, $t\to\infty$, possibly under space-time averaging to be specified \cite{Spohn-book,10.21468/SciPostPhys.4.6.045,doyoncorrelations}.  Eq.~\eqref{eulerheuristic} describes ballistic propagation, with ``propagator" $\mathsf A_i^{~l}$. Proving the linearised Euler equation is a particularly difficult problem. Results have been obtained by adding conservative noise \cite{olla2020equilibrium}, but there are up to now no results for purely hamiltonian microscopic dynamics.

The linearised Euler equation is a consequence of the principle of {\em hydrodynamic projection}, or the Boltzmann-Gibbs principle \cite{Spohn-book}. The linear equation \eqref{contintro} emerges from the generically highly nonlinear microscopic dynamics thanks to the projection of currents onto ballistic waves -- the reduction of the effective space of degrees of freedom at large scales. Determining the set $\{\la q_i\}$ on which this projection occurs is nontrivial. As mentioned, by conventional assumptions, in Galilean gases it comprises the mass, momentum and energy densities, giving the standard Euler equations. Many studies of the emergence of hydrodynamics start with such assumptions (see the seminal works on Mori-Zwanzig projections \cite{zwanzig1961lectures,mori1965transport}). These are broken in integrable systems, where many channels for ballistic transport emerge in classical and quantum gases and chains \cite{Spohn1977,AP03,PhysRevX.6.041065,PhysRevLett.117.207201}, and the set $\{\la q_i\}$ is larger.

Mathematically formulating the universal forms of the linearised Euler equation  and hydrodynamic projection principle, and establishing their emergence from the reversible dynamics of many-body systems, remain challenging tasks.



In this paper, three main statements are proven: an ergodicity statement for time averages of correlation functions, a principle of hydrodynamic projections, and the linearised Euler equation. These are proven in a general framework, applicable to short-range quantum spin chains (with finite local space) of infinite lengths. For the latter, the results are expressed in a self-consistent manner in Sections \ref{sectbasicchain} and \ref{sectresultschain}.

It is known that ergodic states cluster at large distances (see in particular \cite{IsraelConvexity}), and with the Lieb-Robinson bound \cite{LiebRobinson} clustering holds on space-like cones bounded by the Lieb-Robinson velocity $v_{\rm LR}$,
\[
	\omega(\la a(\pm v t,t)\la b(0,0)) \to \omega(\la a)\omega(\la b)\qquad (|t|\to\infty,\ v>v_{\rm LR})
\]
Examples are Kubo-Martin-Schwinger (KMS) states at nonzero temperature \cite{Araki}. We use this along with von Neumann's ergodic theorem \cite{RudinFunctional} and countable dimensionality of the space of observables in order to show clustering under averages along almost all space-time rays (almost-everywhere ergodicity), including outside space-like cones (Theorem \ref{theorelaxchain}). This is we believe the first clustering result in time-like directions in quantum many-body systems.

We then consider correlation functions in an appropriate limit of  large wavelengths and long times, and we show that hydrodynamic projections occur (Theorem \ref{theochain}). The space projected onto is a Hilbert space of  conserved charges $\hicons_0$. This is a subspace of the space of extensive observables considered in \cite{Doyon2017,DoyonDiffusion2019} and defined by a process akin to the Gelfand-Naimark-Segal construction; it is based on the notion of pseudolocal charges \cite{ProsenPseudo1,ProsenPseudo2,ProsenQuasilocal1,ProsenQuasilocal2}. Hydrodynamic projection is seen to be a consequence of almost everywhere ergodicity. The proof requires uniform enough power-law clustering and the Lieb-Robinson bound in order to control time evolution on extensive observables. For thermal states, Araki and others have shown that clustering is in fact uniformly exponential in KMS states \cite{Araki,GN98}. This is the first proof of the Boltzmann-Gibbs principle in Hamiltonian systems. The space $\hicons_0$ is rigorously defined, and is infinite-dimensional in integrable models; showing that it is finite-dimensional in non-integrable models remains an important open problem.

Finally, we show the linearised Euler equations for every local conserved density (Theorem \ref{theoeulerchain}.II). The proof of this statement is based on hydrodynamic projections. It also requires the existence of continuity equations \eqref{contintro}, which is nontrivial. A local conserved density is a local observable whose total sum over the chain gives rise to a conserved charge, and we show that it is always possible to define the space of local observables such that every local conserved density satisfies a continuity equation \eqref{contintro} with a local current (Theorem \ref{theoeulerchain}.I). This latter statement requires clustering faster than any power law (e.g.~KMS states). 

We expect all results to be extendable to higher-dimensional lattice models with finite local spaces at large enough temperatures, and to long-range models with appropriate algebraic decay of correlation functions \cite{kuwahara2020strictly}; part of this extension is done in \cite{ampelogiannis2021almost,ampelogiannis2021ergo}. The only possible exception, where one-dimensionality and strong decay of correlation functions seem to play a role, is the statement that every local conserved density satisfies a continuity equation with a local current.

The paper is organised as follows. In Section \ref{sectbasicchain} we review the algebraic formulation of quantum spin chains of infinite lengths and the theorems which are used to establish our main results. In Section \ref{sectresultschain} we express our main results in the context of quantum spin chains with finite-dimensional local Hilbert space and short-range interactions. In Section \ref{sectvect} we setup basic aspects of our general framework and show almost-everywhere ergodicity. In Section \ref{sectcharges} we further develop the general framework and study the space of conserved charges. In Section \ref{sectproof} we present the hydrodynamic projection result and its proof. In Section \ref{sectcons} we construct conserved currents and present the Euler equation result and proof. In Section \ref{sectchains} we show how to apply the general framework to the particular context of quantum chains. We discuss the results in Section \ref{sectcon}.

\section{Review of the algebraic formulation of quantum spin chains}\label{sectbasicchain}

In this section we review the algebraic formulation of quantum spin chains and express the crucial known results in this context.

\subsection{Algebraic formulation}\label{ssectalgebraicformulation}

The algebraic formulation of quantum statistical mechanics is based on the algebra of observables, on which time evolution and states are defined. The monographs \cite{Ruellebook,IsraelConvexity,BratelliRobinson12,naaijkens2017quantum} overview some of the far-reaching results obtained from the $C^*$ algebra formulation of statistical mechanics. We review the construction for infinite-length quantum spin chains with finite interaction range, see \cite[Chap 6]{BratelliRobinson12}.

A quantum spin chain is an infinite chain of sites, in bijection with $\Z$, each site admitting a finite-dimensional space of degrees of freedom. The linear operators acting on any finite subset $X\subset \Z$ form the $C^*$ algebra of finite matrices $\mathfrak V_X = {\rm End}\big(\otimes_{x\in X}H_x\big),\;H_x\simeq \C^{d_{\rm site}}\ \forall\ x,\,d_{\rm site} \in\N$, with anti-linear $*$-involution the hermitian conjugation $\dag$, and with the operator norm. The normed $*$-algebra
\beq
	\mathfrak V = \mathfrak V_\Z
\eeq
represents all linear operators acting nontrivially on some finite number of sites on the chain, including the identity ${\bf 1}$. It is made into a $C^*$ algebra $\mathfrak U$ by completion,
\beq
	\mathfrak U = \overline{\mathfrak V}.
\eeq
Thus, $\mathfrak U$ also contains limits of Cauchy sequences, which may act nontrivially on unbounded subsets of $\Z$. We denote the norm $||\cdot||_{\mathfrak U}$. Space translations $\iota_x^{\mathfrak U}:\mathfrak U\to\mathfrak U,\,x\in\Z$, with $\iota_x(\mathfrak V_X)=\mathfrak V_{X+x}$, are naturally defined $*$-automorphisms of $\mathfrak U$ forming a representation of the group $\Z$.

We consider a dynamics generated by a finite-range homogeneous Hamiltonian.  In this case, time translations $\tau_t^{\mathfrak U}:\mathfrak U\to\mathfrak U,\,t\in\R$ are $*$-automorphisms forming a strongly continuous representation of $\R$, and commute with $\iota_x^{\mathfrak U}$. The space $\mathfrak V$ lies within the domain of the generator $\delta^{\mathfrak U}$, which is obtained by the commutator with the Hamiltonian: denoting the Hamiltonian density by $\la h\in\mathfrak V$,
\beq\label{deltaU}
	\frc{\dd \tau_t^{\mathfrak U}\la a}{\dd t}\Big|_{t=0} = \delta^{\mathfrak U} \la a = \ri\sum_{x\in\Z}  \,[\iota_x^{\mathfrak U}\la h,\la a]
\eeq
where only a finite number of terms are nonzero. One can show that elements of $\mathfrak V$ are analytic with nonzero radius: $\tau_t^{\mathfrak U}\la a = \sum_{n= 0}^\infty t^n\big(\delta^{\mathfrak U}\big)^n\la a/n!$ has a nonzero radius of convergence for every $\la a\in\mathfrak V$, see \cite[Thm 6.2.4]{BratelliRobinson12}. For $\la a\in\mathfrak U$ we denote $\la a(x,t) = \iota_x^{\mathfrak U} \tau_t^{\mathfrak U} \la a$ the space-time translate of $\la a$.

A state $\omega$ is a continuous, positive linear functional on  ${\mathfrak U}$, which we normalise to $\omega(\1) = 1$. It is bounded as $|\omega(\la a)|\leq ||\la a||_{\mathfrak U}$. It is convenient to define, for $\la a,\la b\in\mathfrak V$, the sesquilinear form
\beq\label{omegainner}
	\bra \la a,\la b\ket = \omega (\la a^\dag \la b) - \omega(\la a^\dag)\omega(\la b).
\eeq
For any finite $\beta\geq 0$, the $(\beta,\tau^{\mathfrak U})$-KMS state $\omega_\beta$ of a finite-range quantum chain is unique \cite{Araki,ArakiUniqueness,KishimotoUniqueness} (see also  \cite[Chap 6]{BratelliRobinson12}). This is the thermal state at temperature $\beta^{-1}$; in particular, the unique normalised trace state is $\Tr=\omega_0$ (normalised as $\Tr(\1) = 1$). 

Quantities that play important roles are the support of a local operator,
\beq
	{\rm supp}(\la a) = \bigcap\{X\subset\Z:\la a\in \mathfrak V_X\}
	\quad (\la a \in \mathfrak V);
\eeq
the distance
\beq
	{\rm dist}(\la a,\la b) = {\rm dist}({\rm supp}(\la a),{\rm supp}(\la b))\quad (\la a,\la b \in \mathfrak V);
\eeq
where for subsets, ${\rm dist}(X,Y) = {\rm min}\{|x-y|:x\in X,\,y\in Y\}$, and we will use a mix notation as well, e.g.~${\rm dist}(\la a,X)$; 
the diameter
\beq
	{\rm diam}(\la a,\la b,\ldots) ={\rm diam}({\rm supp}(\la a)\cup {\rm supp}(\la b)\cup\cdots) \quad (\la a,\la b,\ldots \in \mathfrak V)
\eeq
where for subsets ${\rm diam}(X)= {\rm max}\{|x-y|:x,y\in X\}$;
and the size of local operators
\beq
	|\la a| = |\supp(\la a)|\quad (\la a\in\mathfrak V).
\eeq

Another important notion for our analysis is that of projections onto the observables supported on finite subsets of the chain \cite{BHC06,nachtergaele2019quasi}. Following \cite{nachtergaele2019quasi} the projections $\mathbb P_X:\mathfrak U\to\mathfrak V_X$, $X\subset \Z$ (we only need the cases $|X|<\infty$) may be defined by the property
\beq\label{defP}
	\mathbb P_X (\la a \la b) = \la a \Tr(\la b)\quad
	(\la a\in\mathfrak V_X,\ \la b\in\mathfrak V_{\Z\setminus X}),
\eeq
and by extending by linearity and continuity to $\mathfrak U$. In particular, $\mathbb P_X$ is bounded, and in fact it is a contraction,
\beq\label{boundP}
	||\mathbb P_X \la a||_{\mathfrak U} \leq ||\la a||_{\mathfrak U}
	\quad(\la a\in\mathfrak U).
\eeq
It is worth giving an elementary proof of this. On $\mathfrak V$ it is obtained by elementary means, and then extended to $\mathfrak U$ by continuity. Consider $\la c = \sum_i \la a_i \la b_i$ $(\la a_i\in\mathfrak V_X,\ \la b_i\in\mathfrak V_{\Z\setminus X})$, a generic element of $\mathfrak V$. We see that $\mathbb P_X$ preserves the property of non-negativity, as $\mathbb P_X (\la c^\dag \la c) \geq 0$ because $\Tr (\la b_i^\dag\la b_j)$ is non-negative as a matrix with indices $i,j$. If $\la c$ is hermitian, then $||\la c||_{\mathfrak U}\1 - \la c\geq 0$, and therefore $\mathbb P_X \la c \leq ||\la c||_{\mathfrak U}\,\1$ hence \eqref{boundP} holds for hermitian operators in $\mathfrak V$. Finally, for generic $\la c$ we have $\mathbb P_X (\la c^\dag\la c)-\mathbb P_X (\la c^\dag)\,\mathbb P_X (\la c)\geq 0$ because $\Tr (\la b_i^\dag\la b_j) - \Tr (\la b_i^\dag)\Tr(\la b_j) = \Tr ((\la b_i^\dag-\Tr(\la b_i^\dag)\1)(\la b_j-\Tr(\la b_j)\1))$ is non-negative as a matrix with indices $i,j$. Hence, by the $C^*$ property, $||\mathbb P_X \la c||_{\mathfrak U}^2 = ||\mathbb P_X(\la c^\dag)\mathbb P_X(\la c)||_{\mathfrak U}
\leq ||\mathbb P_X(\la c^\dag\la c)||_{\mathfrak U}
\leq ||\la c^\dag\la c||_{\mathfrak U} = ||\la c||_{\mathfrak U}^2$.

\subsection{Lieb-Robinson bound and clustering}

An important property of the dynamics is the Lieb-Robinson bound \cite{LiebRobinson}. For our purposes, we need a slight improvement of the original theorem, where the exponential bound is explicitly controlled by the size of the operators. We take the Lieb-Robinson bound \cite[Cor 3.1]{SimsReviewLR}, also found earlier in \cite{BHC06}. Further, instead of a bound on commutators, we need a formulation which specifies how much of a time-evolved operator is supported on some set $X$, using the projections $\mathbb P_X$. Such a statement first appeared in \cite{BHC06}, although with a notion of projection which is defined on finite chains. For the notion defined by \eqref{defP} we use instead \cite[Cor 4.4]{nachtergaele2019quasi}. See also \cite{BratelliRobinson12,naaijkens2017quantum}. 
\begin{theorem}\label{theolr} There exists $v_{\rm LR}>0$ and $b,d>0$ such that, for every $\la a\in\mathfrak V$, $t\in\R$ and $X\subset\Z$,
\beq\label{LR}
	||\tau_t^{\mathfrak U}\la a - \mathbb P_X \tau_t^{\mathfrak U}\la a||_{\mathfrak U}
	\leq b\,|\la a|\,||\la a||_{\mathfrak U}\,
	\exp\Big[-d\,\big(\,{\rm dist}(\la a , \Z\setminus X) - v_{\rm LR}|t|\,\big)\Big].
\eeq
\proof This follows immediately from \cite[Cor 3.1]{SimsReviewLR} with \cite[Cor 4.4]{nachtergaele2019quasi}, where in the latter we may take $\epsilon = |\la a|$, and by specialising to the lattice $\Z$. \eproof
\end{theorem}

All KMS states $\omega_\beta$ are space-time translation invariant and exponentially clustering, by an old result of Araki \cite[Thm 2.3]{Araki}. In fact many results exist in various families of states, see the books \cite{BratelliRobinson12,IsraelConvexity} as well as \cite{GN98,Mat02,Hastings04,NS06,BHC06,Hast06,Na07,KGKRE14,Moh15,frolich2015some,Doyon2017}. For our purposes, the most relevant result is \cite[Thm III.2]{GN98}, which extends Araki's to uniform exponential clustering (and also to a larger family of models, but this is not important for us).
\begin{theorem} \label{golodet} {\em \cite[Thm 2.3]{Araki}, \cite[Thm III.2]{GN98}.} Every KMS state $\omega_\beta$, $\beta\geq 0$ is space-time translation invariant, $\omega_\beta\circ\iota_x^{\mathfrak U} = \omega_\beta\circ\tau_t^{\mathfrak U}=\omega_\beta$ for all $x\in\Z,\,t\in\R$, and uniformly exponentially clustering: there exists $c>0$ and $q>0$ such that
\beq\label{expnential}
	|\bra \la a , \la b\ket| \leq c \, ||\la a||_{\mathfrak U} \,||\la b||_{\mathfrak U}\,\exp\Big[-q \,{\rm dist}(\la a,\la b)\Big]
\eeq
for every $\la a,\la b\in\mathfrak V$.
\end{theorem}

\section{Main results in quantum spin chains}\label{sectresultschain}

We now describe the main results of this work, as specialised to the context of quantum spin chains. The results are obtained in a general framework developed in Sections \ref{sectvect} - \ref{sectcons}, and the exact relation between this framework and quantum spin chains is explained in Section \ref{sectchains}. We believe the general framework has far wider applicability, but we leave this for future works.

In this section we take a thermal state $\omega=\omega_\beta$ for some $\beta\geq 0$, with respect to a finite-range  Hamiltonian with density $\la h\in\mathfrak V$. The proofs of all theorems in this section are given in Subsection \ref{ssectproofschain}.

\subsection{The various completions of the space of local spin chain operators}\label{ssectcompletions}

The space of (strictly) local spin chain operators $\mathfrak V$ is physically relevant but not topologically complete. As usual, it is useful to have completeness in order to establish rigorous results. Starting from $\mathfrak V$ and $\omega$, at least three completions are possible, with different physical interpretations.

First, one may construct the $C^*$ algebra $\mathfrak U$ itself, the completion with respect to the operator norm, as reviewed in Section \ref{sectbasicchain}. In a sense, $\mathfrak U$ is the smallest complete algebra of observables that are ``local enough".

Second, one may instead introduce a sesquilinear form on $\mathfrak V$  defined from the state $\omega$ itself, Eq.~\eqref{omegainner},
\[
	\bra \la a,\la b\ket = \omega(\la a^\dag \la b) - \omega(\la a^\dag)\omega(\la b).
\]
It is positive semi-definite as $\bra \la a,\la b\ket = \omega(f(\la a)^\dag f(\la b))$ where $f(\la a) = \la a - {\bf 1}\omega(a)$. On $\mathfrak V$ it possesses a null space $\mathfrak N$ (which contains at least $\C \1$). The equivalence classes $\la a+\mathfrak N$ ($\la a\in\mathfrak V$) form a new vector space $\lo$. For our purposes we do not need to consider algebraic structures on $\lo$; this is simply a vector space. On $\lo$ the sesquilinear form induces a norm $||\cdot||=\sqrt{\bra \cdot,\cdot\ket}$, with respect to which we complete $\lo$ to a Hilbert space $\hi$. Thus, instead of $\mathfrak V$ and $\mathfrak U$, we have $\lo$ and $\hi$. By boundedness of the state, $||\la a||\leq ||f(\la a)||_{\mathfrak U}\leq 2||\la a||_{\mathfrak U}$, and thus to every element of $\mathfrak U$ corresponds an element of $\hi$ (but, generically, there are elements in $\hi$ that do not correspond to elements in $\mathfrak U$). As the space $\mathfrak V$ of local spin chain operators is countable dimensional, so is $\lo$, and thus $\hi$. Further, because the state $\omega$ and the map $f$ are both continuous and linear, $\iota_x^{\mathfrak U}$ and $\tau_t^{\mathfrak U}$ induce invertible isometries, therefore unitary groups, $\iota_x$ and $\tau_t$ on $\hi$, and $\tau$ forms a strongly continuous unitary representation of $\R$.  In particular, $\bra \tau_t\la a,\la b\ket$ is analytic in $t$ in a neighbourhood of 0 for every $\la a,\la b\in\lo$. For $\la a\in\hi$ we denote $\la a(x,t) = \iota_x \tau_t \la a$ its space-time translate.

The Hilbert space $\hi$ is simply related to the Gelfand-Naimark-Segal (GNS) Hilbert space (see e.g.~\cite{BratelliRobinson12}). The GNS construction is a powerful technique for studying $C^*$ algebras and their representations. Here, $\hi$ is to be interpreted as the smallest complete space of observables that have {\em well-behaved two-point correlations with respect to the state $\omega$}; it cannot be smaller than $\mathfrak U$. We refer to elements of $\lo$ as ``local observables".

Finally,  thanks to clustering, Theorem \ref{golodet}, we may construct yet another Hilbert space, denoted $\hi_0$ (Subsection \ref{ssecthilbert}), also studied\footnote{In \cite{Doyon2017} the space $\hi_0$ was denoted $\hi_\omega$, emphasising its dependence on the state $\omega$; in \cite{DoyonDiffusion2019} it was denoted $\hi'$, emphasising that it is first-order, in contrast with the second order space $\hi''$ also considered there. Further in \cite{DoyonDiffusion2019} the space $\hicons_0$ was denoted $\hi_{\rm bal}$, emphasising that it relates to ballistic modes, in contrast with another space $\hi_{\rm dif}$, relating to diffusive modes.}  in \cite{Doyon2017,DoyonDiffusion2019}. For this purpose, we consider the new positive semi-definite \cite[Lem 4.2]{Doyon2017} sesquilinear form on $\lo$
\beq\label{inneroverview}
	\bra \la a,\la b\ket_0 = \sum_{x\in\Z} \bra \la a(x,0), \la b(0,0)\ket
\eeq
which converges by Theorem \ref{golodet}. Again, it may have a null space $\lon_0$, and we construct the vector space $\lo_0$ of equivalence classes $[\la a]_0 = \la a + \lon_0\; (\la a\in\lo)$, on which there is a norm, $||\la a||_0 = \sqrt{\bra \la a,\la b\ket_0}$. We complete it to a Hilbert space $\hi_0$. Thus, instead of $\lo$ and $\hi$, we have $\lo_0$ and $\hi_0$. Translating structures from the GNS Hilbert space $\hi$ to $\hi_0$ is, however, nontrivial. Clearly, space translations act trivially on $\hi_0$ (by space-translation invariance of the state and clustering). Characterising time translations requires more work. But in quantum spin chains, thanks to the Lieb-Robinson bound, we show that time evolution $\tau$ may be defined to form a strongly continuous one-parameter unitary group on $\hi_0$ (Theorems \ref{lemtauthik} and \ref{theobasicchain}).

The inner product \eqref{inneroverview} is naturally interpreted as a susceptibility, and the Hilbert space $\hi_0$ is the smallest complete space of {\em extensive, thermodynamic observables}. Every local observable $\la a\in\lo$ is a ``density" of the extensive observable
$[\la a]_0\in\hi_0$.

Using these structures, we express three results concerning the large-scale dynamics and hydrodynamics in quantum spin chains. These hold (at least) in every quantum spin chain with finite-range interactions. The results and proofs are completely agnostic to the presence or not of integrability or any other specific feature of the interaction. All proofs are provided in Subsection \ref{ssectproofschain}, and based on the general results established for the general framework in Sections \ref{sectvect} - \ref{sectcons}.


\subsection{Ergodicity}\label{ssectchainergo}

The projection onto conserved charges in the Euler scaling limit needs a process of relaxation to occur at large times. A natural, but weak, expression of relaxation, is the equivalence between time averaging and statistical averaging, or ``ergodicity". This is a subject that has been widely discussed, see e.g.~\cite{klein1952,goldstein2010normal}.

Our first main result is a form of ergodicity. It says that the long-time averaging of correlation functions gives the product of averages, along almost every ray (i.e.~velocity) in space-time. Note that this does not imply ergodicity for averaging in time, along ray $v=0$, but that there exist rays $v$ as near to 0 as desired along which ergodicity holds. This is a universal result about relaxation for dynamical correlation functions, valid for any finite-range quantum spin chain. This result is nontrivial, because ergodicity results in the context of $C^*$ algebras are based on the algebra being asymptotically abelian \cite{ruelle1965quantum,kastler1966invariant,robinson1967extremal}, see also \cite{IsraelConvexity}. For time evolution in quantum spin chains, only the Lieb-Robinson bound is available, where abelianness may be obtained in space-like cones only. Technically, von Neumann's ergodic theorem \cite[Thm 12.44]{RudinFunctional} guarantees, under long-time averages, the projection in the GNS Hilbert space onto the invariant subspace of time translation, but we are not aware of any previous result establishing that this subspace is $\C\1$. We use von Neumann's ergodic theorem combined with countable dimensionality of the operator algebra. This is, we believe, the first real-time ergodicity result in strongly interacting models.

The general theorem on which this is based is Theorem \ref{theorelax}.
\begin{theorem} {\em (Almost-everywhere ergodicity)} \label{theorelaxchain}
Let $\la a,\la b\in\mathfrak U$. Then for almost every $v\in\R$ with respect to the Lebesgue measure,
\beq
	\lim_{T\to\infty} \frc1T\int_{0}^T \dd t\, 
	\omega( \la a(\lfloor v t\rfloor,t) \la b(0,0))
	= \omega(\la a)\omega(\la b).
\eeq
\end{theorem}

The result is expected on physical grounds, and encodes relaxation processes of the many-body system. Referring to linear response, it implies that for almost every $v$, the long-time average, along the ray $v$, of the observable $\la a$ after a small perturbation by $\la b$, reverts to its ensemble average before perturbation: the perturbation is negligible over long times. It also implies a vanishing at large $T$ of the variance of the observable $\frc1T \int_0^T\la a(vt,t)$ with respect to $\omega$: long-time averages along almost all rays are non-fluctuating. We refer to \cite{ampelogiannis2021almost,ampelogiannis2021ergo} for a full discussion and extension to arbitrary dimensions and to arbitrary frequencies and wavenumbers.

\subsection{Hydrodynamic projections}

The second set of results concerns hydrodynamic projections. It says that the Euler-scale correlation functions decompose into the conserved charges admitted by the quantum spin chain.

In order to express these results, we introduce the Fourier transforms of connected correlation functions,
\beq\label{inneromega}
	S_{\la a,\la b}(k,t) = \sum_{x\in\Z} \re^{\ri kx}\big( \omega( \la a^\dag(x,t) \la b(0,0)) - \omega(\la a^\dag)\omega(\la b)\big).
\eeq
We are interested in the large-$t$, small-$k$ limit, with $kt$ fixed. Currently we do not know how to show the existence of this limit, or of any Ces\`aro version of it (long-time averages, or time averages of time averages, etc.), in quantum spin chains. However, we can show that $S_{\la a,\la b}(k,t)$ is uniformly bounded for all $(k,t)\in\R^2$ \footnote{We have the identity $S_{\la a,\la b}(k,t) = \bra \tau_t \la a,\la b\ket_k$ using the inner product \eqref{ink}, which is bounded by $|S_{\la a,\la b}(k,t)|\leq ||\la a||_k ||\la b||_k$ and thus uniformly bounded for all $(k,t)\in\R^2$ by Lemma \ref{lemk}.}. Thus we can use the notion of Banach limit, which assigns a ``limit" value to any bounded function. We describe the particular type of Banach limit we need in Appendix \ref{appbanach}, and unless otherwise stated, all results hold independently from the choice of this limit:
\beq\label{defS}
	S_{\la a,\la b}(\kappa) = \balim_{t\to\infty}  S_{\la a,\la b}(\kappa/t,t).
\eeq
There are Banach limits which give the Ces\`aro limit, $\balim_{t\to\infty} f(t) = \lim_{t\to\infty} t^{-1}\int_{0}^t \dd s\,f(s)$, whenever the latter exists. Our Banach limit is chosen as such. If the Ces\`aro limit exists for $S_{\la a,\la b}(0)$, then, by the conventional Kubo formula, this gives the ``Drude weight" for the observables $\la a,\la b$ (see Subsection \ref{ssectdrude}),
\beq
	\mathsf D_{\la a,\la b} = S_{\la a,\la b}(0)\quad\mbox{(Ces\`aro limit).}
\eeq

We then define the space of conserved charges $\hicons_0$ as the set of elements of $\hi_0$ that are $\tau_t$-invariant: $\hicons_0 = \{\la q\in\hi_0 : \tau_t\la q = \la q \;\forall\;t\in \R\}$ (see Subsection \ref{ssectcons}). The space $\hicons_0$ is a closed subspace, and we define the orthogonal projection
\beq\label{introproj}
	\mathbb P : \hi_0 \to \hicons_0.
\eeq
A density $\la q\in\lo$ of $[\la q]_0\in\hicons_0$ is a local conserved density, and such $[\la q]_0$ form the subspace of local conserved charges $\hicons_0^{\rm loc} \subset \hicons_0$. By \cite[Thm 5.2]{Doyon2017}, there is a bijection relating $\hicons_0$ to the space of conserved ``pseudolocal charges" \cite[Def 5.1]{Doyon2017}, quantities discussed in \cite{ProsenPseudo1,ProsenPseudo2,ProsenQuasilocal1,ProsenQuasilocal2} in the context to non-equilibrium states and relaxation, see the review \cite{IlievskietalQuasilocal}; although we do not need this bijection here.

We note that as $\hicons_0$ has at most countable dimensionality, we may choose a basis $\{\la q_i:i\in\N\}\subset \hicons_0$, and form the (possibly infinite-dimensional) matrix $\mathsf C_{ij} = \bra \la q_i,\la q_j\ket_0$, which is positive-definite and invertible, and we have
\beq\label{projCchain}
	\mathbb P\cdot = \sum_{ij} \la q_i \mathsf C^{ij}
	\dbra \la q_j,\cdot\dket_0
\eeq
where $\mathsf C^{ij}$ is the inverse matrix.

By Theorem \ref{theodens0} and Remark \ref{remcons} below, any local spin chain operator $\la q\in\mathfrak V$ that satisfies a continuity equation with local current $\la j$, Eq.~\eqref{contintro}, gives rise to a local conserved density according to our definition. Of course, the quantum chain's energy density $\la h$ is a conserved density; in integrable models, infinitely many conserved densities can be constructed by Bethe ansatz methods \cite{faddeev1996algebraic}, and thus $\hicons_0$ is infinite-dimensional.

Our results Theorem \ref{theochain} show that $\hicons_0$ may indeed be seen as the space of emergent ballistic modes of the theory, which carry correlations between local observables. Again, these are universal results, which we prove in particular for every quantum spin chain with finite range interactions. Using \eqref{projCchain}, Points I and II take their standard forms found in the literature \cite{SciPostPhys.3.6.039,doyoncorrelations}. Points I and II specialise Theorem \ref{theodrude}, and Theorems \ref{theocont} and \ref{main}, respectively. We believe that Point I was already well understood in the literature, however Point II is proven here for the first time.
\begin{theorem} {\em (Hydrodynamic projection)} \label{theochain} The following statements hold:
\bi
\item[I.] For every $\la a,\la b\in\hi_0$, the Drude weight exists (that is, the Ces\`aro limit for $S_{\la a,\la b}(0)$ exists), and satisfies
\beq\label{intro2}
	\mathsf D_{\la a,\la b} = \lim_{t\to\infty} \frc1{t} \int_{0}^t \dd s\,
	S_{\la a,\la b}(0,s) = \mathsf D_{\mathbb P\la a,\mathbb P\la b}.
\eeq
\item[II.] For every $\kappa\in\R$, the sesquilinear function $S_{\la a,\la b}(\kappa)$ of $(\la a,\la b)\in\mathfrak V\otimes \mathfrak V$ is bounded by $||\la a||_0\,||\la b||_0$, and can be extended to a unique sesquilinear function, also denoted $S_{\la a,\la b}(\kappa)$, of $(\la a,\la b)\in\hi_0\times \hi_0$. For every $\la a,\la b\in\hi_0$, the Euler-scale correlation function satisfies
\beq\label{hpf}
	S_{\la a,\la b}(\kappa) = S_{\mathbb P\la a,\mathbb P\la b}(\kappa).
\eeq
\ei
\end{theorem}

The result Point II is a hydrodynamic projection formula. The projection $\mathbb P$ represents the amount of ballistic wave produced by the local observables, and $S_{\mathbb P\la a,\mathbb P\la b}(\kappa)$, which involves only the conserved charges projected onto, represents the ballistic correlations due to the propagating ballistic waves. In particular, we find that Euler-scale correlation functions have good continuity properties with respect to the Hilbert space $\hi_0$. Physically, this means that these correlation functions, for any value of $\kappa$, are determined by the behaviour of the system at zero wavenumber $k=0$.  At $\kappa=0$, we recover the projection formula for the Drude weights: the saturated Mazur bound or Suzuki equality \cite{mazur69,suzuki1971ergodicity,vankampen1971ergodic,CasZoPre95,ZoNaPre97,SciPostPhys.3.6.039}. For $\kappa\neq0$, projection occurs thanks to almost-everywhere ergodicity.

\subsection{Currents and linearised Euler equations}

We prove the linearised Euler equations themselves as an application of the general hydrodynamic projection result. Contrary to traditional linear-response arguments, the linearised Euler equations are seen to arise not by state perturbations, but by hydrodynamic projections  (see the discussion in Subsection \ref{ssectconsdens}).

A nontrivial aspect of obtaining the Euler equations is to establish the existence of currents, with a continuity equation, for every local conserved density. Although local spin chain operators satisfying \eqref{contintro} give local conserved densities, there is no guarantee that a local conserved density $\la q$ (i.e.~$[\la q]\in\hicons_0^{\rm loc}$) possesses a current $\la j\in\lo$ with a continuity equation \eqref{contintro}. We find, however, that it is possible to extend the space of local observables in such a way that every local conserved density indeed has a local current and satisfies \eqref{contintro}. This extension holds at the level of the GNS space; thus we adjoin to $\lo$ elements of $\hi$. This, again, is a universal result, valid for every finite-range quantum spin chain. The general theorem for the existence of currents is Theorem \ref{theocurrent}.

Combining the existence of local currents and the main hydrodynamic projection theorem, we obtain the linearised Euler equation. This is a specialisation of the main aspects of the general Theorem \ref{theoeuler}.
\begin{theorem} {\em (Linearised Euler equation)} \label{theoeulerchain} The following statements hold:
\bi
\item[I.] It is possible to extend the definition of local observables to $\lo^\#$ with $\lo\subset \lo^\#\subset \hi$ in such a way that (1) all time-evolutes are local, $\tau_t (\lo^\#) \subset \lo^\#$ for all $t\in\R$; (2) correlation functions of local observables vanish faster than any power law: for every $\la a,\la b\in\lo^\#,\,p>0$ there is $u>0$ such that $|\bra \iota_x \la a,\la b\ket| \leq u (|x|+1)^{-p}\;\forall x\in\Z$; (3) every  local conserved density $\la q\in\lo^\#$ has an associated local current $\la j\in\lo^\#$, such that a continuity equation holds:
\beq\label{qjchain}
	\frc{\p}{\p t} \la q(x,t) + \la j(x,t) - \la j(x-1,t) = 0.
\eeq
\item[II.] Let $\{\la q_i\}$ form a basis for $\hicons_0$. For every local conserved densities $\la q, \la q'\in\lo^\#$ and every $\kappa\in\R$, the following derivative exists and gives:
\beq\label{dSdkintro}
	\frc{\dd S_{\la q,\la q'}(\kappa)}{\dd \kappa} = \ri S_{\la j,\la q'}(\kappa)
	= \ri \sum_k \mathsf A^{k} S_{\la q_k,\la q'}(\kappa),\qquad
	\mathsf A^{k} = \sum_l\bra \la j,\la q_l\ket_0 \mathsf C^{lk}
\eeq
where $\la j$ is the current associated to $\la q$.
\ei
\end{theorem}
Differentiability in $\kappa$ is a nontrivial part of the statement in Part II. We have not shown that there necessarily exists a local basis $\{[\la q_i]_0\}\subset \hicons_0^{\rm loc}$ for $\hicons_0$. But if there is, then we may apply the result for this basis, with $\la q=\la q_i,\,\la q' = \la q_j,\,\la j = \la j_i$, and we have a closed set of evolution equations for $S_{\la q_i,\la q_j}(\kappa)$. The (possibly infinite-dimensional) matrix
\beq
	\mathsf A_i^{~k} = \sum_l\bra \la j_i,\la q_l\ket_0 \mathsf C^{lk}
\eeq
is the flux Jacobian, and this expression for $\mathsf A_i^{~k}$ is a standard form that agrees with the definition by state variation \cite{SciPostPhys.3.6.039}.

Note that the heuristic version of the linearised Euler equation \eqref{eulerheuristic} essentially implies \eqref{dSdkintro}. We believe our result is the first rigorous formulation and proof of a form of the linearised Euler equation in unitary, interacting many-body quantum systems.

\section{Space-time symmetries and ergodicity} \label{sectvect}

In this section and the following sections, we consider a general framework, in which general results are established. These results ultimately lead to the main results for quantum spin chains expressed above, proved in Section \ref{sectchains}.

We consider a countable-dimensional Hilbert space $\hi$ with inner product $\bra\la a,\la b\ket$ and norm $||\la a|| = \sqrt{\bra \la a,\la a\ket}$ ($\la a,b\in\hi$). As illustrated in the previous section, this Hilbert space has the interpretation as a space of observables within a thermodynamically large physical system, and the inner product, as the connected correlation function of two observables. We express below the minimal requirements on $\hi$, including the presence of groups of unitary space-time translations operators. We then consider the notion of ergodicity. We show that ergodicity along almost all rays in space-time must hold if a weak version of ergodicity along all space-like rays holds. This will play a crucial role in the hydrodynamic projection mechanism described in Section \ref{sectproof}.

\subsection{Space-time translations}\label{ssectspacetime}

Homogeneous and stationary states of thermodynamically large systems are invariant under space and time translations. In the terms of the Hilbert space $\hi$, this translates into the presence of corresponding groups of unitary operators.

We assume that for every $x\in\Z$, there is an invertible linear isometry, equivalently a unitary map, $\iota_x : \hi\to\hi$
\beq\label{homo}
	\bra \la \iota_x\la a, \iota_xb\ket=\bra \la a,\la b\ket\qquad \forall \ \la a,\la b\in\hi,\ x\in\Z.
\eeq
We assume that this set of maps forms a representation of the group $\Z$, with $\iota_{x+y} = \iota_x\iota_y$ and $\iota_0=1$. This is interpreted as the group action of space translations, and \eqref{homo} is the condition of homogeneity of the state.

We likewise assume that for every $t\in\R$ there is a unitary map $\tau_t: \hi\to\hi$,
\beq\label{stat}
	\bra \la \tau_t\la a, \tau_tb\ket=\bra \la a,\la b\ket\qquad \forall \ \la a,\la b\in\hi,\ t\in\R.
\eeq
We also assume that this set of maps forms a representation of the group $\R$, with $\tau_{t+s} = \tau_t\tau_s$ and $\tau_0=1$. This is interpreted as the group of time translations, or time evolution, and \eqref{stat} is the condition of stationarity of the state. For technical reasons, it is convenient to assume that for every $\la a,\,\la b\in\hi$, the function $\bra \tau_t\la a,\la b\ket$ is Lebesgue measurable on $t\in\R$. We also assume that time evolution is homogeneous,
\beq\label{homostat}
	\iota_x\tau_t = \tau_t\iota_x\qquad  \forall\ x\in\Z,\;t\in\R.
\eeq

\begin{rema} \em Here space is taken to be discrete, and time continuous. This is adapted to applications to quantum spin chains. However the theory developed can also be applied to systems with continuous space $\R$, by simply concentrating on a discrete subset $\Z$. Further, many aspects of the theory do not require time to be continuous; however we leave this for future works.Continuity properties in $t$ are discussed in Subsection \ref{ssectstrong}.
\end{rema}

\subsection{An ergodicity theorem} \label{ssectdyn}

We consider the notion of ergodicity in terms of correlation functions: the vanishing of the large-scale averaged connected correlation. We provide a full proof of ergodicity in almost all directions in space-time under the sole condition that ergodicity holds within a space-like neighbourhood. That is, ergodicity in space-like cones, a property which is similar to but much weaker than the Lieb-Robinson bound, is sufficient in order to conclude that ergodicity holds along almost every ray. In particular, the latter conclusion then applies to quantum spin chains (thanks to the Lieb-Robinson bound), see Section \ref{sectchains}.

Ergodicity is naturally studied using spectral theory. In particular, the decomposition structure of ergodic measures translates, in the context of $C^*$ algebras, into a decomposition theory for states of many-body systems. There are strong, general ergodicity results for topological symmetry groups within any extremal state, under the condition that the algebra of observables be asymptotically abelian. See the papers \cite{ruelle1965quantum,kastler1966invariant,robinson1967extremal} as well as the beautiful discussion in the book \cite{IsraelConvexity}. However, in many-body systems, the strongest bound on asymptotic abelianness we are aware of, concerning the time translation group, are those related to the Lieb-Robinson bound \cite{LiebRobinson}. Under this, the algebra is seen to be asymptotically abelian only within space-like cones. The theorem below provide the first result for ergodicity along time-like rays.

Let us denote the extended reals by $\h\R = \R\cup\{\infty\}$.
\begin{defi} \label{defspacelikeergo} The triplet $\hi,\tau,\iota$ (or simply $\hi$) is space-like ergodic if there exists a dense subspace $\lo\subset \hi$ and $v_{\rm c}>0$ such that, for every $\la a,\la b\in\lo$, every $r\in\Z\setminus\{0\}$ and every $v\in\h\R$ with $|v|>v_{\rm c}$, we have
\beq\label{sumproj0}
	\lim_{N\to\infty} \frc1N \sum_{n=1}^N \bra \iota_{r}^n\tau_{v^{-1}r}^n\la a,\la b\ket = 0.
\eeq
\end{defi}
Note that $\iota_{r}\tau_{v^{-1}r}$ is a unitary operator representing a discrete translation in space-time along the ray of velocity $v$ (that is, along the line $x=vt$), and that $v=\infty$ is allowed, and gives translations in space. In applications, the subspace $\lo$ may be taken as the subspace of local observables. In quantum spin chains, the limit of the summand in \eqref{sumproj0} vanishes for every local observables $\la a,\la b$, so that Definition \ref{defspacelikeergo} holds, with any $v_{\rm c}$ larger than the Lieb-Robinson velocity (see Theorem \ref{theobasicchain}).

In Eq.~\eqref{limtheorelax} below, note that the integrand  is integrable over $[0,T]$, as it is measurable (by our framework's assumption) and bounded (as $|\bra \iota_{\lfloor v t\rfloor} \tau_t\la a,\la b\ket|\leq ||\la a||\,||\la b||$). The statement is that the limit exists and vanishes.
\begin{theorem}\label{theorelax}
Assume that $\hi$ is space-like ergodic (Definition \ref{defspacelikeergo}). Let $\la a,\la b\in\hi$. Then for almost all $v\in\R$ with respect to the Lebesgue measure,
\beq\label{limtheorelax}
	\lim_{T\to\infty} \frc1T\int_{0}^T \dd t\, \bra \iota_{\lfloor v t\rfloor} \tau_t\la a,\la b\ket = 0.
\eeq
\end{theorem}
For every $v\in\h\R\setminus\{0\}$, we denote the unitary operator
\beq
	\sigma_v = \iota_1 \tau_{v^{-1}},
\eeq
and we denote $\sigma_0 = \tau_{1}$; again, these represent translations in space-time along the rays $x=v t$. For every $r\in\Z$ and $v\in\h\R$, we denote by $\mathbb P_{\sigma_v^{r}}$ the orthogonal projection onto the null space $\ker({\sigma_v^{r}}-1)$ of $\sigma_v^r-{1}$. Its range $\ran \mathbb P_{\sigma_v^{r}}$ is equal to $\ker(\sigma_v^{r}-1)$, and is different from $\{0\}$ if, and only if, the spectrum of $\sigma_v^r$ contains the unit eigenvalue: there exists $\la a\in\hi$ with $||\la a||=1$ such that $\sigma_v^r \la a = \la a$.
\begin{lemma} \label{lemrelax1} For all $v\in\h\R$ with $|v|>v_{\rm c}$, and all $r\in\Z\setminus\{ 0\}$,
\beq
	\ran \mathbb P_{\sigma_v^{r}} = \{0\}.
\eeq
\end{lemma}
\proof By space-like ergodicity, for every $\la a,\la b\in\lo$ and $|v|> v_{\rm c}$, the following limit exists and vanishes:
\beq\label{sumproj}
	\lim_{N\to\infty} \frc1N \sum_{n=1}^N \bra \sigma_v^{rn}\la a,\la b\ket = 0.
\eeq
By von Neumann's ergodic theorem \cite[Thm 12.44]{RudinFunctional}, the result of the limit gives rise to the orthogonal projection onto the kernel, hence
\beq
	\bra\mathbb P_{\sigma_v^{r}}\la a,\la b\ket = 0.
\eeq
By continuity and density of $\lo\in\hi$ this is extended to all $\la b\in\hi$, thus $\mathbb P_{\sigma_v^{r}}\la a=0$; and by continuity of the projection this is extended to all $\la a\in\hi$.
\eproof

\begin{lemma} \label{lemrelax2} Let $v,w\in\h\R$ with $v\neq w$, and let $\la a\in\ker(\sigma_v-1)$ and $\la b\in\ker(\sigma_w-1)$. Then $\bra \la a,\la b\ket = 0$.
\end{lemma}
\proof Without loss of generality we may choose $w\neq\infty$ and $v\neq 0$. Then $w,v^{-1}\in\R$.

First, assume $w=0$. Then for all $p,q\in\Z$, we have
\beq
	\bra \la a,\la b\ket = \bra \la a,\tau_{p}\la b\ket
	= \bra \tau_{-p}\la a,\la b\ket
	= \bra \iota_q \tau_{qv^{-1}-p}\la a,\la b\ket.
\eeq
For every $v^{-1}\in\R$, it is possible to find $p,q\in\Z$ with $q> 0$ such that $|qv^{-1}-p|<qv_{\rm c}^{-1}$. Indeed, choose $q>v_{\rm c}$, and then choose $p = \lfloor qv^{-1}\rfloor$. Therefore, we find
\beq
	\bra \la a,\la b\ket
	= \bra \sigma_z^q\la a,\la b\ket,\quad
	z^{-1} =v^{-1}-p/q,\quad |z|>v_{\rm c}.
\eeq
By repeated use of this formula, we obtain a normalised sum as in \eqref{sumproj} and use the ergodic theorem. As a consequence,
\beq
	\bra \la a,\la b\ket = \bra \mathbb P_{\sigma_z^q}\la a,\la b\ket = 0
\eeq
where the last equality holds by Lemma \ref{lemrelax1}. 

Second, assume $w\neq 0$. As the rationals $\mathbb Q$ are dense in $\R$, for every $\eta>0$ there exists infinitely many $p/q\in\mathbb Q$ lying in $[wv^{-1},w(v^{-1}+\eta)]$. Choose $0<\ep<\frc{|1-w/v|}{v_{\rm c}+|w|}$ such that
\beq
	w(v^{-1}+\ep) = p/q\in\mathbb Q,
\eeq
with $p,q\in\Z\setminus\{0\}$ and $p\neq q$. Then
\beq
	\bra \la a,\la b\ket = \bra \sigma_v^q \la a,\la b\ket
	= \bra \tau_{-q\ep}\iota_q \tau_{q(v^{-1}+\ep)} \la a,\la b\ket
	= \bra \iota_{q-p}\tau_{-q\ep}\iota_p \tau_{pw^{-1}} \la a,\la b\ket
	= \bra \iota_{q-p}\tau_{-q\ep}\la a,\sigma_w^{-p}\la b\ket
	= \bra \sigma_z^{q-p}\la a,\la b\ket
\eeq
where $z = -(q-p)/(q\ep) = -\ep^{-1} (1-w/v)+ w$. Clearly, $|z|>\ep^{-1}|1-w/v| - |w| > v_{\rm c}$ by the assumptions on $\ep$. Hence
\beq
	\bra \la a,\la b\ket = \bra \mathbb P_{\sigma_z^{q-p}}\la a,\la b\ket = 0
\eeq
where again the last equality follows from Lemma \ref{lemrelax1}.
\eproof

\medskip
\proofof{Theorem \ref{theorelax}} Theorem \ref{theorelax} follows from a set-theoretic argument. Consider the set $\Omega = \{v\in\R : \ran \mathbb P_{\sigma_v}\neq \{0\}\}$. For every $v\in\Omega$, choose $\la a_v\in \ran \mathbb P_{\sigma_v}$ with $||\la a_v||=1$. Clearly, the completion of $\spa\{\la a_v:v\in\Omega\}$ is a closed subspace of $\hi$. Further, by Lemma \ref{lemrelax2}, we have $\bra \la a_v,\la a_w\ket = 0$ for every $v,w\in\Omega$ with $v\neq w$. Hence $\Omega$ parametrises a set of linearly independent unit vectors spanning a closed subspace of $\hi$. As $\hi$ is countable dimensional (by our framework's assumption), the set $\Omega$ must be countable. Therefore, it has Lebesgue measure zero in $\R$. Now choose $v\in\R$, and consider \eqref{limtheorelax}. For the theorem, it is sufficient to assume that $v\neq 0$. Let $\la a,\la b\in\hi$. First take $v>0$. Then
\beqa
	\lim_{T\to\infty} \frc1T\int_{0}^T \dd t\, \bra  \iota_{\lfloor v t\rfloor}\tau_t\la a,\la b\ket
	&=&
	\lim_{\N\ni N\to\infty} \frc1N\int_{0}^N \dd x\, \bra  \iota_{\lfloor x\rfloor}\tau_{xv^{-1}}\la a,\la b\ket \n
	&=&
	\lim_{N\to\infty} \frc1N \sum_{n=0}^{N-1} \int_{0}^1 \dd y\, \bra  \iota_n\tau_{nv^{-1}}\tau_{yv^{-1}}\la a,\la b\ket \n
	&=&
	\lim_{N\to\infty} \frc1N \sum_{n=0}^{N-1} \int_{0}^1 \dd y\, \bra  \sigma_{v}^n\tau_{yv^{-1}}\la a,\la b\ket \n
	&=&
	\int_{0}^1 \dd y\,\bra \mathbb P_{\sigma_v} \tau_{yv^{-1}}\la a,\la b\ket
\eeqa
where the first equality, on whose right-hand side the limit is taken over the integers instead of the reals, follows by boundedness of the integrand. In the second equality we have separated the integral over $[0,N]$ into a sum over $n=0,1,\ldots,N-1$ of integrals over $y\in[0,1)$, writing $x = n+y$; and in the last, we used the bounded convergence theorem in order to exchange the limit of the sum and the integral, and von Neuman's ergodic theorem in order to write the large-$N$ average as a projection. For the case $v<0$, write $\iota_{\lfloor v t\rfloor} = \iota_{\lfloor -v t\rfloor +1}^{-1}$ (valid almost everywhere on $t\in[0,T]$). This gives $\int_{0}^1 \dd y\,\bra \mathbb P_{\sigma_{v}} \iota_{-1}\tau_{y|v|^{-1}}\la a,\la b\ket$. The result is zero whenever $v\in\R\setminus \Omega$, by definition of $\Omega$. As $\Omega$ has measure zero, the theorem follows.
\eproof

\section{Clustering and the space of conserved charges}\label{sectcharges}

The Hilbert space structure of Section \ref{sectvect} is not sufficient in order to describe the space of conserved charges onto which hydrodynamic projection occurs. We need an additional structure: that of a dense subspace
\beq
	\lo\subset\hi.
\eeq
Physically, the subspace $\lo$ is that of observables, within $\hi$, which possess stronger locality properties. In the context of quantum spin chains, these may be identified with the set of operators whose supports are finite subsets of $\Z$: the local operators (see Sections \ref{sectresultschain} and \ref{sectchains}). They may also be identified with quasi-local operators (see e.g.~\cite{IlievskietalQuasilocal}). The Hilbert space $\hi$ is the completion of $\lo$ with respect to $||\cdot||$.

The choice of the subspace $\lo$ affects the various structures constructed below, but the theory holds for any such choice which satisfies all the required properties. In this section, we describe the requirements on $\lo$ and $\hi$ and define the basic structures, including the family of new Hilbert spaces $\hi_k,\,k\in\R$, and the space of conserved charges $\hicons_0$. We express a variety of fundamental results that will be of use afterwards.

The main requirements on $\lo$ and $\hi$ are expressed in Subsection \ref{ssectclus}: observables in $\lo$ should cluster strongly enough in space. If strong continuity and stronger properties of clustering hold on time evolution, then stronger results can be obtained; these properties are explained in Subsection \ref{ssectstrong}. In Subsection \ref{ssecthilbert} we study the various Hilbert spaces $\hi_k$ used to describe large-scale behaviours, and in Subsection \ref{ssectcons} we define, using these, the subspace of conserved quantities. The conserved quantities will be identified with the emergent degrees of freedom in the hydrodynamic projection formulae. All requirements on $\hi,\lo$ are established in the context of quantum spin chains in Section \ref{sectchains}.

For brevity, we refer to $\hi$, $\lo$, $\{\tau_t:t\in\R\}$ and $\{\iota_x:x\in\Z\}$, or simply $\hi, \lo$, with the properties that $\lo\subset\hi$ is dense and that all requirements expressed around \eqref{homo} - \eqref{homostat} are satisfied, as a {\em dynamical system}.

\begin{rema}\em
In applications, space translations often map local observables into themselves, $\lo\to\lo$, and one naturally defines them simply as invertible isometries on $\lo$. Since an invertible isometry $\iota_x:\lo\to\lo$ is continuous on $\lo$, and therefore has a unique continuous extension to $\hi$, in \eqref{homo} we define it on $\hi$ already. [For the continuity statement: if $\iota_x:\lo\to\lo$ is an invertible isometry, let $\la a = \lim_n \la a_n$ and $\la a,\la a_n,\la b\in\lo$. Then $\lim_n \bra \iota_x\la a_n,\la b\ket = \lim_n \bra \la a_n,\iota_x^{-1}\la b\ket = \bra \la a,\iota_x^{-1}\la b\ket = \bra \iota_x\la a,\la b\ket$ and since $||\iota_x\la a_n|| = ||\la a_n||$ is bounded and $\lo$ is dense this implies $\lim_n \iota_x\la a_n = \iota_x \la a$, thus $\iota_x$ is continuous].
\end{rema}

\begin{rema}\label{remdense}\em
In applications, time evolution often does not map local observables into themselves. But $\tau_t\la a\in\hi$ for every $\la a\in\lo$, and our assumptions imply that $\lo$ is dense in $\hi$. That is, although $\tau_t\la a$ might not be a local observable, we assume that it is possible to approximate it with arbitrary precision by local observables.
\end{rema}

\subsection{Clustering}\label{ssectclus}

On the the dynamical system $\hi,\lo$, we will require certain  clustering properties with respect to space translations. We define clustering in a weak enough fashion, as follows.
\begin{defi} \label{defclus0} We say that a pair $(\la a,\la b)\in\hi\times\hi$ is $p$-clustering, respectively a subset $\mathcal C \subset\hi\times\hi$ is uniformly $p$-clustering, for some $p>0$, if there exists $c>0$ such that
\beq\label{clusform}
	|\bra \iota_x\la a,\la b\ket| \leq  c(|x|+1)^{-p}\quad \forall\ x\in\Z,\ \mbox{resp.}\ 
	\forall\ x\in\Z \mbox{ and }(\la a,\la b)\in\mathcal C.
\eeq
If $(\la a,\la b)\in\hi\times\hi$ is $p$-clustering for some $p>1$, then for every $k\in\R$ we define
\beq\label{ink}
	\dbra \la a,\la b\dket_k = \sum_{x\in\Z} \re^{\ri kx} 
	\bra \iota_x\la a,\la b\ket
\eeq
(the series converges).
\end{defi}
In most situations, one might expect that it be too strong to ask for every pair in $\hi\times\hi$ to be $p$-clustering for $p>1$, as elements of $\hi$ that are obtained by completion from local elements might not have good enough clustering properties anymore. However, for time translations of local elements in $\lo$, clustering is a natural condition. The space $\lo$ must in fact be a ``good" subspace of local elements: appropriate uniformity of clustering is important for $\tau_t$ to give rise to a unitary group on the Hilbert spaces we construct below. Therefore, let us define
\beq\label{Vhat}
	\h\lo= {\rm span}\{\tau_t\la a:\la a\in\lo,\,t\in\R\}\subset\hi.
\eeq
We assume that {\em there exists a strict lower bound $p_{\rm c}\geq 1$ for the clustering power law}, such that every pair in $\h\lo$ is $p$-clustering for some $p>p_{\rm c}$. Further, we assume that for every element $\la a\in\h\lo$, we may associate a converging sequence $\sigma_n\la a\in\lo$, with $\lim_n \sigma_n\la a = \la a$, such that for every pair of elements in $\h\lo$, $p$-clustering, for some $p>p_{\rm c}$, holds uniformly on the set of pairs of elements of their associated converging sequences ($\sigma_n$ may be seen as a linear operator $\h\lo\to\lo$). All this is expressed in the following definition.
\begin{defi}\label{defclus} We say that the dynamical system $\hi,\lo$ is $p_{\rm c}$-clustering if, to every $\la a\in\h\lo$, we may associate a sequence $\sigma_n\la a\in\lo$ with
\[
	\lim_n \sigma_n\la a = \la a,
\]
such that for every $\la a,\la b\in\h\lo$, the set of pairs $\{(\sigma_n{\la a},\sigma_m\la b)\}$ is uniformly $p$-clustering for some $p>p_{\rm c}$. For every $\la a\in\lo$ we take $\sigma_n \la a = \la a\;\forall\;n$.
\end{defi}
Clearly, if the dynamical system is $p_{\rm c}$-clustering, then it is $p_{\rm c}'$-clustering for every $p_{\rm c}'\in[0,p_{\rm c}]$. We will say that it is $\infty$-clustering if it is $p_{\rm c}$-clustering for every $p_{\rm c}\in[0,\infty)$. In the rest of this section, we assume that:
\[
	\mbox{the dynamical system $\hi,\lo$ is $1$-clustering.}
\]

We now establish simple fundamental lemmas from the above structure.
\begin{lemma}\label{lemform} Let $k\in\R$, let $\la a\in \hi$, and assume that the pair $(\la a,\la a)$ is $p$-clustering for some $p>1$. Then $\dbra \la a,\la a\dket_k$ is non-negative:
\beq\label{pos}
	\dbra \la a,\la a\dket_k \geq 0.
\eeq
\end{lemma}
\proof See for instance the proof of \cite[Lem 4.2]{Doyon2017}. Consider the observables $\la b = \sum_{x=-L}^L \re^{\ri k x} \iota_x \la a\in\hi$, for $L\in\N$. Then we have
\beq
	0\leq\bra \la b,\la b\ket = \sum_{x,y=-L}^L
	\re^{-\ri k (x-y)}\bra
	\la \iota_x \la a, \la \iota_ya\ket 
	= \sum_{x=-2L}^{2L}
	(2L+1-|x|)\,
	\re^{-\ri k x}
	\bra \la \iota_x\la a, \la a\ket.
\eeq
We write $1-|x| = 2-(|x|+1)$. Clearly,
\beq
	\Big|\sum_{x=-2L}^{2L} 2\, \re^{-\ri kx}\bra \iota_x \la a,\la a\ket\Big|\leq 2c\sum_{x=-\infty}^{\infty} (|x|+1)^{-p} = O(L^0).
\eeq
Further,
\beq
	\Big|\sum_{x=-2L}^{2L} (|x|+1)\, \re^{-\ri kx}\bra \iota_x \la a,\la a\ket\Big| \leq c\sum_{x=-2L}^{2L} (|x|+1)^{-p+1} = O(L^{2-p}).
\eeq
Therefore, $0\leq \bra \la b,\la b\ket = 2L\bra \la a,\la a\ket_k + O(L^{2-p},L^0)$. The result \eqref{pos} is obtained by dividing by $L$ and taking the limit $L\to\infty$.
\eproof

\begin{lemma}\label{lemk} Let $(\la a, \la b)\in\hi\times \hi$ be $p$-clustering for some $p>1$. Then $\bra \la a,\la b\ket_k$ is uniformly bounded on $k\in\R$, and
\beq
	\lim_{k\to0} \dbra \la a,\la b\dket_k = \dbra \la a,\la b\dket_0.
\eeq
\end{lemma}
\proof For every $x\in\Z$, we have $\lim_{k\to0} \re^{\ri kx}\bra\iota_x\la a,\la b\ket = \bra\iota_x\la  a,\la b\ket$. We bound the summand in $\sum_{x\in\Z} \re^{\ri kx}\bra\iota_x\la a,\la b\ket$ by $|\bra \iota_x\la a,\la b\ket|$. The latter is summable by $p$-clustering for $p>1$. Hence by the bounded convergence theorem, we have $\lim_{k\to0} \sum_{x\in\Z} \re^{\ri kx}(\iota_x\la a,\la b) = \sum_{x\in\Z} (\iota_x\la a,\la b)$.
\eproof

\begin{rema}\label{remstable}\em
It is simple to see that $\h\lo$ itself can be chosen as the space of local observables, and $\hi$, $\h\lo$, $\{\tau_t:t:\in\R\}$, $\{\iota_x:x:\in\Z\}$ is a dynamical system. In this case, the set of local observables is stable under time evolution. In particular, for this system to be $p_{\rm c}$-clustering, Definition \ref{defclus}, it is sufficient to require $p$-clustering pairwise (there is no need for uniformity), as in this case one can choose $\sigma_n{\tau_t \la a} = \tau_t \la a$ for all $n$. Using $\hi,\h\lo$ for the dynamical system simplifies the discussion, and we will take $\lo = \h\lo$ in Section \ref{sectcons}. This however affects certain structures, for instance changing the meaning of the Hilbert spaces $\hi_k$ introduced below. In Sections \ref{sectcharges} and \ref{sectproof} we keep the separation between $\lo$ and $\h\lo$ for generality.
\end{rema}

\subsection{Strongly continuous one-parameter groups}\label{ssectstrong}

Besides the assumptions expressed in Subsections \ref{ssectspacetime} and \ref{ssectclus}, it is often the case that finer properties of time evolution holds, and this leads to finer statements about the Hilbert spaces constructed below, and the subspace of conserved charges. Although these are not necessary in order to establish our main projection theorem, it is useful to consider such finer properties.

Assume that $\tau_t\la a$ is continuous in $t$ with respect to the norm topology for every $\la a\in\hi$ and $t\in\R$. Then $\{\tau_t:t\in\R\}$ forms what is called a strongly continuous one-parameter unitary group -- we will simply say that ``$\tau$ is strongly continuous".  As a consequence, Stone's theorem  \cite[Thm 13.35]{RudinFunctional} implies that there is an anti-self-adjoint operator $\delta$, the generator of the group, which is not necessarily continuous, with (dense) domain $\lo'\subset\hi$ such that $\tau_t\la a\in \lo'$ and $\tau_t\la a$ is differentiable for every $\la a\in\lo'$, and
\beq\label{derivativetaut}
	\frc{\dd}{\dd t} \tau_t\la a = \tau_t\delta \la a = \delta \tau_t\la a.
\eeq
Stationarity then implies
\beq\label{sym}
	\bra \delta \la a,\la b\ket = -\bra \la a,\delta \la b\ket\quad
	\forall\ \la a,\la b\in\lo'
\eeq
Below, when assuming that $\tau$ is strongly continuous, we will also assume that 
\beq\label{deltaass}
	\lo\subset \lo'\quad \mbox{and}\quad\delta(\lo)\subset \lo.
\eeq

In order for the presence of a strongly continuous one-parameter unitary group to lead to finer results below, strong enough clustering properties are required. It is convenient to give precise definitions here, that can be proven in particular situations (such as in spin chains); this will make the requirements for the finer results below clearer. We consider two such stronger clustering properties: {\em continuous clustering} and {\em differentiable clustering}.
\begin{defi}\label{deficlusgroup}
The following definitions apply with respect to the $p_{\rm c}$-clustering dynamical system $\hi,\lo$, with time evolution $\tau$.

We say that $\tau$ is continuously clustering if for every $\la a,\la b\in\lo$, there is $\ep>0$ such that the family $\{(\tau_t\la a,\la b):t\in [-\ep,\ep]\}$ is uniformly $p$-clustering for some $p>p_{\rm c}$.

We say that $\tau$ is differentiably clustering if it is strongly continuous (in particular Eq.~\eqref{deltaass} holds) and continuously clustering, and if for every $\la a,\la b\in\lo$, there is $\ep>0$ such that both families $\{(t^{-1}(\tau_t-1)\la a,\la b):t\in [-\ep,\ep]\}$ and $\{(t^{-2}(\tau_t+\tau_{-t}-2)\la a,\la a):t\in [-\ep,\ep]\}$ are uniformly $p$-clustering for some $p>p_{\rm c}$.
\end{defi}
Note that, by strong continuity, $t^{-1}(\tau_t-1)\la a$ and $t^{-2}(\tau_t+\tau_{-t}-2)\la a$ for $t\in[-\ep,\ep]$ are continuous families of elements of $\hi$, hence the condition for differentiable clustering makes sense; in particular, at $t=0$ we get $\delta \la a$ and $\delta^2\la a$, respectively. Note also that differentiable clustering does not follow from continuous clustering: for instance, if $\tau$ is continuously clustering, then so is $\tau-1$, and thus there is a uniform power $p>p_{\rm c}$ for the clustering of $(t^{-1}(\tau_t-1)\la a,\la b)$; however, because of the factor $t^{-1}$, there isn't necessarily a uniform coefficient $c$ (see Eq.~\ref{clusform}). Finally, note that the second condition in differentiable clustering involves a pair formed out of $\la a$ only; this is indeed sufficient for our purposes.

One particularly useful lemma for applications gives differentiable clustering from a stronger, but sometimes more natural, condition.
\begin{lemma}\label{lemmaanalytic}
Let $\tau$ be strongly continuous. Suppose that for every $\la a,\la b\in\lo$, the function $\bra\tau_t\la a,\la b\ket$ can be analytically continued in $t$ to a neighbourhood of 0, and there is $\ep>0$ such that the family $\{(\tau_t\la a,\la b):t\in\C,|t|<\ep\}$ is uniformly $p$-clustering for some $p>p_{\rm c}$. Then $\tau$ is differentiably clustering.
\end{lemma}
\proof
Under the assumptions of the lemma, $\bra \tau_t\iota_x\la a,\la b\ket$ is analytic in a neighbourhood $|t|<\ep$, and there exists $c>0$ and $p>p_{\rm c}$ such that
\beq
	|\bra \tau_t \iota_x\la a,\la b\ket| \leq c(|x|+1)^{-p}
\eeq
for all $|t|<\ep$ and $x\in\Z$. By analyticity, for all $|t|<\ep/4$,
\beqa
	t^{-1}\bra (\tau_t-1)\iota_x\la a,\la b\ket
	&=&\oint_{|s|=|t|/2} \frc{\dd s}{2\pi \ri} \frc{1}{s(t+s)} \bra(\tau_{t+s}-1)\iota_x\la a,\la b\ket\n
	&=&\oint_{|s|=\ep/2} \frc{\dd s}{2\pi \ri} \frc{1}{s(t+s)} \bra(\tau_{t+s}-1)\iota_x\la a,\la b\ket
\eeqa
where in the second line we used the fact that, in particular, there is no singularity at $t+s=0$. Therefore
\beq\label{diffclus1}
	|t^{-1}\bra (\tau_t-1)\iota_x\la a,\la b\ket|\leq
	\frc{1}{2\pi}\times \frc2\ep \times \frc4\ep\times 2c(|x|+1)^{-p}
\eeq
for all $|t|<\ep/4$, which shows the first part of the definition of differentiable clustering. Likewise, for all $|t|<\ep/4$,
\beq
	t^{-2}\bra (\tau_t+\tau_{-t}-2)\iota_x\la a,\la b\ket
	=\oint_{|s|=\ep/2} \frc{\dd s}{2\pi \ri} \frc{1}{s(t+s)^2} \bra(\tau_{t+s}+\tau_{-t-s}-2)\iota_x\la a,\la b\ket
\eeq
and therefore
\beq\label{diffclus2}
	|t^{-2}\bra (\tau_t+\tau_{-t}-2)\iota_x\la a,\la b\ket|\leq
	\frc{1}{2\pi}\times \frc2\ep \times \frc{16}{\ep^2}\times 4c(|x|+1)^{-p}.
\eeq
This shows (a slightly stronger version of) the second part of the definition of differentiable clustering.
\eproof

Below we will show that such strong clustering properties are sufficient to map $\{\tau_t:t\in\R\}$ into a strongly continuous one-parameter unitary group on the new Hilbert spaces constructed.

\subsection{Hilbert spaces $\hi_k$}\label{ssecthilbert}

For $k\in\R$ and $\la a\in \hi$ such that $(\la a,\la a)$ is $p$-clustering with $p>1$, we denote
\beq
	||\la a||_k = \sqrt{\dbra \la a,\la a\dket_k}.
\eeq
Our assumption that the dynamical system $\hi,\lo$ is $1$-clustering (see Definitions \ref{defclus0} and \ref{defclus}) implies that $\dbra \la a,\la b\dket_k$ exists for all $\la a,\la b\in\lo$, $k\in\R$, and thus gives a positive semi-definite sesquilinear form on $\lo$. There may be a nontrivial null space $\lon_k = \{\la a \in \lo: ||\la a||_k=0\}$; by the Cauchy-Schwarz inequality $\dbra \la a,\la b\dket_k=0$ for all $\la a\in\lon_k, \la b\in
\lo$ (we discuss the space $\lon_0$ in Section \ref{sectcons}). The space of equivalence classes is denoted $\lo/\lon_k=\lo_k$, on which $||\cdot||_k$ is a norm. The completion of $\lo_k$ with respect to $||\cdot||_k$ gives rise to a Hilbert space, denoted $\hi_k$. We still denote by $\bra\cdot,\cdot\ket_k$ the inner product on this Hilbert space. When confusion may arise, we will denote by
\beq\label{classk1}
	[\la a]_k = \la a + \lon_k\in\lo_k\quad(\la a\in\lo)
\eeq
the equivalence class of $\la a$. These are the ``local elements" in $\hi_k$. A basic property of antisymmetric linear operators on such equivalence classes is as follows.
\begin{lemma}\label{lemdeltak}
Let $\delta:\lo\to\lo$ be a linear map such that the antisymmetry \eqref{sym} holds. Then
\beq\label{symk}
	\dbra \delta \la a,\la b\dket_k =
	-\dbra \la a,\delta \la b\dket_k
\eeq
for all $\la a,\la b\in\lo$. Further, $\delta$ is well defined on $\lo_k$ for every $k\in\R$, and $\delta[\la a]_k = [\delta \la a]_k$ for every $\la a\in\lo$.
\end{lemma}
\proof The first part is immediate. For the second part, let $\la a\in\lo$ with $||\la a||_k=0$. Then by antisymmetry we find $\dbra \delta \la a, \delta \la a\dket_k = -\dbra \delta ^2 \la a,\la a\dket_k \leq ||\delta^2 \la a||_k\,||a||_k = 0$, wherefore $||\delta \la a||_k=0$.
\eproof

Clustering implies, in fact, that $\bra \la a,\la b\ket_k$, as defined in \eqref{ink}, exists for all $\la a,\la b\in\h\lo$ (see Eq.~\eqref{Vhat}). That is, we can time-evolve elements of $\lo$, and consider the form $\bra\cdot,\cdot\ket_k$ on such time-evolved elements.
We would like to assess if $\tau_t \la a$ can be identified, in an appropriate way, with an element of $\hi_k$. More precisely, given $\la a\in\lo$ and $t,k\in\R$, is there a unique element $\la c^{\hi_k}\in\hi_k$ -- a Cauchy-converging sequence with respect to $||\cdot||_k$ in the space of equivalence classes of local observables $\lo_k$ -- such that $\bra \la c^{\hi_k} ,\la b\ket_k = \bra \la c ,\la b\ket_k$ for $\la c = \tau_t \la a$ and every $\la b\in\lo$ (by density arguments, it is sufficient to consider $\la b\in\lo$)? By the Riesz  representation theorem, it is simple to show that such an element must exist, and that it is bounded by $||\la a||_k$. We may denote it $\tau_t^{\hi_k} [\la a]_k$, and $\tau_t^{\hi_k}$ is a continuous linear functional on $\hi_k$, with $||\tau_t^{\hi_k} [\la a]_k||\leq ||\la a||_k$. However, this is not sufficient in order to establish unitarity and the group property of $\tau_t^{\hi_k}$. As we show below, the stronger conditions of uniform clustering on converging sequences $\sigma_n\tau_t\la a$, as in Definition \ref{defclus}, allows us to establish unitarity and the group property. Further, strong continuity also translates to $\tau_t^{\hi_k}$ if the differentiable clustering holds (Definition \ref{deficlusgroup}). A partial proof is given for instance in the last part of the proof of \cite[Thm 6.3]{Doyon2017}. We give here a full statement and proof. The following theorem is the main structural result for this paper.
\begin{theorem}\label{lemtauthik} For every $t,k\in\R$, there is a unique unitary map $\tau_t^{\hi_k}:\hi_k\to\hi_k$ such that
\beq\label{tauthik}
	\bra \tau_t^{\hi_k} [\la a]_k,\tau_s^{\hi_k}[\la b]_k\ket_k = 
	\bra \tau_t\la a,\tau_s \la b\ket_k\quad(\la a,\la b\in\lo,\,t,s\in\R).
\eeq
For every $k\in\R$ and $\la a,\la b\in\hi_k$, the function $t\mapsto \bra \tau_t^{\hi_k} \la a,\la b\ket_k$ is Lebesgue measurable on $\R$. Further, for every $k\in\R$:
\bi
\item[I.] the group property holds, $\tau_t^{\hi_k}\tau_s^{\hi_k} = \tau_{t+s}^{\hi_k}$ for all $t,s\in\R$;
\item[II.] let $\la a\in\lo$, $t\in\R$, then the following limit exists in $\hi_k$ and gives $\lim_n [\sigma_n\tau_t \la a]_k = \tau_t^{\hi_k}[\la a]_k$;
\item[III.] if $\tau$ is strongly continuous and continuously clustering (Definition \ref{deficlusgroup}), then $\tau_t^{\hi_k}$ forms a  strongly continuous one-parameter unitary group, and continuity is uniform on $k\in\R$; and
\item[IV.] if $\tau$ is differentiably clustering (Definition \ref{deficlusgroup}), then the generator $\delta^{\hi_k}$ of $\tau_t^{\hi_k}$ satisfies
\beq\label{deltahik}
	\delta^{\hi_k}[\la a]_k = [\delta \la a]_k = \delta[ \la a]_k
\eeq
for all $\la a\in\lo$.
\ei
\end{theorem}
\proof First, once \eqref{tauthik} is established, measurability is obtained by the measurability assumption of Subsection \ref{ssectspacetime}: for every $\la a,\la b\in\lo$, the function $\bra \tau_t\la a,\la b\ket_k$ is measurable as it is the point-wise limit of a sequence of measurable functions (the finite sums), and for every $\la a,\la b\in\hi_k$, by density there is $\la a_n,\,\la b_n\in\lo_k$ with $\lim_n \la a_n = \la a$ and $\lim_n \la b_n = \la b$, and thus $\bra \tau_t\la a,\la b\ket_k = \lim_{m,n} \bra \tau_t\la a_m,\la b_n\ket_k$ is measurable.

Here and in other proofs below, we use the following two facts. Consider a Hilbert space $\hi$ and a dense subspace $\lo$. If the sequence $\bra \la a_n,\la b\ket$ converges for all $\la b\in \lo$, and if $||\la a_n||$ is uniformly bounded, then, by the Riesz representation theorem, there exists $\la b\in\hi$ such that weak convergence $\la a_n\rightharpoonup \la b$ holds. If $\la a_n\rightharpoonup \la b$ and $||\la a_n||\to||\la b||$, then $\la a_n \to \la b$ (norm topology in $\hi$).

Fix $k\in\R$. Let $\la a,\la b\in\lo$ and $t\in\R$. By definition, $\lim_n \sigma_n\tau_t \la a = \tau_t\la a$ (convergence on $\hi$). But also, $\lim_n \bra \iota_x\sigma_n\tau_t \la a,\la b\ket=\bra  \iota_x \tau_t\la a,\la b\ket$ by continuity of $\iota_x$; and the quantity $\bra \iota_x\sigma_n\tau_t \la a,\la b\ket$ is uniformly (over $n$) bounded by a summable function of $x$, by the uniformity requirement of Definition \ref{defclus}. Hence, by the bounded convergence theorem, the series $\sum_{x\in\Z} \re^{\ri kx} \bra \la \iota_x\sigma_n\tau_t \la a,\la b\ket$ converges to that of the pointwise limit of the summand. Therefore, passing to the quotient space, for every $\la b\in\lo_k$ the limit $\lim_n \bra \sigma_n\tau_t \la a,\la b\ket_k$ exists and gives $\bra\tau_t\la a,\la b\ket_k$ (as evaluated by the series \eqref{ink}). A similar argument shows that $||\sigma_n\tau_t \la a||_k$ is uniformly bounded over $n$. Since $\lo_k$ is dense in $\hi_k$, we conclude that $[\sigma_n\tau_t \la a]_k$ converges weakly in $\hi_k$. In fact, we find $\lim_n||\sigma_n\tau_t \la a||_k = ||\tau_t\la a||_k = ||\la a||_k $ (using stationarity \eqref{stat}), thus
\beq\label{wkroep}
	[\sigma_n\tau_t \la a]_k\rightharpoonup\tau_t^{\hi_k}[\la a]_k,
\eeq
with $||\tau_t^{\hi_k}[\la a]_k||\leq ||\la a||_k$. This defines a bounded linear map $\tau_t^{\hi_k}:\lo_k\to\hi_k$ which satisfies \eqref{tauthik} with $s=0$. This map extends by continuity to $\tau_t^{\hi_k}:\hi_k\to\hi_k$.

Using weak convergence in $\hi_k$, for $\la a,\la b \in \lo$ and $s,t\in\R$ we have $\bra \tau_t^{\hi_k}[\la a]_k,\tau_s^{\hi_k}[\la b]_k\ket_k = \lim_n \bra\sigma_n\tau_t \la a,\tau_s^{\hi_k}[\la b]_k\ket_k = \lim_n \lim_m\bra\sigma_n\tau_t \la a,\sigma_m\tau_s \la b\ket_k$. Thanks to the uniform clustering assumption, the right-hand side can be evaluated by pointwise convergence by the bounded convergence theorem, giving $\bra\tau_t\la a,\tau_s\la b\ket_k$. This gives \eqref{tauthik} in its generality. By density and continuity, and by the one-parameter group property of $\tau_t$, the group property $\tau_t^{\hi_k} \tau_s^{\hi_k} = \tau_{t+s}^{\hi_k}$ follows, and this implies that $\tau_t^{\hi_k}$ is unitary (as it is then an invertible isometry). In particular, $||\tau_t^{\hi_k} [\la a]_k||_k = ||\la a||_k = \lim_n ||\sigma_n\tau_t \la a||_k$, therefore, combined with weak convergence \eqref{wkroep}, we have $\lim_n [\sigma_n\tau_t \la a]_k = \tau_t^{\hi_k} [\la a]_k$ in $\hi_k$.

Assume that for every $\la a\in\lo$, we have $\lim_{t\to0} \tau_t\la a = \la a$ on $\hi$ (strong continuity at $t=0$), and that for every $\la a,\la b\in\lo$, there exists $\ep>0$ such that $\{(\tau_t \la a,\la b):t\in[-\ep,\ep]\}$ is uniformly $p$-clustering for some $p>1$ (continuous clustering). Fix $\la a,\la b\in\lo$. Then by the result just established,
\[
	\lim_{t\to0} \bra \tau_t^{\hi_k}[\la a]_k,[\la b]_k\ket_k = \lim_{t\to0} \bra \tau_t\la a,\la b\ket_k = \lim_{t\to0}\sum_x \re^{\ri kx}\bra \iota_x\tau_t \la a,\la b\ket.
\]
We have the bound
\[
	|\bra \tau_t \la a,\la b\ket_k - \bra \la a,\la b\ket_k|
	\leq \sum_x |\bra \iota_x\tau_t \la a,\la b\ket
	-\bra \iota_x\la a,\la b\ket|	
\]
By strong continuity, the summand on the right-hand side converges to zero pointwise. By uniform $p$-clustering on the interval $t\in[-\ep,\ep]$, the summand is further uniformly bounded by a summable sequence. Hence, by the bounded convergence theorem, the limit of the series exist and gives zero. As the right-hand side does not depend on $k$, the limit exists uniformly in $k$. Therefore,
\beq\label{limproofhik}
	\lim_{t\to0} \bra \tau_t^{\hi_k}[\la a]_k,[\la b]_k\ket_k = \bra \la a,\la b\ket_k
\eeq
for all $\la a,\la b\in\lo$, uniformly on $k\in\R$. Note that $||\tau_t^{\hi_k}[\la a]_k||_k = ||\la a||_k$ is uniformly bounded on $k\in\R$ by Lemma \ref{lemk}. Therefore, $||\tau_t^{\hi_k}[\la a]_k||_k$ is uniformly bounded in any neighbourhood of $t=0$, and this, uniformly on $k\in\R$. As $\lo_k$ is dense in $\hi_k$, the limit \eqref{limproofhik} for all $\la b\in\lo$ and the bound on $||\tau_t^{\hi_k}[\la a]_k||_k$, both uniform in $k$, imply that $\tau_t^{\hi_k}[\la a]_k \rightharpoonup [\la a]_k$ as $t\to0$ uniformly on $k\in\R$. Since $||\tau_t^{\hi_k}[\la a]_k|| = ||\la a||_k$, we conclude $\lim_{t\to 0} \tau_t^{\hi_k}[\la a]_k = [\la a]_k$ on $\hi_k$, uniformly on $k\in\R$. Hence $\{\tau_t^{\hi_k}:t\in\R\}$ forms a strongly continuous one-parameter unitary group, and this shows Point III. Stone's theorem can be applied (note that continuity at $t=0$ is sufficient); we denote the associated generator by $\delta^{\hi_k}$.

Finally, assume further that \eqref{deltaass} holds, and that for every $\la a,\la b\in\lo$, there exists $\ep>0$ such that $\{(t^{-1}(\tau_t-1) \la a,\la b):t\in[-\ep,\ep]\}$ is uniformly $p$-clustering for some $p>p_{\rm c}$ (the first statement of differentiable clustering). We use a similar line of arguments. Fix $\la a,\la b\in\lo$. Then
\[
	\lim_{t\to0} t^{-1}\bra (\tau_t^{\hi_k}-1)[\la a]_k,[\la b]_k\ket_k = \lim_{t\to0} t^{-1}\bra (\tau_t-1)\la a,\la b\ket_k = \lim_{t\to0}\sum_x \re^{\ri kx} t^{-1} \bra \iota_x (\tau_t-1) \la a,\la b\ket.
\]
By differentiability, the limit exists pointwise in the series and gives $\lim_{t\to 0} \re^{\ri kx} t^{-1} \bra \iota_x(\tau_t-1) \la a,\la b\ket = \re^{\ri kx}\bra \iota_x\delta \la a,\la b\ket$. By uniform $p$-clustering on the interval $t\in[-\ep,\ep]$, the summand $\re^{\ri kx}t^{-1}\bra \iota_x(\tau_t-1) \la a,\la b\ket$ is uniformly bounded by a summable sequence. Hence, by the bounded convergence theorem, the limit of the series exist and we find
\[
	\lim_{t\to0} t^{-1}\bra (\tau_t^{\hi_k}-1)[\la a]_k,[\la b]_k\ket_k = \bra \delta \la a,\la b\ket_k
\]
for all $\la a,\la b\in\lo$. A similar line of arguments, which we do not repeat, gives, by the second statement of differentiable clustering, a uniform bound for
\beq\label{taupr4}
	||t^{-1}(\tau_t^{\hi_k}-1)[\la a]_k||_{k}^2 = t^{-2}\bra (2 - \tau_t^{\hi_k}-\tau_{-t}^{\hi_k})[\la a]_k,[\la a]_k\ket_k
\eeq
on $t\in[-\ep,\ep]$. Since $\lo_k$ is dense in $\hi_k$ and we have a uniform bound, we find the following weak limit in $\hi_k$
\[
	t^{-1} (\tau_t^{\hi_k}-1)[\la a]_k \rightharpoonup
	[\delta \la a]_k\quad (t\to 0).
\]
Taking the limit in \eqref{taupr4} we further obtain $\lim_{t\to 0} ||t^{-1}(\tau_t^{\hi_k}-1)[\la a]_k||_{k}^2 = -\bra \delta^2\la a,\la a\ket_k = \bra \delta \la a,\delta \la a\ket_k$ (Lemma \ref{lemdeltak}), and thus we have convergence in $\hi$, hence differentiability,
\[
	\lim_{t\to0} t^{-1} (\tau_t^{\hi_k}-1)[\la a]_k = [\delta \la a]_k.
\]
Therefore, $[\la a]_k$ lies within the domain of $\delta^{\hi_k}$ for all $\la a\in\lo$, and we have \eqref{deltahik} (the last equality is by Lemma \ref{lemdeltak}) .
\eproof

For lightness of notation, below we denote $\tau_t^{\hi_k}$ simply by $\tau_t$ (and, in the strongly continuous case, $\delta^{\hi_k}$ by $\delta$), as by Theorem \ref{lemtauthik} there should be no confusion. We denote
\beq\label{hlo}
	\h\lo_k = {\rm span}\{\tau_t^{\hi_k}[\la a]_k:a\in\lo,t\in\R\}.
\eeq
We note that Theorem \ref{lemtauthik} gives a surjective map $\h\lo\to\h\lo_k$, and we extend the symbol $[\cdot]_k$ from \eqref{classk1} to $\h\lo$ in order to denote this map:
\beq\label{classk}
	[\cdot]_k: \h\lo \to \h\lo_k.
\eeq
\begin{rema} \em In applications to quantum chains, the spaces $\hi_k$, for $k\in\R$, are a generalisation of the Hilbert spaces constructed in \cite{Doyon2017,DoyonDiffusion2019}, where the case $k=0$ was considered.
\end{rema}

\begin{rema}\label{remhvv}\em If we choose $\lo$ large enough so that $\h\lo = \lo$ (see Remark \ref{remstable}), then $\tau_t(\lo)\subset \lo$ and $\tau_t(\lo_k)\subset \lo_k$, and the action of time evolution simplifies to
\beq\label{taumapex}
	\tau_t^{\hi_k}[\la a]_k = [\tau_t \la a]_k.
\eeq
In this case, if $\la a\in\lon_k$, then $\tau_t\la a\in \lon_k$ for all $t\in\R$.
\end{rema}

\subsection{The subspace of conserved charges}\label{ssectcons}

With the above constructions, we are now able to define the most important objects for hydrodynamic projections.

We define the {\em subspace of conserved charges} $\hicons_0$ as all elements $\la q\in\hi_0$ that are invariant under the time evolution unitaries $\tau_t$:
\beq\label{conserved}
	\hicons_0 = \{\la q\in\hi_0 : \tau_t\la q = \la q \;\forall\;t\in \R\}.
\eeq
By the group property of $\tau_t$, it is sufficient to require $\tau_t\la q = \la q$ for all $t$ in a non-empty interval. If $\tau$ is strongly continuous, then it is sufficient to require $\tau_t\la q = \la q$ for all $t$ in a subset of $\R$ with a finite accumulation point. In general, since $\tau_t$ is a continuous map for every $t$, then $\ker (\tau_t-1)$ is closed for every $t$ by the open mapping theorem, hence $\cap_{t\in\R}\ker (\tau_t-1)$ is closed, so that $\hicons_0$ is a closed subspace.

It will be convenient to define the space of local conserved charges as the space of conserved charges that are equivalence classes of local elements, that is
\beq
	\hicons_0^{\rm loc}= \hicons_0 \cap \lo_0
	\qquad\mbox{(subspace of local conserved charges).}
\eeq
Further, we will refer to {\em local conserved density} any representative of such an equivalence class, $\la q\in\hla q$ with $\hla q \in \hicons_0^{\rm loc}$. The space is equivalently described as
\beq\label{locdens}
	\hicons^{\rm loc} = \{\la q\in\lo: [q]_0 \in \hicons_0\} \qquad\mbox{(subspace of local conserved densities).}
\eeq

One question is as to the relation of $\hicons_0$ with the generator $\delta$ of time evolution, such as arises in the strongly continuous case, Subsection \ref{ssectstrong}. This is established in the following theorems.
\begin{theorem}\label{theodens0} Let time translation $\tau$ be differentiably clustering (Definition \ref{deficlusgroup}). Then
\beq\label{inclusion}
	\hicons_0^{\rm loc} = \ker \delta_{\lo_0},\quad \overline{\ker \delta_{\lo_0}} \subset \hicons_0 \subset \overline {\im\delta_{\lo_0}}^\perp
\eeq
where $\ker\delta_{\lo_0} = \{\la a\in\lo_0: \delta \la a=0\}$ and $\im\delta_{\lo_0} = \{\delta \la a: \la a\in\lo_0\}$ (recall, by Lemma \ref{lemdeltak}, that $\delta$ can be seen as acting on $\lo_0$).
\end{theorem}
\proof 
For $\hicons_0^{\rm loc} \subset \ker \delta_{\lo_0}$, we note that if $\la q\in \hicons_0^{\rm loc}$, then by strong continuity and Theorem \ref{lemtauthik}.IV, $\dd \tau_t\la q / \dd t\big|_{t=0} = \delta \la q$. Hence by conservation, $ \delta \la q=0$, thus $\la q \in \ker \delta_{\lo_0}$. For $\ker \delta_{\lo_0}\subset \hicons_0^{\rm loc}$, we let $\la q \in\ker\delta_{\la 0}$, and we have $\delta\la q = 0$. By strong continuity, Theorem \ref{lemtauthik}.III, we get $\dd \tau_t \la q / \dd t = \tau_t \delta\la q = 0$, and integrating, $\tau_t\la q - \la q = \int_0^t \dd s\,\dd \tau_s \la q / \dd s = 0$ for all $t\in\R$. Thus $\la q \in \hicons_0^{\rm loc}$.

Let $\la q \in \overline{\ker \delta_{\lo_0}}$. Then $\la q = \lim_n \la q_n$ with $\la q_n \in \ker\delta_{\lo_0}$. Again by Theorem \ref{lemtauthik}.IV, we have $\dd  \tau_t\la b/\dd t =\tau_t\delta \la b$ for all $\la b\in\lo_0,\, t\in\R$. Integrating this equation with $\la b = \la q_n$, we find $(\tau_t-1)\la q_n=0$. By continuity, we obtain $(\tau_t-1)\la q=0$, hence $\la q \in \hicons_0$. On the other hand, assume that $\la q\in \hicons_0$. Then, for all $\la a\in\lo_0$ and $t\in \R$, we have $0 = \dbra \tau_{t}\la q- \la q,\la a\dket_0= \dbra  \la q,\tau_{-t}\la a - \la a\dket_0$, and taking the derivative at $t=0$, this gives $\dbra  \la q,\delta \la a\dket_0 = 0$. By continuity, $\dbra  \la q,\la b\dket_0 = 0$ for all $\la b\in \overline{\im\delta_{\lo_0}}$, thus $\la q\in \overline{\im\delta_{\lo_0}}^\perp$.\eproof

One can say a bit more if $\lo$ is chosen large enough to contain all time-evolved local observables, $\h\lo=\lo$ (see Remarks \ref{remstable} and \ref{remhvv}).

\begin{theorem} \label{theodens} Let time translation $\tau$ be differentiably clustering. If $\tau_t(\lo_0)\subset\lo_0$ for all $t\in \R$, then
\beq
	\hicons_0 = \overline{\im\delta_{\lo_0}}^\perp.
\eeq
\end{theorem}
\proof Thanks to Theorem \ref{theodens0}, we only need to prove that $\overline{\im\delta_{\lo_0}}^\perp\subset \hicons_0$. Hence, assume that $\la q\in\overline{\im\delta_{\lo_0}}^\perp$. By density of $\lo_0$, we can represent $\la q = \lim_n \la q_n$ for $\la q_n\in\lo_0$. Therefore, for every $\ep>0$ there exists $N>0$, independent of $\la a\in \lo_0$, such that $|\dbra \delta \la q_n,\la a\dket_0| = |\dbra \la q_n,\delta \la a\dket_0| = 
|\dbra \la q - \la q_n,\delta \la a\dket_0| < \ep ||\delta \la a||_0$ for all $n>N$, where the second equality follows from $\la q\in\overline{\im\delta_{\lo_0}}^\perp$. In particular, $|\dbra \delta \la q_n,\tau_t\la a\dket_0|<\ep ||\delta \tau_t\la a||_0 = \ep || \tau_t\delta \la a||_0 =\ep || \delta \la a||_0$ for all $n>N$ and $t\in \R$. By differentiability, we have, for every $\ep>0$ and $t\in \R$
\beqa
	|\dbra \tau_t\la q-\la q,\la a\dket_0|
	&=& \lim_n
	|\dbra \tau_t\la q_n-\la q_n,\la a\dket_0|
	= \lim_n
	\Big|\int_0^t \dd s\,\dbra \tau_s\delta \la q_n,\la a\dket_0\Big|\n
	&=& \lim_n
	\Big|\int_0^t \dd s\,\dbra \delta \la q_n,\tau_{-s}\la a\dket_0\Big|
	<\ep
	\int_0^t \dd s\,||\delta \la a||_0 \n
	&=& \ep t ||\delta \la a||_0
\eeqa
for all $\ep>0$, and therefore $\dbra \tau_t\la q-\la q,\la a\dket_0=0$. Since $\lo_0$ is dense, this implies $\tau_t\la q-\la q=0$.
\eproof

\begin{rema}\label{remastrict} \em It is expected to happen, in some cases, that the first inclusion in \eqref{inclusion} be strict. That is, it may be that the space of conserved charges $\hicons_0$, is {\em not} the completion of the space of all local conserved charges $\hicons_0^{\rm loc}$. Indeed, in integrable quantum chains, with local observables being as usual those supported on finitely many sites, it is known that there can be quasi-local conserved densities that are not obtained as limits of local conserved densities, see \cite{IlievskietalQuasilocal}. However, we expect that one can circumvent this by extending the space $\lo$ (and thus $\lo_0$) to include quasi-local observables.
\end{rema}

\begin{rema} \label{remcons}\em Let $\la q\in\lo$. We claim that a time-independent $[\la q]_0\in\hi_0$ should indeed be interpreted as a conserved charge, in the terminology normally used in many-body physics; and that $\la q$ is an associated conserved density. In order to see this, we recall that the inner product may be defined in terms of a statistical mechanics state $\omega$ as in \eqref{inneroverview}. The quantity $\dbra \la q,\la a\dket_0$ is the expectation, in $\omega$, of the total charge $\sum_x \iota_x\la q$ times the zero-average local observable $\la a - \omega(\la a)$. It should thus indeed be invariant under time evolution of $\la q$ -- that is, $[\la q]_0$ should be time independent -- if the total charge is conserved. In this sense, then, the null space $\lon_0$ corresponds to the ambiguity in the definition of the local density for a given conserved charge. The elements of $\hi_0$ are shown in \cite{Doyon2017} to be in bijection with the pseudolocal charges \cite{ProsenPseudo1,ProsenPseudo2}, and are framed within the theory of linearly extensive charges in \cite{DoyonDiffusion2019}.

Another way of understanding the conserved densities is to recall that every element of the form $\iota_{x+1}\la a-\iota_x\la a$ for $\la a\in\lo$ has image under the quotient map $[\cdot]_0$ that vanishes, simply by using telescopic summation and clustering in the definition \eqref{ink} of $\bra \cdot,\cdot\ket_0$ with $k=0$. If $\la  q\in\lo$ is a conserved density, and if $\tau_t\la q$ is differentiable in $t$, then in local models one expects a continuity equation,
\beq
	\frc{\dd }{\dd t}\tau_t \la q + \iota_{x}\la j - \iota_{x-1}\la j = 0
\eeq
for some current $\la j\in\hi$. If $\la j$ has appropriate clustering properties, then in $\bra\cdot,\cdot\ket_0$ the derivative vanishes:  $\tau_t[\la q]_0$ is indeed independent of $t$ if $\la q$ satisfies a continuity equation. A general theorem on the existence of local currents is obtained in Section \ref{sectcons}.
\end{rema}

\section{The hydrodynamic projection formula}\label{sectproof}

In this section, we express and prove the general hydrodynamic projection formula for Euler-scale correlation functions. The necessary assumptions are those expressed in Subsections \ref{ssectspacetime} and \ref{ssectclus}, along with Property \ref{propspacelike} below. This property simply asks for almost-everywhere ergodicity, as shown in Theorem \ref{theorelax} from space-like ergodicity, and for correlations functions to have an appropriate behaviour as functions of the ray. For simplicity, the dynamical system $\hi,\lo$ is tacitly assumed to be $1$-clustering (see Definition \ref{defclus}), and it is explicitly stated when stronger conditions are required. All assumptions, including Property \ref{propspacelike}, are shown to hold in KMS states of quantum spin chains in Section \ref{sectchains}.

As an introduction to hydrodynamic projections, in Subsection \ref{ssectdrude}, we show how the simpler and well known projection formula for the Drude weights arise from our general framework. In Subsection \ref{ssectspacelike}, we state Property \ref{propspacelike}, and the relation it bears with a stronger space-like clustering property of the Lieb-Robinson type; this is useful in applications to spin chains. In Subsection \ref{ssectban} we consider the problem of the existence of the Euler scaling limit. In Subsection \ref{ssecteuler} we define the Euler map, which is used to obtain appropriate continuity properties. Finally, in Subsection \ref{ssectproj} the main projection results are obtained.

\subsection{Drude weights}\label{ssectdrude}

Before obtaining the general projection results for Euler-scale correlation functions, a much simpler result can immediately be obtained, which requires only the structures of Subsections \ref{ssectspacetime} and \ref{ssectclus}. This is the hydrodynamic projection formula for the Drude weights.

As per the conventional Kubo formula, a Drude weight is the quantity
\beq\label{drudedef}
	\mathsf D_{\la j_1,\la j_2} = \lim_{t\to\infty} (t-t_0)^{-1}\int_{t_0}^t \dd s\,\bra \tau_s\la j_1,\la j_2\ket_{0}
\eeq
for $\la j_1,\la j_2$ being the currents associated to conserved densities. Theorem \ref{theocurrent} in Section \ref{sectcons}  shows the existence of local currents for every local conserved density in systems that are ``complete" (as per Definition \ref{deficomplete}). Once local currents are given, for the general result presented here it is not necessary for the system to be complete.

Instead of local currents $\la j_1,\la j_2$, we may in fact take generic local observables $\la a,\la b\in\lo$. Further, we may extend the concept of Drude weights to $\la a,\la b\in\hi_0$ by continuity, as, if the limit exists (proven below), the right-hand side of \eqref{drudedef} makes sense for such an extension. Thus, we see $\mathsf D_{\la a,\la b}$ as a ``generalised'' Drude weight for any extensive observable $\la a,\la b\in\hi_0$.

The projection formula for the Drude weight, and for the Euler-scale correlation functions below, involves the orthogonal projection
\beq
	\mathbb P : \hi_0 \to \hicons_0
\eeq
onto the subspace of conserved charges $\hicons_0$, see \eqref{conserved}. The projection formula for the Drude weight has been studied for a long time \cite{mazur69,CasZoPre95,ZoNaPre97,SciPostPhys.3.6.039}, and it is in fact a quite direct consequence of von Neumann's mean ergodic theorem as applied to unitary operators \cite[Thm 12.44]{RudinFunctional}. In particular, the expression Eq.~\eqref{projCchain} for $\mathbb P$ holds, as we assume countable dimensionality of $\lo$, which implies countable dimensionality of $\hi_0$. Using this expression, the standard projection formula for the Drude weight \cite{SciPostPhys.3.6.039} is an immediate consequence of \eqref{drudeproj}.

\begin{theorem}\label{theodrude} For every $t_0\in\R$ and every $\la a,\la b\in\hi_0$, the Drude weight $\mathsf D_{\la a,\la b}$ exists and is obtained by projecting onto the space of conserved charges,
\beq\label{drudeproj}
	\mathsf D_{\la a,\la b}
	 = 
	\bra \mathbb P \la a,\la b\ket_0
	= \mathsf D_{\mathbb P\la a,\mathbb P\la b}.
\eeq
\end{theorem}
\proof We wish to show that
\beq\label{cesaro0}
	\mathsf D_{\la a,\la b} = \lim_{t\to\infty} (t-t_0)^{-1}\int_{t_0}^t \dd s\,\bra \tau_s\la a,\la b\ket_{0} = \bra \mathbb P \la a,\la b\ket_0.
\eeq
Since the integrand is bounded and measurable (by Theorem \ref{lemtauthik}), and the measure is finite for $t$ finite, then the Lebesgue integral exists. Since the integrand is bounded, then it is sufficient to take $t_0=0$, and to take the limit over $t=n\ep\in\N\ep$ for some $\ep>0$, as if this limit exists, then it does over $t\in\R$ as well, and gives the same result. Then, we have
\beq\label{Dabproof}
	\mathsf D_{\la a,\la b} = \lim_{\N\ni n\to\infty}
	\frc1{n}\int_{0}^n \dd s\,\bra \tau_{\ep s}\la a,\la b\ket_{0} =
	\lim_{n\to\infty} \int_0^1\dd s\,\frc1n  \sum_{m=0}^{n-1}\bra 
	\tau_\ep^m\tau_{\ep s}\la a,\la b\ket_0.
\eeq
Note that $\tau_\ep$ is unitary on $\hi_0$ (Theorem \ref{lemtauthik}). By von Neumann's mean ergodic theorem  \cite[Thm 12.44]{RudinFunctional}, the limit over $n$ exists on the integrand and gives $\bra \tau_{\ep s} \mathbb P_{\tau_\ep}\la a,\la b\ket_0$, where $\mathbb P_{\tau_\ep}$ is\footnote{By \cite[Thm 12.44]{RudinFunctional}, the result is $\mathbb P_{\tau_\ep} = f(\tau_\ep)$ with $f(1)=1$ and $f(x)=0$ for all complex $x\neq 1$. If the spectral decomposition of $\tau_\ep$ is $E$, this gives $\mathbb P_{\tau_\ep}=0$ if $1$ is not in the spectrum of $\tau_\ep$, and $\mathbb P_{\tau_\ep} = E(\{1\})$ otherwise; the statement then follows by \cite[Thm 12.29]{RudinFunctional}.} the orthogonal projection onto the null space of $\tau_\ep-1$. This is Lebesgue measurable. Since the integrand in \eqref{Dabproof} is uniformly bounded over $s$ and $n$ by the quantity $||\la b||_0 ||\la a||_0$, which is integrable, by the bounded convergence theorem the limit over $n$ and the integral over $s$ can be interchanged. The result exists and is $\int_0^1 \dd s\,\bra \tau_{\ep s} \mathbb P_{\tau_\ep}\la a,\la b\ket_0$. Since this holds for all $\ep>0$, we may replace $\la a$ by $\mathbb P\la a$ (note that $\ker(\tau_\ep-{1}) = \ker(\tau_{-\ep}-{1})$, so it is indeed sufficient to consider $\ep>0$). Thus we obtain
\beq
	\mathsf D_{\la a,\la b} = \int_0^1 \dd s\,\bra \tau_{\ep s} \mathbb P\la a,\la b\ket_0 = \int_0^1 \dd s\,\bra \mathbb P\la a,\la b\ket_0
	= \bra\mathbb P\la a,\la b\ket_0.
\eeq
\eproof

\subsection{Relaxation from space-like clustering}\label{ssectspacelike}

Our main projection theorem will need two additional conditions: almost everywhere ergodicity, proven in Theorem \ref{theorelax} from space-like ergodicity (Definition \ref{defspacelikeergo}), and appropriate integrability over the set of rays, uniform in time. These two additional conditions may be interpreted as demanding a certain relaxation property for the system, indeed a natural requirement for hydrodynamic projection to occur.
\begin{assum}\label{propspacelike} The dynamical system $\hi,\lo$ satisfies the following two conditions:
\bi
\item[a.] Let $\la a,\la b\in\hi$. Then for almost all $v\in\R$ with respect to the Lebesgue measure,
\beq\label{limproprelax}
	\lim_{T\to\infty} \frc1T\int_{0}^T \dd t\, \bra \iota_{\lfloor v t\rfloor} \tau_t\la a,\la b\ket = 0.
\eeq
\item[b.] $\hi,\lo$ is $2$-clustering, and for every $\la a,\la b\in\lo$, there exist $T>0$ and a Lebesgue integrable function $f: \R\to\R_+$ such that
\beq\label{boundass1}
	|\bra \iota_{\lfloor v t\rfloor} \tau_t\la a,\la b\ket |\leq f(v)\quad\forall v\in\R,\,t>T\quad \mbox{with the property that} \quad
	\int_\R \dd v \,(|v|+1) f(v)<\infty.
\eeq
\ei
\end{assum}
Note that Property \ref{propspacelike}a is a property of the Hilbert space $\hi$, and not of the choice of local observables $\lo$.

Interestingly, thanks to Theorem \ref{theorelax}, a somewhat weak condition of uniform clustering along all rays within space-like cones is sufficient for the above relaxation property to hold. This condition is akin to the Lieb-Robinson bound, but weaker, and will be shown to hold in quantum spin chains (Theorem \ref{theobasicchain}).
\begin{defi}\label{deflr} We say that the dynamical system $\hi,\lo$ is space-like $p_{\rm c}$-clustering with velocity $v_{\rm c}>0$, if it is $p_{\rm c}$-clustering, and if for every $\la a,\la b\in\lo$, there exist $p>p_{\rm c}$, $0<v<v_{\rm c}$ and $c>0$ such that
\beq\label{boundlr}
	|\bra \iota_{ x} \tau_{t}\la a,\la b\ket |\leq
	c(|x| +1)^{-p}
	\quad\forall\ x\in\Z,\ t\in v^{-1}[-|x|,|x|].
\eeq
\end{defi}
\begin{lemma}\label{lemmaspacelike} If the dynamical system $\hi,\lo$ is space-like $2$-clustering, then it satisfies Property \ref{propspacelike}.
\end{lemma}
\proof Property \ref{propspacelike}a is a consequence of Theorem \ref{theorelax}. For Property \ref{propspacelike}b, with the bound \eqref{boundlr}, we choose $p>2$ and $T>0$, and
\beq
	f(v) = \lt\{\ba{ll} 
	||\la a||\,||\la b||& (|v| < V) \\
	c (|vT|+1)^{-p}& (|v|\geq V).
	\ea\rt.
\eeq
\eproof

\begin{rema}\label{remaspacelike} \em
Recall that one may take $\hi,\h\lo$ as a dynamical system (that is, including all time-evolutes within the space of local observables, see Remark \ref{remstable}). If the dynamical system $\hi,\lo$ is space-like $p_{\rm c}$-clustering with velocity $v_{\rm c}$, then so is $\hi,\h\lo$. Indeed, for any given $\la a,\la b\in\lo$, in \eqref{boundlr} we may replace $\la a$ by $\tau_s\la a$ for any $s\in\R$, and $V$ by $V+\ep$ for any $\ep>0$, and the bound stays true\footnote{This follows, as for every $s\in\R$ and $\ep>0$, there is $X>0$ large enough such that, for every $v$ with $|v|\geq V+\ep$, there exists $v'$ with $|v'|\geq V$, such that $|v^{-1} + s/x|^{-1} > |v'|$ for all $|x|>X$.}.
\end{rema}

\subsection{Banach limits}\label{ssectban}

In the hydrodynamic projection formula \eqref{hpf}, the Euler-scale correlation function \eqref{defS} is involved, where a certain large-time limit is taken. Establishing properties for the dynamics of observables at large time is a particularly difficult problem in many-body systems. For instance, showing that the Euler scaling limit in \eqref{defS} exists as an ordinary limit requires subtle relations between space translations $\iota_x$ and time translations $\tau_t$. However, it turns out that we do not need the limit to exist.

Fix $\kappa\in\R$. Thanks to the Hilbert space structures $\hi_k$ and to Theorem \ref{lemtauthik}, for every $\kappa,t\in\R$ and for every $\la a,\la b\in\hi_{\kappa/t}$, the quantity $\bra \tau_t \la a,\la b\ket_{\kappa/t}$ is bounded, $|\bra \tau_t \la a,\la b\ket_{\kappa/t}|\leq ||\la a||_{\kappa/t}\,||\la b||_{\kappa/t}$, as per the Cauchy-Schwartz inequality. Further, by Lemma \ref{lemk}, for every local observables and their time evolutes $\la a,\la b\in\h\lo$, the bound is uniform for all $t>0$. Yet, the large-time limit might not exist, for instance the function might be oscillating indefinitely. Physically, in order to obtain the Euler scale, one expects that it be needed to take appropriate fluid-cell averages, in order to ``wash out" potential oscillations in space and time. Although the Fourier transform in $x$ provides a space averaging, an additional time averaging may be required in order to avoid such oscillations.

In order to average in time, one may look, for instance, at the Ces\`aro limit (or Ces\`aro mean)
\beq\label{cesaro}
	\lim_{t\to\infty} \frc1{t-t_0}\int_{t_0}^t \dd s\,\bra \tau_s\la a,\la b\ket_{\kappa/s},
\eeq
which is independent of $t_0$. There are strong results for time-averaged quantities in statistical mechanics, as part of ergodic theory. However, as far as we are aware, the above is still not guaranteed to exist, and neither do any higher-order Ces\`aro limit (the recursive time-averages of time-averages). We do not know at present what is the minimal procedure for fluid cell averaging.

Instead, a universal concept is that of {\em Banach limits} (or Mazur-Banach limits), see \cite{bookconway}. For our purposes, a Banach limit $\phi$ is a continuous linear functional on the Banach space of bounded functions $f: \R_+\to\C$ (with norm $||f|| =\limsup_{t\to\infty} |f(t)|$), with the properties of positivity, if $f(t)\geq0$ for all $t>0$ then $\phi(f)\geq0$, invariance under affine transformations $t\mapsto a t+b$ for $a>0$ and $b\in\R$, and compatibility with the limit, $\phi(f) = \lim_{t\to\infty} f(t)$ if the limit exists. Banach limits exist thanks to the Hahn-Banach theorem (a  proof is given in Appendix \ref{appbanach} for our specific definition), and satisfy $|\phi(f)|\leq ||f||$. There isn't a unique Banach limit: many limits might be attributable to a given bounded function. If a function does have a unique Banach limit, then it is said to be almost-convergent. If the $n$th order Ces\`aro limit of the function exists, then there is a Banach limit that gives it (although even in this case, the function does not necessarily have a unique Banach limit).

We will show that the hydrodynamic projection formula can be established for {\em any composition of a Banach limit with a Ces\`aro limit}. Thus, there is no necessity for the limit in \eqref{defS} to actually exist, although if it does, then the projection formula holds for the result of the limit. Below we {\em choose one Banach limit $\phi$ throughout}, and we denote the composition with the Ces\`aro limit as
\beq
	\balim_{t\to\infty} f(t) = \phi(F),\quad
	F(t) = \frc1t \int_0^t\dd s\,f(s).
\eeq
See Appendix \ref{appbanach}. By the above discussion, it has the property
\beq\label{babounded}
	|\balim\limits_{t\to\infty} f(t)| \leq
	\limsup_{t\to\infty} \Big|\frc1t \int_0^t\dd s\,f(s)\Big|
	\leq \limsup_{t\to\infty} |f(t)|.
\eeq
We then define, for every $\la a,\la b\in\h\lo$ (every local observables and their time-evolutes) and every $\kappa\in\R$,
\beq\label{Sab}
	S_{\la a,\la b}(\kappa) = \balim\limits_{t\to\infty}
	\bra \tau_t \la a,\la b\ket_{\kappa/t}.
\eeq
In particular, we note that the Drude weights \eqref{drudedef} are $\mathsf D_{\la a,\la b} =  S_{\la a,\la b}(0)$, the Euler-scale correlation function at $\kappa=0$; in this case by Theorem \ref{theodrude} the result does not depend on the choice of the Banach limit $\balim_{t\to\infty}$.

\subsection{The Euler map}\label{ssecteuler}

In this subsection, we show that there exists a continuous linear map $\li_\kappa:\hi_0\to\hi_0$ representing the Euler scaling limit of correlation functions \eqref{Sab}. The existence of this continuous map will then allow for an appropriate projection argument leading to \eqref{hpf}. Below, we will use the notation $S_{\la a,\la b}(\kappa)$ in a flexible way, with
\beq\label{hS}
	S_{\la a,\la b}(\kappa) = \bra \li_\kappa \la a,\la b\ket_0
\eeq
if $\la a$ and $\la b$ lie in $\hi_0$ or in $\h V$; thanks to the following theorem, there is no ambiguity with Eq.~\eqref{Sab}.

\begin{theorem}\label{theocont}
For every $\kappa\in\R$, there exists a unique continuous linear map $\li_\kappa:\hi_0\to\hi_0$ such that
\beq
	S_{\la a,\la b}(\kappa) =\bra \li_\kappa [\la a]_0,[\la b]_0\ket_0
\eeq
for all $\la a,\la b\in\h\lo$, with the map  $[\cdot]_0$ as defined in \eqref{classk1} and \eqref{classk}. The family of maps $\{\li_\kappa:\kappa\in\R\}$ is equicontinuous.
\end{theorem}
\proof Let $\la a,\la b\in\h\lo$. Then, for every $\kappa\in\R$,
\beq\label{resin}
	\limsup_{t\to\infty} |\dbra \tau_t\la a,\la b\dket_{\kappa/t}|
	\leq \limsup_{t\to\infty} ||\tau_t\la a||_{\kappa/t}
	||\la b||_{\kappa/t} = \lim_{k\to 0} ||\la a||_k ||\la b||_k= ||\la a||_0 ||\la b||_0
\eeq
where the last step is by Lemma \ref{lemk}. With the bound \eqref{babounded}, the result \eqref{resin} says, by specialising to $\la b\in\lo$, that $S_{\la a,\cdot}(\kappa)$ can be seen as a linear map $\lo_0\to\C$. This linear map is bounded with respect to $||\cdot||_0$:
\beq\label{ressup}
	\sup_{\la b\in\lo\setminus\{0\}} \frc{|S_{\la a,\la b}(\kappa)|}{||\la b||_0} \leq ||\la a||_0<\infty.
\eeq
Thus it is a continuous linear functional on $\lo_0$, and can be extended by continuity to $\hi_0$. By the Riesz representation theorem, there exists an element, which we denote $\li_{\kappa} \la a\in\hi_0$, such that
\beq\label{eula2}
	S_{\la a,\la b}(\kappa) = \bra \li_{\kappa} \la a,[\la b]_0\ket_0\quad(\la a\in\h\lo,\la b\in\lo).
\eeq
Further, by \eqref{ressup}, the linear map $\la a\to \li_{\kappa} \la a$ can be seen as acting on $\lo_0$, and is bounded as
\beq
	||\li_\kappa|| = \sup_{\la a\in\lo\setminus\{0\}} \frc{||\li_\kappa [\la a]_0||}{||\la a||_0}\leq 1.
\eeq
Hence it is continuous and can be extended by continuity to $\hi_0$. Thus, there exists a unique continuous linear map $\li_\kappa:\hi_0\to\hi_0$ such that
\beq\label{interm}
	S_{\la a,\la b}(\kappa) = \bra \li_\kappa [\la a]_0,[\la b]_0\ket_0\quad(\la a,\la b\in\lo).
\eeq
Equicontinuity of the family of maps on $\kappa\in\R$ is clear from the bounds established.

The right-hand side of \eqref{interm} can be extended by continuity to $\hi_0$, that is, by replacing $[\la a]_0$ and/or $[\la b]_0$ by elements of $\hi_0$. In particular, we may take $\la a,\la b\in\h\lo$ and consider $[\la a]_0,[\la b]_0\in\h\lo_0$, see \eqref{hlo} and the map  \eqref{classk}. The left-hand side can be evaluated for $\la a,\la b\in\h\lo$ by taking the large-time limit \eqref{Sab} for such elements. However, we need to establish that the result of the left-hand side for $\la a,\la b\in\h\lo$ agrees with the result of the right-hand side for $[\la a]_0,[\la b]_0\in\h\lo_0\subset\hi_0$.

Let $\la c=\tau_s \la a\in\h\lo$ and $\la d=\tau_{s'}\la b\in\h\lo$, and recall that all elements of $\h\lo$ are of this form. As $|\bra \tau_{t}(\la c- \sigma_m{\la c}),\la d\ket_{\kappa/t}|\leq ||\la c- \sigma_m{\la c}||_{\kappa/t}\,||\la d||_{\kappa/t}$, and as $||\la d||_{\kappa/t}$ is uniformly bounded on $t\in\R$ (Lemma \ref{lemk}), Theorem \ref{lemtauthik}.II implies that $\lim_{m} \bra \tau_{t}(\la c- \sigma_m{\la c}),\la d\ket_{\kappa/t}=0$ uniformly in $t$. Similarly, $\lim_{n} \bra \tau_{t}\la c,\la d- \sigma_n{\la d}\ket_{\kappa/t}=0$ and $\lim_{m,n} \bra \tau_{t}(\la c-\sigma_m{\la c}),\la d- \sigma_n{\la d}\ket_{\kappa/t}=0$ uniformly in $t$. Hence, $\lim_{m,n} \bra\tau_t \sigma_m{\la c},\sigma_n{\la d}\ket_{\kappa/t} = \bra \tau_t \la c,\la d\ket_{\kappa/t}$ uniformly in $t$. Therefore, we can exchange the limits on $t$ and on $m,n$:
\beqa
	\lefteqn{|\,
	\balim_{t\to\infty} \bra\tau_t \la c,\la d\ket_{\kappa/t}
	-
	\lim_{m,n}\balim_{t\to\infty} \bra\tau_t \sigma_m{\la c},\sigma_n{\la d}\ket_{\kappa/t}\,|}&&\n
	&= &
	\lim_{m,n}\, |\balim_{t\to\infty}\big(\bra \tau_t \la c,\la d\ket_{\kappa/t}
	- \bra \tau_t \sigma_m{\la c},\sigma_n{\la d}\ket_{\kappa/t}\big)| \n
	&\leq&
	\lim_{m,n} \,\limsup_{t\to\infty}
	\big|\bra \tau_t \la c,\la d\ket_{\kappa/t}
	- \bra \tau_t \sigma_m{\la c},\sigma_n{\la d}\ket_{\kappa/t}\big|\n
	&=& 0
\eeqa
where in the last line, the uniform limit statement has been used. As a consequence,
\beqa
	\balim_{t\to\infty} \bra \tau_t \la c,\la d\ket_{\kappa/t}
	&=& \lim_{m,n}\balim_{t\to\infty} \bra\tau_t \sigma_m{\la c},\sigma_n{\la d}\ket_{\kappa/t} \n
	&=& \lim_{m,n} S_{\sigma_m{\la c},\sigma_n{\la d}}(\kappa) \n
	&=& \bra \li_\kappa [\la c]_0,[\la d]_0\ket_0
\eeqa
where in the last line we used Eq.~\eqref{interm}, Theorem \ref{lemtauthik}.II, and the map \eqref{classk}.
\eproof

\subsection{Projection onto the subspace of conserved charges}\label{ssectproj}

The main lemma, which uses Property \ref{propspacelike}, says that taking the Euler scaling limit after making a finite time-shift of one of the elements of the correlation function, does not depend on this time-shift. This is natural, as at the Euler scale, the long-time limit has been taken; but it is nontrivial, as this limit is taken simultaneously with the long wavelength limit. The result is at the basis of the projection mechanism. 

\begin{lemma} \label{lemgt} Assume that the dynamical system $\hi,\lo$ satisfies Property \ref{propspacelike}. Then for every $\la a,\la b\in\lo$ and $s\in\R$, there exists $T_0>0$ such that for every $\kappa\in\R$, the following holds:
\beq\label{toshowinv}
	\lim_{T\to\infty} \frc1{T} \int_{T_0}^T \dd t\,g(t) = 0,\quad g(t) = \sum_{x\in\Z} \Big(\re^{\ri \kappa x/t} - \re^{\ri \kappa x/(t+s)}\Big)\bra \tau_t \iota_x \la a,\la b\ket.
\eeq
\end{lemma}
\proof We evaluate
\beqa
	g(t)
	&=& \sum_{x\in\Z}
	2\ri\exp\Big[
	\frc{\ri \kappa x}2 \lt(\frc1t + \frc1{t+s}\rt)
	\Big]\, \sin\Big[ \frc{\kappa x}2\lt(\frc1t - \frc1{t+s}\rt)\Big] \,
	\bra \tau_t \iota_x \la a,\la b\ket \n
	&=& \sum_{v\in t^{-1}\Z} \frc{\ri \kappa v}t\,
	\exp\Big[
	\frc{\ri \kappa v}2 \lt(1 + \frc t{t+s}\rt)
	\Big]\, 
	\frc{2t}{\kappa v}\sin\Big[ \frc{\kappa  v}{2t}\lt(t - \frc{t^2}{t+s}\rt)
	\Big] \,\bra \tau_t \iota_{vt} \la a,\la b\ket\n
	&=& \int_\R \dd  v \,\ri \kappa  v_t
	\exp\Big[
	\frc{\ri \kappa  v_t}2 \lt(1 + \frc t{t+s}\rt)
	\Big]\, 
	\frc{2t}{\kappa v_t}\sin\Big[ \frc{\kappa  v_t}{2t}\lt(t - \frc{t^2}{t+s}\rt)
	\Big] \,\bra \tau_t \iota_{\lfloor vt\rfloor} \la a,\la b\ket.
	\label{gtinter}
\eeqa
In the second line we defined $ v = x/t$, and in the third line we used the notation
\beq
	 v_t = \lfloor  v\rfloor_{t^{-1}}
\eeq
where, for $\ep>0$,
\beq\label{floorep}
	\lfloor y\rfloor_\ep = \ep\lt\lfloor \frc{y}\ep \rt\rfloor
\eeq
is the ``$\ep$-part'' of $y$. 

Using the assumption of ray-integrable lineshapes, Property \ref{propspacelike}b, we now show that the absolute value of the integrand in \eqref{gtinter} is uniformly bounded, for $t$ large enough, by an integrable function. A simple analysis shows that for every $u>0$ the bound $|t - t^2/(t+s)|\leq |s|+u$ holds for all $t$ large enough, and for every $\gamma>0$ the bound $y^{-1}|\sin\gamma y|\leq \gamma$ holds for all $y\in\R\setminus\{0\}$. Therefore, there exists $u>0$ such that
\beq
	\Big|\frc{2t}{\kappa v}\sin \frc{\kappa  v}{2t}\lt(t - \frc{t^2}{t+s}\rt)\Big| \leq
	|s|+u
\eeq
for all $t$ large enough. We can assume $t>1$, and thus $| v_t|\leq | v|+1$. As a consequence, the absolute value of the integrand in \eqref{gtinter} is bounded by
\beq
	(|s|+u) \kappa  (| v|+1)|\bra \tau_t \iota_{\lfloor vt\rfloor} \la a,\la b\ket| \leq (|s|+u) \kappa  (| v|+1) f(v)
\eeq
and, by Property \ref{propspacelike}b, this gives a finite integral over $v\in\R$. Therefore, there exists $T_0>0$ such that $|g(t)|$ is uniformly bounded for $t>T_0$ and $\kappa$ in any compact subset of $\R$, and in particular, for every $\kappa\in\R$, the function $g(t)$ is integrable on any compact subset of the region $t>T_0$.

We apply the Ces\`aro limit $\lim_{T\to\infty} T^{-1}\int_{T_0}^T \dd t$ on the right hand side of \eqref{gtinter}. As the integrand is uniformly bounded by an integrable function of $ v$, we may use the bounded convergence theorem and apply the limit on the integrand\footnote{The Ces\`aro limit involves a Lebesgue integral. The limit definition of the Lebesgue integral can also be applied to the integrand of the $ v$ integral by another use of the bounded convergence theorem.}. We use the fact that the following ordinary limits exist for every $ v\in\R$:
\beq
	\lim_{t\to\infty} \frc{2t}{\kappa v_t}\sin\Big[ \frc{\kappa  v_t}{2t}\lt(t - \frc{t^2}{t+s}\rt)
	\Big] = s,\quad
	\lim_{t\to\infty}  v_t
	\exp\Big[
	\frc{\ri \kappa  v_t}2 \lt(1 + \frc t{t+s}\rt)
	\Big]
	=  v
	\re^{\ri \kappa  v}
\eeq
and we obtain
\beq
	\lim_{T\to\infty} \frc1T\int_{T_0}^T \dd t\,g(t)
	= \int_\R \dd  v\,\ri \kappa s  v
	\re^{\ri \kappa  v}
	\Big(\lim_{T\to\infty} \frc1T\int_{T_0}^T \dd t\,
	\bra \tau_t \iota_{\lfloor vt\rfloor} \la a,\la b\ket\Big).
\eeq
By almost-everywhere ergodicity Property \ref{propspacelike}a (along with the uniform bound on $\bra \tau_t \iota_{\lfloor  vt\rfloor} \la a,\la b\ket$), we have
\[
	\lim_{T\to\infty} \frc1{T}\int_{T_0}^T \dd t\,\bra \tau_t \iota_{\lfloor  vt\rfloor} \la a,\la b\ket=0
\]
a.e.~on $ v\in\R$, and therefore the result vanishes.
\eproof

This is now sufficient in order to obtain the hydrodynamic projection formula \eqref{hpf}, which uses the orthogonal projection $\mathbb P: \hi_0 \to \hicons_0$. Recall that by Theorem \ref{theocont}, $S_{\la a,\la b}(\kappa)$ is the continuation to $\la a,\la b\in\hi_0$ of the Banach limit of Fourier-transforms of correlation functions of local observables and their time evolutes, Eq.~\eqref{Sab}. Recall also that $S_{\la a,\la b}(0)$ is related to the Drude weights, Subsection \ref{ssectdrude}.
\begin{theorem}\label{main}
Assume that the dynamical system $\hi,\lo$ satisfies Property \ref{propspacelike}. For every $\la a,\la b\in\hi_0$ and $\kappa\in\R$,
\beq\label{maineq}
	S_{\la a,\la b}(\kappa) = S_{\mathbb P\la a,\mathbb P\la b}(\kappa).
\eeq
Property \ref{propspacelike} is not required for the case $\kappa=0$.
\end{theorem}
\proof Lemma \ref{lemgt} (which holds trivially for $\kappa=0$ without the need for Property \ref{propspacelike}) implies that, for every $\la a,\la b\in\lo$ and $\kappa,s\in\R$,
\beqa
	0 &=& \balim_{t\to\infty} \lt(
	\bra \tau_t \la a,\la b\ket_{\kappa/t} 
	-
	\bra \tau_t \la a,\la b\ket_{\kappa/(t+s)} \rt) \n
	&=& 
	\balim_{t\to\infty} 
	\bra \tau_t \la a,\la b\ket_{\kappa/t} 
	-
	\balim_{t\to\infty} 
	\bra \tau_t \la a,\la b\ket_{\kappa/(t+s)}  \n
	&=&
	\balim_{t\to\infty} 
	\bra \tau_t \la a,\la b\ket_{\kappa/t} 
	-
	\balim_{t\to\infty} 
	\bra \tau_{t-s}\la a,\la b\ket_{\kappa/t}\n
	&=&
	\balim_{t\to\infty} 
	\bra \tau_t \la a,\la b\ket_{\kappa/t} 
	-
	\balim_{t\to\infty} 
	\bra \tau_t \la a,\tau_{s}\la b\ket_{\kappa/t}.
	\label{deriv33}
\eeqa
The first line follows from the first bound in \eqref{babounded} and Lemma \ref{lemgt}; the second from linearity and the third from shift invariance of the Banach limit; the fourth from the group property and unitarity of $\tau_s$, Theorem \ref{lemtauthik}.  Therefore, by Theorem \ref{theocont}, for every $\la a,\la b\in\lo_0$ and $s\in\R$, 
\beq
	\dbra \li_\kappa (\tau_{-s}-1) \la a,\la b\dket_0 = \dbra \li_\kappa \la a,(\tau_s-1) \la b\dket_0 = 0.
\eeq
By density of $\lo_0$ and continuity of $\li_\kappa$, this extends to $\la a,\la b\in\hi_0$.
Hence, for every $t,\kappa\in\R$, we have the inclusions $\im(\tau_t-1)\subset\ker\li_\kappa$ and $\im\li_\kappa\subset(\im(\tau_t-1))^\perp= \ker(\tau_{-t}-1)$, the last equality by unitarity. With the orthogonal decompositions $\hi_0 = \ker(\tau_{-t}-1) \oplus \im(\tau_t-1)$, for every $t\in\R$, we conclude that
\beq
	\li_\kappa = \mathbb P\li_\kappa \mathbb P.
\eeq
\eproof

As $\hicons_0$ has at most countable dimensionality, we can choose a basis $\{\la q_i\}$ for $i$ in some countable set, with positive-definite, invertible infinite-dimensional matrix $\mathsf C_{ij} = \dbra \la q_i,\la q_j\dket_0$, and we have
\beq\label{PC}
	\mathbb P = \sum_{ij} \la q_i \mathsf C^{ij}
	\dbra \la q_j,\cdot\dket_0
\eeq
where $\mathsf C^{ij}$ is the inverse infinite-dimensional matrix (that is, $\sum_k \mathsf C_{ik} \mathsf C^{kj} = \delta_i^{~j}$). We conclude:
\begin{corol}\label{cormain} Under the conditions of Theorem \ref{main},
\beq\label{countable}
	S_{\la a,\la b}(\kappa) = \sum_{ijkl}\bra \la a, \la q_i \ket_0 \,\mathsf C^{ij}
	S_{\la q_j,\la q_k}(\kappa) \mathsf C^{kl} \bra \la q_l,\la b \ket_0.
\eeq
\end{corol}
We emphasise that there does not always exist a basis of conserved charges that lies entirely within the set of {\em local elements} $\lo_0$. See Remark \ref{remastrict}.

\section{Conserved currents and linearised Euler equations}\label{sectcons}

In many-body systems, one often argues that if the sum over all positions of a local observable vanishes in some sense, ``$\sum_{x\in\Z} \la b(x,t) = 0$", then the local observable must be a ``total derivative", $\la b = \p \la a$. Here we define the discrete derivative by $\p = 1 - \iota_{-1}$, that is
\beq\label{der}
	\p \la a(x,t) = \la a(x,t) - \la a(x-1,t).
\eeq
Indeed, one argues that, in this case, the sum specialises to the ``boundary terms" at infinity, which  heuristically don't contribute.  We will denote such an element $\la a$ as $\la a = \p^{-1} \la b$, and refer to it as the anti-derivative of $\la b$.

The precise sense in which the series $\sum_{x\in\Z} \la b(x,t)$ vanishes can naturally be taken to be $||\la b||_0=0$. In this section, we show that indeed, in this case there exists an anti-derivative $\p^{-1} \la b$ with strong enough clustering properties. This has two important consequences.

First, this gives a {\em characterisation of the set of null elements} $\lon_0$, and thus of the equivalence classes from which the Hilbert space $\hi_0$ is constructed. That is, the null elements are exactly the total derivatives of observables which can be viewed as local, and thus the Hilbert space $\hi_0$ is the Cauchy completion of the space of ``local elements $\lo$ up to total derivatives".

Second, this gives a (small part of) the Noether theorem in this very general context, without the need for Lagrangians or Hamiltonians. Indeed, assume that there is a generator $\delta$ for time translations as in \eqref{derivativetaut}. If $\la b= \la q $ is a local conserved density, then $||\delta \la q||_0=0$ (see \eqref{inclusion}). Hence, $\delta \la q + \p \la j =0$ for the current $\la j=-\p^{-1}\delta \la q$, and thus there is a continuity equation (see Remark \ref{remcons}): to every local conserved density, there is an associated continuity equation.

In fact, continuity equations are in turn a powerful tool in order to evaluate correlation functions. Combined with hydrodynamic projections, they give rise to hydrodynamic equations. Obtaining the linearised Euler equations for correlation functions of conserved densities from our hydrodynamic projection result, Theorem \ref{main}, is perhaps its most important application.

The goal of this section is, first to determine precisely the notion of anti-derivative and when it exists (Subsection \ref{ssectanti}), and the context in which all local conserved densities have an associated local current (Subsection \ref{ssectconteq}); and second, to show that the linearised Euler equation are obtained from Theorem \ref{main}, using continuity equations emerging from the existence of anti-derivatives (Subsection \ref{ssectconsdens}).

For simplicity, in Subsections \ref{ssectconteq} and \ref{ssectconsdens} we assume that the space of local observables $\lo$ is stable under time evolution, $\tau_t(\lo)\subset \lo$, that is,
\beq\label{lolo}
	\h\lo = \lo.
\eeq
This can always be achieved by adjoining to $\lo$ all time-evolutes of local observables; see Remarks \ref{remstable} and \ref{remhvv}. In the context of Gibbs states in quantum spin chains, \eqref{lolo} could also be achieved by taking, instead of the space of local spin-chain operators $\mathfrak V$ from which $\lo$ is constructed (Sections \ref{sectresultschain} and \ref{sectchains}), the space of quasi-local operators \cite{IlievskietalQuasilocal}; however we will not explicitly use this construction here. Although in quantum spin chains, under \eqref{lolo}, the operators represented by $\lo$ are no longer supported on finite numbers of sites, we still refer to $\lo$ as the space of local observables.

As we will see, the existence of anti-derivatives and of continuity equations require stronger clustering properties than $1$-clustering (used in the previous sections, see Definition \ref{defclus}). The strongest conclusions are reached if the dynamical system $\hi,\lo$ is $\infty$-clustering ($p_{\rm c}$-clustering for $p_{\rm c}$ arbitrarily large). Note that this requirement is satisfied in the construction of $\hi$ based on Gibbs states in quantum spin chains, as shown in Section \ref{sectchains}, because exponential clustering is proven for such states; and thus in this context, the strongest conclusions apply.

\subsection{Anti-derivatives of null elements}\label{ssectanti}

In this subsection, for lightness of notation, we denote $\iota_x\la a= \la a_x$.

Null elements are elements of $\lon_0 = \{\la b\in\lo: ||\la b||_0=0\}$: the null subspace of $\lo$ under $\bra\cdot,\cdot\ket_0$. This is the space moded out to form $\hi_0$, see Subsection \ref{ssecthilbert}. It is convenient here to extend it by adjoining elements of $\hi$ that cluster fast enough and that are null under $||\cdot||_0$. The main general lemma  expresses the fact that every such element must be expressible has a derivative: it possesses an ``anti-derivative" in $\hi$. This anti-derivative can be defined, naturally, as a sum from $-\infty$. In Lemma \ref{lemanti} we also extend clustering properties to anti-derivative, including, and, in Lemma \ref{lemantispacelike}, extend space-like clustering. These lemmas do not require the simplifying assumption \eqref{lolo}.
\begin{lemma} \label{lemanti} Let $\lon_0^{\rm c}\subset\hi$ be the space of ``clustering null element": all $\la b\in\hi$ such that (1) the pair $(\la b,\la b)$ is $p$-clustering for some $p>2$, (2) for every $\la a\in\lo$, the pair $(\la a,\la b)$ is $p$-clustering for some $p>1$, and (3) $||\la b||_0=0$.
\bi
\item[I.] Let $\la b\in\mathcal \lon_0^{\rm c}$. Its anti-derivative is the unique element $\p^{-1}\la b\in \hi$ obtained as a result of the weak convergence
\beq\label{antider}
	\sum_{y=-z}^0 \la b_y \rightharpoonup \sum_{y=-\infty}^0 \la b_y = \p^{-1}\la b \quad(z\to\infty).
\eeq
Further, $\p\la b\in\lon_0^{\rm c}$, and $\p\p^{-1}\la b = \p^{-1} \p\la b = \la b$. In particular, if $\la b = \p \la a $ for some local element $\la a\in\lo$, then $\p^{-1}\la b = \la a$. Finally, if $\tau_t\la b$ satisfies Points (1) and (2) of the definition of $\lon_0^{\rm c}$, then $\tau_t\la b\in\lon_0^{\rm c}$, and
\beq\label{exchdt}
	\tau_t \p^{-1} \la b = \p^{-1} \tau_t\la b.
\eeq
\item[II.] Let $\mathcal C \subset \hi\times \lon_0^{\rm c}$ be a uniformly $p$-clustering family for $p>1$. The family $\{(\la a,\p^{-1} \la b):(\la a,\la b)\in\mathcal C\}$ is uniformly $(p-1)$-clustering.
\item[III.] Let $\mathcal C\subset \lon_0^{\rm c}\times\lon_0^{\rm c}$ be a uniformly $p$-clustering family for $p>2$. The family $\{(\p^{-1} \la b,\p^{-1} \la b'):(\la b,\la b')\in\mathcal C\}$ is uniformly $(p-2)$-clustering.
\ei
\end{lemma}

\proof By the Riesz representation theorem, in order to prove weak convergence, Eq.~\eqref{antider}, to a unique element $\p^{-1} \la b \in \hi$, since $\lo$ is dense in $\hi$, it is sufficient to show that $\lim_{z\to\infty} \sum_{y=-z}^0 \bra \la b_y, \la a\ket$ exists in $\C$ for all $\la a\in\lo$, and that $||\sum_{y=-z}^0 \la b_y||$ is uniformly bounded for $z\in\N$. The former holds by the clustering assumption of the theorem. For the latter, we use the fact that $||\la b||_0=0$ in order to write
\beq
	\Big|\sum_{y=-z}^0\sum_{y'=-z}^0 \bra \la b_y, \la b_{y'}\ket\Big|
	\leq \Big|\sum_{y=-z}^0\sum_{y'=-\infty}^{-z-1} \bra \la b_y, \la b_{y'}\ket\Big| + \Big|\sum_{y=-z}^0 \sum_{y'=1}^{\infty} \bra\la b_y,\la b_{y'}\ket\Big|.
\eeq
We then use translation invariance as well as the assumed clustering form \eqref{clusform}: there exists $c>0$ and $p>2$ such that
\beqa
	\Big|\sum_{y=-z}^0\sum_{y'=-z}^0 \bra \la b_y, \la b_{y'}\ket\Big|
	&\leq & \Big|\sum_{y=-z}^{0}\sum_{y'=-\infty}^{-z-1-y} \bra \la b, \la b_{y'}\ket \Big| + \Big|\sum_{y=-z}^0\sum_{y'=1-y}^{\infty}  \bra \la b, \la b_{y'}\ket \Big| \n
	&\leq& \lt(\sum_{y'=-\infty}^{-1}\sum_{y=-z}^{\min\{-z-1-y',0\}}   + \sum_{y'=1}^\infty\sum_{y=\max\{-z,1-y'\}}^{0} \rt) c(|y'|+1)^{-p}\n
	&\leq& \lt(\sum_{y'=-\infty}^{-1}   + \sum_{y'=1}^\infty \rt) |y'| c(|y'|+1)^{-p}< \infty
	\label{boundnormproof}
\eeqa
where finiteness follows from $p>2$. Since the bound is independent of $z$, it is uniform.

It is clear that $\la b_x\in\lon_0^{\rm c}$ for every $x\in\Z$, and thus $\p \la b\in \lon_0^{\rm c}$. By the weak limit definition \eqref{antider} of $\p^{-1}$ and unitarity of the space translation operator $\iota_x$, it is clear that $ \p^{-1}\la b_x = (\p^{-1}\la b)_x \in\hi$ for every $x\in\Z$. Further, $\la b_{-z}\rightharpoonup 0$ ($z\to\infty$), as $\lo$ is dense, by clustering the limit vanishes weakly with respect to $\lo$, and $\la b_{-z}$ has norm uniformly bounded by $||\la b||$. Therefore,
\beq
	\p \big(\p^{-1}\la b\big) = \p^{-1} \big(\p\la b\big) \leftharpoonup
	\la b + \sum_{y=-z}^{-1} \la b_y - \sum_{y=-z-1}^{-1} \la b_y = \la b - \la b_{-z-1}
	\rightharpoonup \la b\quad (z\to\infty).
\eeq
Now, assume $\la b = \p \la a$ for $\la a\in\lo$. Then, again by telescopic summation and clustering, we have $\p^{-1} \la b = \la a$. Finally,  it is clear that $||\tau_t\la b||=||\la b|| =0$, hence if $\tau_t\la b$ satisfies Points (1) and (2), then it is in $\lon_0^{\rm c}$. Then, the equality
\beq
	\tau_t \p^{-1} \la b = \p^{-1} \tau_t\la b
\eeq
follows by the weak limit definition \eqref{antider} of $\p^{-1}$ and unitarity of the operator $\tau_t$.  Thus we have shown Point I.

In order to show the clustering property of the pair $(\la a,\p^{-1} \la b)$, we use again the form \eqref{clusform}. First we let $x\geq 1$, and find that there exist $c>0$ and $p>1$ such that
\beq\label{pranti1}
	|\bra\la a_x,\p^{-1}\la b\ket| =  \Big|\sum_{y=-\infty}^0\bra\la a_x, \la b_y\ket\Big| 
	\leq  \sum_{y=-\infty}^0c(x-y+1)^{-p}
	\leq  \int_{-\infty}^1 \dd y\,c(x-y+1)^{-p} = \frc{c}{p-1} x^{-p+1}.
\eeq
As $x\geq 1$ and $p>1$, we have $x^{-p+1} \leq 2^{p-1}(x+1)^{-p+1}$.
Second, we let $x\leq 0$ and use $||\la b||_0=0$, and find that there exist $c>0$ and $p>1$ such that
\beq\label{pranti2}
	|\bra\la a_x,\p^{-1}\la b\ket|= \Big|\sum_{y=1}^\infty\bra\la a_x, \la b_y\ket\Big|
	\leq  \sum_{y=1}^\infty c(y-x+1)^{-p}
	\leq \int_{0}^\infty \dd y\,c(y-x+1)^{-p} = \frc{c}{p-1} (|x|+1)^{-p+1}.
\eeq
Therefore, the clustering form \eqref{clusform} holds with the new coefficient $c'=2^{p-1} c/(p-1)$ and the exponent $p'=p-1$. Hence $(\la a,\p^{-1} \la b)$ is $(p-1)$-clustering, and Point II holds. Further, the explicit formulae for the power $p'$ and the coefficient $c'$ make the corresponding uniform clustering statement clear.

In order to show the clustering property of $(\p^{-1} \la b,\p^{-1} \la b')$, note that the quantity $\bra\p^{-1}\la b_x,\p^{-1}\la b'\ket$ can be evaluated, by weak convergence, as the converging double series
\beq\label{pranti3}
	\bra\p^{-1}\la b_x,\p^{-1}\la b'\ket =
	\sum_{y=-\infty}^x  \bra\la b_y,\p^{-1}\la b'\ket =
	-\sum_{y=x+1}^\infty  \bra\la b_y,\p^{-1}\la b'\ket =
	-\lim_{z\to\infty} \sum_{y=x+1}^z
	\sum_{y'=-\infty}^0
	\bra\la b_y,\la b_{y'}'\ket
\eeq
where the second equality holds as $||\la b||_0=0$. Let us assume $x\geq 1$. Using the clustering form \eqref{clusform} with $p>2$, this is bounded, again by bounding sums by integrals, as
\beq
	|\bra\p^{-1}\la b_x,\p^{-1}\la b'\ket| \leq
	\lim_{z\to\infty} \sum_{y=x+1}^z
	\sum_{y'=-\infty}^0
	c(y-y'+1)^{-p} \leq
	 \sum_{y=x+1}^\infty
	\frc{c}{p-1}y^{-p+1} \leq
	\frc{c}{(p-1)(p-2)} x^{-p+2}.
\eeq
Further, as $x\geq 1$ and $p\geq 2$, we have $x^{-p+2}\leq 2^{p-2}(x+1)^{-p+2} $. By a similar set of arguments, we have
\beq
	\bra\p^{-1}\la b_x,\p^{-1}\la b'\ket =
	-\lim_{z\to\infty} \sum_{y'=1}^z
	\sum_{y=-\infty}^x
	\bra\la b_y,\la b_{y'}'\ket
\eeq
which, for $x\leq 0$, is bounded as
\beq\label{pranti4}
	|\bra\p^{-1}\la b_x,\p^{-1}\la b'\ket| \leq
	\lim_{z\to\infty} \sum_{y'=1}^z
	\sum_{y=-\infty}^x
	c(y'-y+1)^{-p} \leq
	\sum_{y'=1}^\infty
	\frc{c}{p-1}(y'-x)^{-p+1} \leq
	\frc{c}{(p-1)(p-2)}|x|^{-p+2}
\eeq
and again we may use $|x|^{-p+2} \leq 2^{p-2}(|x|+1)^{-p+2}$. As a result, Point III holds, with the clustering form \eqref{clusform} with the new coefficient $c' = 2^{p-2}c/((p-1)(p-2))$ and exponent $p'=p-2$. Again, the explicit formulae make the corresponding uniform clustering statement clear.
\eproof

\begin{lemma} \label{lemantispacelike}
Assume that the dynamical system $\hi,\lo$ is space-like $2$-clustering with velocity $v_{\rm c}>0$. Let $\la a\in\lo$ and $\la b,\la b'\in\lon_0 \subset\lon_0^{\rm c}$. There exists $p>p_{\rm c}$, $0<V<v_{\rm c}$ and $c>0$ such that
\beq\label{boundlranti}
	|\bra \iota_{\lfloor x\rfloor} \tau_{xv^{-1}}\la a,\p^{-1} \la b\ket | 
	\leq
	c(|\lfloor x\rfloor|+1)^{-p},\quad
	|\bra \iota_{\lfloor x\rfloor} \tau_{xv^{-1}}\p^{-1}\la b,\p^{-1} \la b'\ket |\leq
	c(|\lfloor x\rfloor|+1)^{-p}
	\quad\forall\ |v|\geq V,\; x\in\R.
\eeq
\end{lemma}
\proof We use the bound \eqref{boundlr}. This implies, in particular, that if $y,z\in\Z$, $x\in\R$ and $\ep\in[0,1)$, with $z+\ep\geq x>0, y\leq 0$, then
\beqa\label{pranti7}
	|\bra \tau_{xv^{-1}}\la a_z,\la b_y\ket |
	&=& |\bra \tau_{xv^{-1}}\la a_{z-y},\la b\ket | \n
	&=& |\bra \tau_{(z-y+\ep)(\frc{x}{z-y+\ep})v^{-1} }\la a_{z-y},\la b\ket | \n
	&\leq&
	c(z-y+1)^{-p}
	\quad\forall\ |v|\geq V
\eeqa
and if $y,z\in\Z$, $x\in\R$ and $\ep\in[0,1)$, with $z+\ep\leq x\leq 0, y> 0$, then
\beq\label{pranti8}
	|\bra \tau_{xv^{-1}}\la a_z,\iota_y\la b\ket |\leq
	c(y-z+1)^{-p}
	\quad\forall\ |v|\geq V
\eeq
with similar inequalities involving $\la b,\la b'$ instead of $\la a,\la b$. We make the same arguments as those made around Eqs.~\eqref{pranti1}-\eqref{pranti4}, but replacing, in \eqref{pranti1} and \eqref{pranti2}, $\la a$ by $\tau_{(x+\ep)v^{-1}}\la a$, and in \eqref{pranti3}-\eqref{pranti4}, $\la b$ by $\tau_{(x+\ep)v^{-1}}\la b$, for any $\ep\in[0,1)$ (representing the fractional part of $x$ in \eqref{boundlranti}). Thanks to $p_{\rm c}$-clustering, $\tau_t\la b,\tau_t\la b'\in\lon_0^{\rm c}$ for any $t\in\R$, hence by Lemma \ref{lemanti}.I the arguments can be applied under these replacements. The same bounds hold thanks to \eqref{pranti7} and \eqref{pranti8}, and \eqref{boundlranti} is obtained.
\eproof
\subsection{Continuity equations}\label{ssectconteq}

In this subsection we assume that the space of local observables include all time-evolutes, so that Eq.~\eqref{lolo} holds, $\h\lo=\lo$. Recall that this may be done simply by augmenting the space appropriately, Remark \ref{remstable}. It is important to point out that the stronger continuous and differentiable clustering properties of Definition \ref{deficlusgroup}, that may be true with respect to $\hi,\lo$ (before augmentation), are not automatically carried through to $\hi,\h\lo$. However, in the particular case of quantum spin chains, we show in Section \ref{sectchains} both differentiable clustering with respect to $\hi,\lo$ (where $\lo$ represents the space of operators supported on a finite number of sites), and with respect to $\hi,\h\lo$.

The existence of anti-derivatives of null elements shown in Lemma \ref{lemanti} implies the existence of appropriate continuity equations. The simplest and most general form is as follows. Let $\la q\in\hi$ be such that its discrete time derivative $(\tau_t -1)\la q$, for some given $t\in\R$, is a clustering null element, as per the definition in Lemma \ref{lemanti}. Then $\la j^{(t)}=\p^{-1}\big((\tau_t-1)\la q\big)\in\hi$ is such that
\beq
	(\tau_t -1)\la q + (1-\iota_{-1}) \la j^{(t)} = 0
\eeq
in $\hi$ (here we have written explicitly the discrete space derivative $\p=1-\iota_{-1}$ for symmetry of the equation). That is, a discrete continuity equation exists in $\hi$. If $\tau$ is strongly continuous, $\la q$ lies in $\lo$ (which is in the domain of the generator $\delta$, see Subsection \ref{ssectstrong}), and $\delta \la q$ is a clustering null element, then we may set $\la j = -\p^{-1}\delta \la q\in\hi$ such that
\beq\label{conteq}
	\delta \la q + \partial \la j = 0.
\eeq
Note that by Theorem \ref{theodens0}, if in fact $\tau_t$ is differentiably clustering (Definition \ref{deficlusgroup}), the element $\la q$ is a conserved density, as the corresponding equivalence class $[\la q]_0\in\hi_0$ is a conserved charge, $[\la q]_0 \in \hicons_0$. That is, we have a continuous-time continuity equation for every local conserved density with appropriate clustering property.

The above discussion hides one subtlety: even if $\la q\in\lo$, the currents lie in $\hi$, but not necessarily in $\lo$. Consider the second case discussed above, with Eq.~\eqref{conteq}. In general, even if $\la q$ is a local conserved density as defined in Eq.~\eqref{locdens}, it is not guaranteed that $\delta \la q$ is the discrete derivative of a local element; Theorem \ref{lemanti} is a formal construction of $\p^{-1}\delta \la q$ as a weakly converging limit in $\hi$, and does not require it to be local. Of course, it may be that, from explicit calculations in a quantum chain for instance, $\delta \la q$ is manifestly the discrete derivative of a local element, $\delta \la q = \p \la a$ for some $\la a\in\lo$, in which case (by Theorem \ref{lemanti}.I)  $\la j = \p^{-1}\delta \la q=-\la a$  is local. However this is not guaranteed. Therefore, a priori, it is not clear how to construct elements of the Hilbert space $\hi_0$ associated with $\p^{-1}\delta \la q$: the construction of this Hilbert space is based on the local observables $\lo$, since it is the completion of the space of equivalence classes $\lo_0 = \lo/\lon_0$. As our main theorems in Sections \ref{sectcharges} and \ref{sectproof} for the dynamical system $\hi,\lo$ necessitate the Hilbert space $\hi_0$, in order to express continuity equations in the most useful fashion, we need to solve this problem.

One can further see the difficulty by the fact that if $\la b\in\lo$ is a null element, $||\la b||_0=0$, then it is not possible to directly construct in $\hi_0$ its anti-derivative $[\p^{-1}\la b]_0$. Indeed, the definition \eqref{antider} does not make sense if interpreted in $\hi_0$, as it involves an infinite sum over space-translates, while space-translation acts trivially in $\hi_0$ (thus this would be multiplication by infinity); and the element to which $\la b$ maps in $\hi_0$, the equivalence class $[\la b]_0$, is the zero element $\lon_0$ (that is, $\la b\equiv0$ in $\hi_0$). It is of course this combination of zero times infinity that gives, in the right construction, a nontrivial element of a new space $\hi_0$.

We simply need to choose an adequate space of local observables $\lo\subset\hi$ in which lie all anti-derivatives of null elements. This is done by (possibly) enlarging our initial choice of $\lo$ so as to include anti-derivatives. We emphasise again that this gives rise to a non-trivial change of the resulting Hilbert space $\hi_0$. 

Assume that the dynamical system $\hi,\lo$ is $3$-clustering (Definition \ref{defclus}). As $\h\lo=\lo$, this simply says that each pair of elements in $\lo$ is $p$-clustering for some $p>3$.  Then $\lon_0 \subset \lon_0^{\rm c}$: every null element $\la b\in\lon_0$ is a clustering null element, as per the definition in Lemma \ref{lemanti}.  Let us adjoin all anti-derivatives of such elements that are not already in $\lo$: we enlarge $\lo$ to $\lo^+ = \lo \cup {\rm span}\{\p^{-1}\la b:\la b\in\lon_0\}\subset\hi$. Then, it is clear that $\hi,\lo^+$ is a new dynamical system according to the discussion in Subsection \ref{ssectspacetime}. In fact, by Lemma \ref{lemanti}, this dynamical system retains all of the properties of $\hi,\lo$, except for a weaker clustering.
\begin{lemma}\label{lemext} Assume $\h \lo = \lo$, and let the dynamical system $\hi,\lo$ be $p_{\rm c}$-clustering for some $p_{\rm c}\geq 3$. Let $\lo^+ = \lo \cup {\rm span}\{\p^{-1}\la b:\la b\in\lon_0\}$. Then the dynamical system $\hi,\lo^+$ is $(p_{\rm c}-2)$-clustering, and $\widehat{\lo^+} = \lo^+$. If $\tau$ is  continuously (differentiably) clustering with respect to $\hi,\lo$, then it also is with respect to $\hi,\lo^+$. If $\hi, \lo$ is space-like $p_{\rm c}$-clustering with velocity $v_{\rm c}$, then $\hi,\lo^+$ is space-like $(p_{\rm c}-2)$-clustering with velocity $v_{\rm c}$.
\end{lemma}
\proof As mentioned, direct consequence of Lemma \ref{lemanti} is that $\lon_0\subset\lon_0^{\rm c}$ (thus the definition of $\lo^+$ makes sense), and that the dynamical system $\hi,\lo^+$ is $(p_{\rm c}-2)$-clustering. Further, thanks to Lemma \ref{lemantispacelike}, the conclusion about space-like clustering also holds. Assume \eqref{lolo}. Then $\lo^+$ is also stable under time evolution, $\widehat{{\lo}^+} = \lo^+$. Indeed, if $||\la b||_0=0$ ($\la b\in\lo$), then $||\tau_t\la b||_0=0$ for all $t\in\R$, and thus $\p^{-1}\tau_t\la b\in\lo^+$ exists. Further, any element in $\widehat{\lo^+}/\lo^+$ must be of the form
\beq
	\tau_t \p^{-1} \la b +\lo^+ = \p^{-1} \tau_t\la b + \lo^+ = \lo^+,\quad \la b\in\lo
\eeq
where we used \eqref{exchdt}. Thus there are no elements in $\widehat{\lo^+}$ that are not in $\lo^+$. Further, if the finer clustering properties of Definition \ref{deficlusgroup} hold for the dynamical system $\hi,\lo$, then they hold for $\hi,\lo^+$. Indeed suppose $\tau$ is continuously clustering with respect to $\hi,\lo$. Any element in $\lo^+$ that has been adjoined to $\lo$ is of the form $\p^{-1}\la b$ for some $\la b\in\lon_0$. By the uniformity statement of Lemma \ref{lemanti}, Point II, and the assumption of $p_{\rm c}$-clustering, for any $\la a\in\lo$, $\la b\in \lon_0$, the family $\{(\la a,\p^{-1}(\tau_t \la b)):t\in[-\ep,\ep]\} = \{(\la a,\tau_t \p^{-1}\la b):t\in[-\ep,\ep]\}$ is uniformly $p$-clustering for $p>p_{\rm c}-1$, and by Point III, for any $\la b\in\lon_0$, $\la b'\in \lon_0$, the family $\{(\p^{-1}\la b,\p^{-1}(\tau_t \la b')):t\in[-\ep,\ep]\} = \{(\p^{-1}\la b,\tau_t \p^{-1}\la b'):t\in[-\ep,\ep]\}$ is uniformly $p$-clustering for $p>p_{\rm c}-2$. Thus $\tau$ is continuously clustering with respect to the new dynamical system $\hi,\lo^+$. Similar arguments hold, with strong continuity, for the families involving $\p^{-1}\big( t^{-1}(\tau_t-1)\la b\big)$ and $\p^{-1}\big( t^{-2}(\tau_t+\tau_{-t}-2)\la b\big)$. \eproof

Clearly, in $\lo^+$ there may be new elements with zero norm; that is, $\lon_0^+=\{\la b\in \lo^+: ||\la b||_0=0\}$ may be larger than $\lon_0=\{\la b\in \lo: ||\la b||_0=0\}$. However, if the system is $p_{\rm c}$-clustering for $p_{\rm c}$ large enough, we may repeat the process. Either after a finite number of steps stability is reached, or an infinite number of steps must be executed. In both cases, the result is a space that contains all its anti-derivatives. Therefore, under $\infty$-clustering, it is always possible to enlarge the space of local observables in order to ensure that all anti-derivatives are present. This shows the following theorem.
\begin{theorem}\label{theocomplete} Assume $\h \lo = \lo$. If the dynamical system $\hi,\lo$ is $\infty$-clustering, then it is possible to enlarge $\lo$ to a new space $\lo^{\#}\subset\hi$ such that
\bi
\item[I.] $\hi,\lo^\#$ is $\infty$-clustering;
\item[II.] $\widehat{\lo^\#} = \lo^{\#}$ (that is, $\tau_t(\lo^\#) \subset \lo^\#$ for all $t\in\R$);
\item[III.] if $\la b\in\lo^\#$ and $||\la b||_0=0$, then $\p^{-1}\la b\in\lo^\#$;
\item[IV.] if $\tau$ is continuously (differentiably) clustering with respect to $\hi,\lo$, then it also is with respect to $\hi,\lo^\#$; and
\item[V.] if $\hi,\lo$ is space-like $\infty$-clustering with velocity $v_{\rm c}$, then so is $\hi,\lo^\#$.
\ei
\end{theorem}

With this enlargement, we then have a discrete-time continuity equation for all local conserved densities -- those that lie within $\lo^\#$ --, and all currents are local. With differentiable clustering, then the more usual continuous-time continuity equation holds. This is expressed in the following theorem.

Points II and III of Theorem \ref{theocomplete} have interesting consequences, thus it is useful, first, to give a name for any dynamical system $\hi,\lo^\#$ that satisfies them.
\begin{defi}\label{deficomplete}
If a dynamical system $\hi,\lo^\#$ is $3$-clustering, and is such that $\lo^\#$ satisfies Points II and III of Theorem \ref{theocomplete} (that is, the space of local observables $\lo^\#$ contains all time-evolutes and all anti-derivatives), we say that it is complete.
\end{defi}

\begin{theorem}\label{theocurrent} Assume that the dynamical system $\hi,\lo$ is complete (Definition \ref{deficomplete}). Let $\la q\in \hicons^{\rm loc}$ be a local conserved density (Eq.~\eqref{locdens}). Then for every $t\in\R$, a discrete-time continuity equation holds
\beq\label{contdisc}
	(\tau_t-1)\la q + (1-\iota_{-1}) \la j^{(t)}=0
\eeq
with local ``time-integrated current"
\beq\label{jt}
	\la j^{(t)} = -\p^{-1}\big((\tau_t-1) \la q\big)\in\lo.
\eeq
If $\tau$ is differentiably clustering (Definition \ref{deficlusgroup}), with generator $\delta$ (Eq.~\ref{derivativetaut}), then there is a continuous-time continuity equation (recall that $\p = 1-\iota_{-1}$ is the discrete space derivative)
\beq\label{contcont}
	\delta \la q + \p \la j =0
\eeq
with local current
\beq\label{j}
	\la j = -\p^{-1}\delta \la q\in\lo.
\eeq
In this case,
\beq\label{contint}
	\la j^{(t)} = \int_0^t \dd s\,\tau_s \la j
\eeq
where the integral exists in $\hi$.
\end{theorem}
\proof The first part, Eq.~\eqref{contdisc}, follows from Lemma \ref{lemanti} with Definition \ref{deficomplete}. The second part, Eq.~\eqref{contcont}, follows further from Theorem \ref{lemtauthik}.IV, which shows that for $\hla q= [\la q]_0 \in \hicons_0^{\rm loc}$, we have $0 = \dd \tau_t\hla q/\dd t|_{t=0} = \delta\hla q = [\delta \la q]_0$, implying $||\delta \la q||_0=0$. Finally, thanks to differentiability of $\tau_s\delta\la q$ as a function of $s$, we have $(\tau_t-1)\delta \la q = \int_0^t \dd s\, \tau_s \delta \la q$. Therefore, for all $\ep>0$ we have
\[
	\int_0^t \dd s\, \tau_s \la j = -\int_0^t \dd s\, \tau_s \p^{-1} \delta \la q = -\int_0^t \dd s\, \p^{-1} \tau_s \delta \la q
\]
by \eqref{exchdt}. As $||\p^{-1} \tau_s \delta \la q|| = ||\p^{-1}\delta \la q||$ is uniformly bounded on $s$,  by the bounded convergence theorem, the weak limit defining $\p^{-1}$ and the $s$-integral can be exchanged,
\[
	\int_0^t \dd s\, \tau_s \la j = - \p^{-1} \int_0^t \dd s\,\tau_s \delta \la q = - \p^{-1} (\tau_t-1)\delta\la q.
\]
This shows Eq.~\eqref{contint}. \eproof

The relation between $\la j^{(t)}$ and $\la j$ can also be expressed as follows:
\begin{lemma}\label{lemjtj}
In the context of Theorem \ref{theocurrent}, if $\tau$ is differentiably clustering, then \eqref{contint} holds as an equality in $\hi_0$, and
\beq\label{limjtj}
	\lim_{t\to0} t^{-1} \la j^{(t)} = \la j\qquad\mbox{in $\hi_0$ (that is, with respect to the metric induced by $||\cdot||_0$).}
\eeq
See Eqs.~\eqref{jt} and \eqref{j}.
\end{lemma}
\proof On $\hi$, we have
\beq
	\la j^{(t)} = \int_0^{t} \dd s\,\tau_s \la j
	= \lim_{\ep\searrow0} \int_0^{t} \dd s\,\tau_{\lfloor s\rfloor_\ep} \la j
\eeq
where we recall that $\lfloor y\rfloor_\ep$ is the ``$\ep$-part'' of $y$, Eq.~\eqref{floorep}. Noting that for every $\ep>0$ the integral is a finite sum of elements in $\lo$,  we use linearity of the quotient map $[\cdot]_0$, as well as Eq.~\eqref{taumapex}, in order to obtain the equality in $\hi_0$, with $\tau_{\lfloor s\rfloor_\ep}$ acting on $\hi_0$. Strong continuity of $\tau_t$ on $\hi_{0}$ (Theorem \ref{lemtauthik}.III) then implies that the limit on $\ep$ can be taken and gives the integral $\int_0^{t} \dd s\,\tau_s \la j$ on $\hi_0$ (Eq.~\eqref{contint} on $\hi_0$). As the integrand is continuous, the result is differentiable in $t$, and we obtain \eqref{limjtj}.
\eproof

\subsection{Two-point functions of conserved densities}\label{ssectconsdens}

The most important application of the continuity equations, for our purposes, is to obtain the Euler-scale hydrodynamic equation for two-point functions of local conserved densities $\la q_i\in\hicons^{(\rm loc)}$ (the linearised Euler equation). 

First, let us review the standard formulation. By linear response arguments, two-point correlation functions of conserved densities are argued to satisfy a dynamical equation at large scales. Consider a state $\omega(\cdot)$ (see Sections \ref{sectresultschain} and \ref{sectchains}). Then this takes the form
\beq\label{eulertwopoint}
	\frc{\p}{\p t}\omega\big(\la q_i(x,t) \la b(0,0)\big)
	+ \sum_j \mathsf A_i^{~j}
	\frc{\p}{\p x}  \omega\big(\la q_j(x,t) \la b(0,0)\big)  = 0
\eeq
for arbitrary local observable $\la b$, where the flux Jacobian is
\beq\label{Aij}
	\mathsf A_i^{~j}
	= \sum_k \bra \la j_i,\la q_k\ket_0 \mathsf C^{kj}.
\eeq
The quantities $\la q_i$ are local conserved densities, representing the ``slow modes" of the system. In the sum in \eqref{Aij}, they are to be identified with a basis of $\hicons_0$. For the static covariance matrix $\mathsf C_{ij}$ and its inverse $\mathsf C^{ij}$, see the discussion around \eqref{PC}. 

Eq.~\eqref{eulertwopoint} can be argued for as follows. The continuity equations for local conserved densities give a similar equation where the second term in \eqref{eulertwopoint} is, instead, $\frc{\p}{\p x}\omega\big(\la j_i(x,t)\la b(0,0)\big)$, with $\la j_i$ the current associated to $\la q_i$. By linear response, one interprets the insertion of $\la j_i$ via a small variation of the state. One evaluates how this insertion affects the state by assuming that $\omega(\cdots)$ is completely characterised by one-point averages $\omega(\la q_j)$. Under this assumption, one obtains \eqref{eulertwopoint}, with
\beq\label{Aijlin}
	\mathsf A_i^{~j} = \frc{\p \omega(\la j_i)}{\p \omega(\la q_j)}.
\eeq
One-point averages of conserved densities characterise the state, as they determine the Lagrange parameters $\beta^i$, or thermodynamic potentials, that in turn determine the state in the usual Gibbs form, $\Tr ( \re^{-\sum_{i,x}\beta^i \la q_i(x,0)}\cdots) / \Tr ( \re^{-\sum_{i,x}\beta^i \la q_i(x,0)})$. Derivatives with respect to Lagrange parameters give integrated two-point functions\footnote{Here, using the inner products \eqref{omegainner} and \eqref{inneroverview}, we must assume all total charges $Q_i = \sum_x \la q_i(x,0)$ to commute with the density matrix. If this is not the case, the present argument would lead to the Kubo-Mori-Bogoliubov inner product  instead. Our general theory is insensitive to the choice of the inner product, as long as all basic properties are satisfied. Importantly, in finite-range quantum spin chains, the linearised Euler equation is rigorously proven below with the chosen inner products \eqref{omegainner} and \eqref{inneroverview}, without the condition that the $Q_i$s commute with the density matrix, even though the na\"ive linear-response argument fails in this case. Intuitively, in the linear-response argument, one would have to define abstract deformation directions $\beta^i$ associated to $\bra \la q_i,\la a\ket_0$.}:
\beq
	\frc{\p\omega(\la a)}{\p\beta^i} = \bra \la q_i,\la a\ket_0.
\eeq
Using the chain rule, one gets \eqref{Aij}. See e.g.~\cite{Spohn-book,DoyonLecture2020}. 

Eqs.~\eqref{eulertwopoint} and \eqref{Aij}, as presented here, are heuristic, and establishing these in a rigorous fashion is nontrivial. In particular, they are only expected to hold in some sense, after appropriate fluid-cell averaging, at large space and time separations.

In this subsection, we show rigorously a version of Eq.~\eqref{eulertwopoint}, for the Fourier transforms of correlation functions at large wavelength and large time: we prove the Euler equation for $S_{\la q_i,\la q_j}(\kappa)$, where $\la q_i,\la q_j$ are local conserved densities,
\beq\label{dSdkbasis1}
	\frc{\dd S_{\la q_i,\la q_j}(\kappa)}{\dd \kappa} = \ri \sum_k \mathsf A_i^{~k} S_{\la q_k,\la q_j}(\kappa).
\eeq
The proof relies entirely on hydrodynamic projections, and does not require the linear response idea. Noticeably, the emergence of an Euler equation, which involves the derivative in $\kappa$,  {\em does not require} the time evolution to be strongly continuous (hence differentiable): uniform enough properties of clustering are sufficient. Even without microscopic differentiability of the time evolution, at large scales, differentiability is recovered.

Slightly stronger results can be obtained if we assume the existence of a {\em local} basis for the space of conserved charges $\hicons_0$ (see the discussion around \eqref{locdens}); that is, if $\hicons_0^{\rm loc}$ is dense in $\hicons_0$. In this case, we may replace $\la q_j$ above with any element of $\hi_0$, by hydrodynamic projection and basis decomposition. Thus for this result we will appeal to this property:
\begin{assum}\label{ass3} $\hicons_0^{\rm loc}$ is dense in $\hicons_0$. That is, there exists a countable basis of local conserved charges $\{[q_i]_0\}\subset\hicons_0^{\rm loc}$ for the space of conserved charges $\hicons_0$, with associated local densities $\{q_i\}\subset \hicons^{\rm loc}$.
\end{assum}
If the space of local observables $\lo$ is taken to be large enough, then this is expected to hold in most systems. For instance, in non-integrable systems, where only a finite number of local conserved charges exist, this would be immediate -- although proving that a system admits only a finite number of conserved charges is a nontrivial task. In integrable systems, one may include all quasi-local observables within the space $\lo$, and quasi-local charges are expected to form a basis for $\hicons_0$. A full proof, however, would need much further analysis.

Establishing \eqref{dSdkbasis1} is one step towards the exact Euler-scale correlation function. If the space of conserved charges is finite dimensional, as expected in most non-integrable models, then one can solve \eqref{dSdkbasis1} easily. However, in the infinite-dimensional case, further analysis is required. If Property \ref{ass3} holds, then a solution is the action, on $\mathsf C$, of the strongly continuous one-parameter group generated by the (not necessarily bounded) operator $\mathsf A$, whose domain includes the dense subspace $\hicons_0^{\rm loc}\subset \hicons_0$; formally
\beq
	S(\kappa) = \re^{\ri \kappa \mathsf A}\mathsf C.
\eeq
This solution, though, is not necessarily unique.

The first set of results, Lemmas \ref{lemeuler},  \ref{lemmaLconteuler} and \ref{lemmaconteuler}, are technical results concerning continuity of the Euler-scale correlation function $S_{\la a,\la b}(\cdot)$. The second set, Theorem \ref{theoeuler}, is our main theorem concerning the existence of a linearised Euler equation.
\begin{lemma}\label{lemeuler}
Assume that the dynamical system $\hi,\lo$ is complete (Definition \ref{deficomplete}), and that $\tau$ is continuously clustering (Definition \ref{deficlusgroup}). Let $\la q\in\hicons^{\rm loc}$. For every $\la b\in\lo$, $\ep>0$, $\kappa\in\R$ and $\eta>-1$, and every bounded functions $t\in\R \mapsto a_t\in[-a,a]$ and $t\in\R \mapsto b_t\in[-b,b]$ ($a,b>0$), we have the following bounds (for both signs):
\beq\label{eulerbound}
	\limsup_{t\to\pm\infty} |\bra \tau_t (\tau_{\eta t + b_t}-\tau_{a_t})\la q,\la b\ket_{\kappa/t}| \leq \ep^{-1} |\kappa \eta|\,||\la j^{(\ep)}||_0\,||\la b||_0
\eeq
and
\beq\label{eulerbound2}
	\limsup_{t\to\pm\infty} |\bra \tau_t\la q,\la b\ket_{\kappa/t}
	- \bra \tau_t\la q,\la b\ket_{0}| \leq \ep^{-1} |\kappa|\,||\la j^{(\ep)}||_0\,||\la b||_0
\eeq
where $\la j^{(\ep)}$ is the current associated to $\la q$ as per \eqref{jt}.
\end{lemma}
\proof We note that the reverse-time dynamical system $\hi,\lo,\t\tau,\iota$, with $\t\tau=\tau^{-1}$, satisfies the same properties as $\hi,\lo,\tau,\iota$. Hence if \eqref{eulerbound} and \eqref{eulerbound2} hold for one sign, then they hold for both. Thus, without loss of generality we take $t>0$.

Using
\beqa
	\limsup_{t\to\infty} |\bra \tau_t (\tau_{\eta t + b_t}-\tau_{a_t})\la q,\la b\ket_{\kappa/t}|
	&=&\limsup_{t\to\infty}|\bra \tau_{(1+\eta)t} (\tau_{-\eta t + a_t}-\tau_{b_t})\la q,\la b\ket_{\kappa/t}|\n
	&=&\limsup_{t\to\infty}|\bra \tau_{t} (\tau_{-\frc{\eta}{1+\eta} t + a_{t/(1+\eta)}}-\tau_{b_{t/(1+\eta)}})\la q,\la b\ket_{\kappa(1+\eta)/t}|\no
\eeqa
we also assume $\eta>0$ without loss of generality.

We have
\beqa
	\bra \tau_t (\tau_{\eta t + b_t}-\tau_{a_t})\la q,\la b\ket_{\kappa/t} &=& -\sum_{x\in\Z} \re^{\ri \kappa x/t} 
	\bra \tau_t \iota_x\p (\la j^{(t\eta+b_t)}-\la j^{(a_t)}),\la b\ket \n
	&=& \Big(\re^{\ri \kappa /t} -1 \Big) \sum_{x\in\Z}
	\re^{\ri \kappa x/t} 
	\bra  \tau_t\iota_x (\la j^{(t\eta+b_t)}-\la j^{(a_t)}),\la b\ket \label{iiter}
\eeqa
where we used the continuity equation \eqref{contdisc} of Theorem \ref{theocurrent}. In the last step we used ``summation by parts'': the fact that if $(\la a,\la b)$ is $p$-clustering for $p>1$, the following telescopic sum vanishes,
\beq
	\sum_{x\in\Z} \Big(\re^{\ri \kappa (x+1)/t} - \re^{\ri \kappa x/t}\Big) \bra\iota_x\la a,\la b\ket
	+ \sum_{x\in\Z} \re^{\ri \kappa x/t} \bra\iota_x\p\la a,\la b\ket
	=0.
\eeq
Let $\ep>0$. We have the finite-sum representation
\beqa
	\la j^{(t\eta+b_t)} - \la j^{(a_t)} &=& \sum_{s=0\atop \Delta s = \ep/t}^{\lfloor \eta\rfloor_{\ep/t}-\ep/t}
	\tau_{ts} j^{(\ep)} +  \tau_{\lfloor t\eta\rfloor_{\ep}} \la j^{(\{t\eta\}_{\ep})} + \tau_{t\eta} \la j^{(b_t)} - \la j^{(a_t)}\n
	&=& t \int_0^\eta \dd s\,
	\tau_{\lfloor ts\rfloor_\ep}\la j_\ep + \la c,\qquad
	\la j_\ep= \ep^{-1} \la j^{(\ep)}
\eeqa
which holds by telescopic summation from the definition \eqref{jt}, linearity of $\p^{-1}$ and the relation \eqref{exchdt}. Here, we use the notation \eqref{floorep}, as well as $\{x\}_\ep = x-\lfloor x\rfloor_\ep$, and we define
\beq\label{cc}
	\la c =
	\tau_{\lfloor t\eta\rfloor_{\ep}} \la j^{(\{t\eta\}_{\ep})}
	- \{ t\eta\}_{\ep} \tau_{\lfloor t\eta\rfloor_{\ep}} \la j_\ep + \tau_{t\eta} \la j^{(b_t)} - \la j^{(a_t)}.
\eeq

Continuing, we obtain, from \eqref{iiter},
\beq
	\bra \tau_t (\tau_{\eta t}-\tau_{a_t})\la q,\la b\ket_{\kappa/t}
	= t\big(\re^{\ri \kappa /t} -1 \big) 
	\int_0^{\eta} \dd s\,\bra \tau_{t+\lfloor ts\rfloor_\ep} \la j_\ep,\la b\ket_{\kappa/t}
	+ \big(\re^{\ri \kappa /t} -1 \big) 
	\bra \tau_{t} \la c,\la b\ket_{\kappa/t}
	. \label{interm11}
\eeq
We bound the factor in the second term on the right-hand side as
\[
	|\bra \tau_{t} \la c,\la b\ket_{\kappa/t}|\leq ||\la c||_{\kappa/t}\, ||\la b||_{\kappa/t}\leq \big(||\la j^{(\{t\eta\}_{\ep})}||_{\kappa/t} + \ep ||\la j_\ep||_{{\kappa/t}} + ||\la j^{(b_t)}||_{\kappa/t} +  ||\la j^{(a_t)}||_{\kappa/t} \big)\,||\la b||_{\kappa/t}.
\]
By continuous clustering, Definition \ref{deficlusgroup}, it is clear that the family of pairs $(\,(\tau_t -1)\la q,(\tau_t -1)\la q\,):t\in I$ is uniformly $p$-clustering for some $p>3$ for any compact subset $I\subset \R$. Therefore, by the uniform clustering statement of Lemma \ref{lemanti}.III, the family $(\,\la j^{(t)},\la j^{(t)}\,):t\in I$ is uniformly $(p-2)$-clustering. Hence, with an argument as in Lemma \ref{lemk}, the quantities $||\la j^{(\{t\eta\}_{\ep})}||_{\kappa/t}$, $||\la j^{(a_t)}||_{\kappa/t}$ and $||\la j^{(b_t)}||_{\kappa/t}$ are all bounded on $t\in\R$ (as $\{t\eta\}_{\ep}\in[0,\ep]$, $a_t\in[-a,a]$ and $b_t\in[-b,b]$ lie in compact subsets). Further, by Lemma \ref{lemk}, both $||\la j_\ep||_{\kappa/t}$ and $||\la b||_{\kappa/t}$ are also bounded for $t\in\R$. Therefore, $\limsup_{t\to\infty} |\bra \tau_{t} \la c,\la b\ket_{\kappa/t}|<\infty$, and thus the usual limit can be taken on the prefactor $\big(\re^{\ri \kappa /t} -1 \big) $ in the second term in \eqref{interm11}, giving 0:
\beq\label{interm110}
	\limsup_{t\to \infty} \Big|\big(\re^{\ri \kappa /t} -1 \big) 
	\bra \tau_{t} \la c,\la b\ket_{\kappa/t}\Big|
	=0.
\eeq
Thus,
\beq
	\limsup_{t\to\infty} \big|\bra \tau_t (\tau_{\eta t}-\tau_{a_t})\la q,\la b\ket_{\kappa/t}\big|
	= \limsup_{t\to\infty}\Big| t\big(\re^{\ri \kappa /t} -1 \big)
	\int_0^{\eta} \dd s\,\bra \tau_{t+\lfloor ts\rfloor_\ep} \la j_\ep,\la b\ket_{\kappa/t}\Big|
	. \label{interm1}
\eeq

Thanks to the bound
\beq\label{boundinter}
	\Big| \int_0^{\eta} \dd s\,\bra \tau_{t+\lfloor ts\rfloor_\ep} \la j,\la b\ket_{\kappa/t}\Big|
	\leq \int_0^{\eta} \dd s\,|| \la j_\ep||_{\kappa/t} \,||\la b||_{\kappa/t} = \eta\,|| \la j_\ep||_{\kappa/t} \,||\la b||_{\kappa/t},
\eeq
as well as Lemma \ref{lemk}, the supremum limit can be bounded in \eqref{interm1}. Taking the ordinary limit on the pre-factor, this gives
\beq\label{lipcontS}
	\limsup_{t\to\infty} \big|\bra \tau_t (\tau_{\eta t}-\tau_{a_t})\la q,\la b\ket_{\kappa/t}\big|
	\leq |\kappa|\, \eta\, ||\la j_\ep||_0\,||\la b||_0.
\eeq
This shows \eqref{eulerbound}.

Eq.~\eqref{eulerbound2} is shown similarly. Again we may restrict to $t>0$. We write
\beq
	\limsup_{t\to\infty} |\bra \tau_t\la q,\la b\ket_{\kappa/t}
	- \bra \tau_t\la q,\la b\ket_{0}|
	= \limsup_{t\to\infty} |\bra (\tau_t-1) \la q,\la b\ket_{\kappa/t}
	- \bra (\tau_t-1) \la q,\la b\ket_{0}|
\eeq
using Lemma \ref{lemk}, which implies $\lim_{t\to\infty} \bra \la q,\la b\ket_{\kappa/t} =\bra \la q,\la b\ket_{0}$. Further,
\beqa
	\bra (\tau_t-1) \la q,\la b\ket_{\kappa/t}
	- \bra (\tau_t-1) \la q,\la b\ket_{0}&=& -\sum_{x\in\Z} \Big(\re^{\ri \kappa x/t} -1\Big)
	\bra \iota_x\p \la j^{(t)},\la b\ket \n
	&=& \Big(\re^{\ri \kappa /t} -1 \Big) \sum_{x\in\Z}
	\re^{\ri \kappa x/t} 
	\bra  \iota_x \la j^{(t)},\la b\ket \n
	&=&  t\big(\re^{\ri \kappa /t} -1 \big) 
	\int_0^{1} \dd s\,\bra \tau_{\lfloor ts\rfloor_\ep} \la j_\ep,\la b\ket_{\kappa/t}
	+ \big(\re^{\ri \kappa /t} -1 \big) 
	\bra \la c,\la b\ket_{\kappa/t} \no
\eeqa
where $\la c$ is \eqref{cc} with $b_t=a_t=0$ and $\eta=1$. With the bounds on $\la c$ made above, the result \eqref{eulerbound2} follows.
\eproof

\begin{lemma}\label{lemmaLconteuler}
In the context of Lemma \ref{lemeuler}, for every $\la b\in\hi_0$, the functions $S_{\la q,\la b}(\cdot)$ and $S_{\la b,\la q}(\cdot)$ are Lipschitz continuous on $\R$.
\end{lemma}
\proof
Let $\kappa\in\R$, $u>0$, and $\la b\in\lo$. We write
\beqa
	S_{\la q,\la b}(u \kappa ) - S_{\la q,\la b}(\kappa)
	&=& \balim_{t\to\infty} 
	\big(\bra \tau_t \la q,\la b\ket_{u\kappa/t} - 
	\bra \tau_t \la q,\la b\ket_{\kappa/t}\big) \n
	&=& \balim_{t\to\infty} 
	\bra \tau_t (\tau_{t(u-1)}-1)\la q,\la b\ket_{\kappa/t}
	\label{iiter0}
\eeqa
where we used linearity and invariance of the Banach limit under scale transformations. Therefore, by \eqref{babounded} and \eqref{eulerbound} (with $a_t=b_t=0$), we have
\beq\label{boundSiter}
	|S_{\la q,\la b}(u\kappa) - S_{\la q,\la b}(\kappa)|
	\leq \ep^{-1} |\kappa|\,|u-1|\,||\la j^{(\ep)}||_0\,||\la b||_0
\eeq
and thus Lipschitz continuity for $S_{\la q,\la b}(\cdot)$ on $\R\setminus\{0\}$. Further, by \eqref{eulerbound2}, we have
\beq\label{boundSiter0}
	|S_{\la q,\la b}(\kappa) - S_{\la q,\la b}(0)|
	\leq \ep^{-1} |\kappa|\,||\la j^{(\ep)}||_0\,||\la b||_0,
\eeq
thus Lipschitz continuity for $S_{\la q,\la b}(\cdot)$ at 0.

As the Lipschitz bound is controlled by $||\la b||_0$, and $S_{\la q,\la b}(\kappa)$ can be extended by continuity to $\la b\in\hi_0$ by Theorem \ref{theocont}, so can Lipschitz continuity. The same result applies as well to the reverse-time Euler-scale correlation function $\t S_{\la q,\la b}(\kappa)$. By definition we have $S_{\la b,\la q}(\kappa)=\t S_{\la q,\la b}(\kappa)^*$, which shows Lipschitz continuity for $S_{\la b,\la q}(\cdot)$.
\eproof

\begin{lemma}\label{lemmaconteuler}
In the context of Lemma \ref{lemeuler}, if Properties \ref{propspacelike} and \ref{ass3} hold, for every $\la a,\la b\in\hi_0$, the function $ S_{\la a,\la b}(\cdot)$ is continuous on $\R$.
\end{lemma}
\proof
Since Property \ref{propspacelike} is assumed, the main projection theorem holds, and in particular its Corollary \ref{cormain}. Thus we have
\beq
	S_{\la a,\la b}(\kappa) = \sum_{j}
	c^j S_{\la q_j,\la b}(\kappa),\qquad
	c^j = \sum_i \bra \la a, \la q_i \ket_0 \,\mathsf C^{ij}.
\eeq
Since $\sum_j c^j \la q_j$ converges in $\hi_0$ and the family $\{S_{\la a,\la b}(\kappa):\kappa\in\R\}$ is $\hi_0$-equicontinuous (Theorem \ref{theocont}), for every $\eta>0$ there exists $N\in\N$ such that for all $\kappa\in\R$,
\beq
	\Big| S_{\la a,\la b}(\kappa) - \sum_{j=1}^{N} c^jS_{\la q_j,\la b}(\kappa) \Big| \leq \eta
\eeq
(assuming without loss of generality that the index set for $\la q_j$ is $j\in\N$). Therefore, using \eqref{boundSiter},
\beq
	| S_{\la a,\la b}(u\kappa) - S_{\la a,\la b}(\kappa) |
	\leq 2\eta + |\kappa|\, |u-1|\, \sum_{i=1}^{N} c^i \ep^{-1} ||\la j_{i}^{(\ep)}||_0\,||\la b||_0.
\eeq
For every $\eta>0$ and $N\in\N$, there exists a neighbourhood $X$ of 0 such that for all $\kappa \,(u-1)\in X$, we have $|\kappa|\, |u-1|\, \sum_{i=1}^{N} c^i \ep^{-1}||\la j_{i}^{(\ep)}||_0\,||\la b||_0<\eta$. Hence, for every $\eta>0$, there exists a neighbourhood $X$ of 0 such that for all $\kappa \,(u-1)\in X$,
\beq
	| S_{\la a,\la b}(u\kappa) - S_{\la a,\la b}(\kappa) | < 3\eta.
\eeq
This shows continuity.
\eproof

\begin{theorem}\label{theoeuler}
Assume that the dynamical system $\hi,\lo$ is complete (Definition \ref{deficomplete}), and that $\tau$ is continuously clustering (Definition \ref{deficlusgroup}).

For every local conserved densities $\la q,\la q'\in \hicons^{\rm loc}$, the function $S_{\la q,\la q'}(\cdot)$ is continuously differentiable on $\R$, and its derivative is
\beq\label{dSdk}
	\frc{\dd S_{\la q,\la q'}(\kappa)}{\dd \kappa} = \ri S_{{\la k},\la q'}(\kappa).
\eeq
where we may take
\beq\label{kcur}
	\la k = \lt\{\ba{ll}
	\ep^{-1} \la j^{(\ep)} & \mbox{as per Eq.~\eqref{jt}, for any $\ep>0$, or}\\
	\la j & \mbox{as per Eq.~\eqref{j}, if $\tau$ is differentiably clustering.}
	\ea\rt.
\eeq

If Property \ref{propspacelike} holds, then
\beq\label{dSdkbasis}
	\frc{\dd S_{\la q,\la q'}(\kappa)}{\dd \kappa} = \ri \sum_i \mathsf A^{i} S_{\la q_i,\la q'}(\kappa),\qquad
	\mathsf A^{i} = \sum_j\bra \la k,\la q_j\ket_0 \mathsf C^{ji}
\eeq
where $\{\la q_i\}$ forms a basis for $\hicons_0$. If in addition Property \ref{ass3} holds, then for every $\la b\in\hi_0$,
\beq\label{Sqb}
	\frc{\dd S_{\la q,\la b}(\kappa)}{\dd \kappa} = \ri S_{\la k,\la b}(\kappa) = \ri \sum_i \mathsf A^{i} S_{\la q_i,\la b}(\kappa).
\eeq
If, further, the space of conserved charges $\hicons_0$ is finite dimensional, then
\beq \label{matAtheo}
	S_{\la q_i,\la b}(\kappa) = \sum_j \big(\re^{\ri \kappa \mathsf A}\big)_i^{~j}
	\bra \la q_j,\la b\ket_0,\qquad
	\mathsf A_i^{~j} = \sum_k\bra \la k_i,\la q_k\ket_0 \mathsf C^{kj}.
\eeq

\end{theorem}
\proof
We consider $\kappa\in\R$ and $u>0$. We may use \eqref{iiter0} along with \eqref{interm11} (with $a_t=1$) from Lemma \ref{lemeuler}. As the second term in \eqref{interm11} has vanishing supremum limit at $t\to\infty$, Eq.~\eqref{interm110}, and as the factor multiplying the integral in the first term of \eqref{interm11}  has the finite limit $\ri \kappa$ as $t\to\infty$, by the bound \eqref{babounded} we find
\beq
	S_{\la q,\la q'}(u \kappa ) - S_{\la q,\la q'}(\kappa)
	= \ri\kappa \balim_{t\to\infty}
	\int_0^{u-1} \dd s\,\bra \tau_{t+\lfloor ts\rfloor_\ep} \la j_\ep,\la q'\ket_{\kappa/t}.
	\label{interm2}
\eeq

We would like to apply the Banach limit on the integrand, in order to extract the Euler-scale correlation function. If the Banach limit were in fact an ordinary limit, or a Ces\`aro limit of any finite order, then we could use the bounded convergence theorem, as the integrand is bounded. However, the bounded convergence theorem does not hold for Banach limits in general. It is possible to show, instead, that one can interchange the Banach limit and the integral if the integrand is, in addition, {\em continuous}, or at least has appropriate continuity properties asymptotically as $t\to\infty$. Thanks to Lemma \ref{lemeuler}, using the fact that $\la q'$ is a local conserved density, this is the case here. We provide the explicit steps using Lemma \ref{lemeuler} for the precise situation at hand; the general statement about exchanging integrals and Banach limits under continuity requirements may be obtained similarly.

For the derivation below, let us define: $\eta = |u-1|/n$ for some $n\in\N$, $\gamma = -\{s-1\}_{\eta}/s$ and $a_{s,t} = -\{ts\}_\ep$, with  $|\gamma|\leq \eta/s$ (note that $s>0$ in the third step and onwards) and $|a_{s,t}|\leq \ep$. Then we write
\beqa
	\lefteqn{\Big|\balim_{t\to\infty}
	\Big(
	\int_0^{u-1} \dd s\,\bra  \tau_{t + \lfloor t s\rfloor_\ep } \la j_\ep,\la q'\ket_{\kappa/t}
	- \int_0^{u-1} \dd s\,\bra  \tau_{t + t \lfloor s\rfloor_{\eta} } \la j_\ep,\la q'\ket_{\kappa/t}\Big)\Big|} && \n
	&\leq &
	\int_0^{u-1} \dd s\,
	\limsup_{t\to\infty}
	\Big|
	\bra  \tau_{t+\lfloor t s \rfloor_\ep } \la j_\ep,\la q'\ket_{\kappa/t}
	- \bra  \tau_{t + t \lfloor s\rfloor_{\eta} t} \la j_\ep,\la q'\ket_{\kappa/t}\Big|\n
	&= &
	\int_0^{u-1} \dd s\,
	\limsup_{t\to\infty}
	\Big|
	\bra  \tau_{t+ts}(\tau_{-t\{s\}_{\eta} }-\tau_{a_{s,t}}) \la j_\ep,\la q'\ket_{\kappa/t}
	\Big|\n
	&= &
	\int_1^u \dd s\,
	\limsup_{t\to\infty}
	\Big|
	\bra  \tau_{t}(\tau_{\gamma t}-\tau_{a_{s-1,t/s}}) \la j_\ep,\la q'\ket_{\kappa s/t}
	\Big|\n
	&= &
	\int_1^u \dd s\,
	\limsup_{t\to-\infty}
	\Big|
	\bra  \tau_{t}(\tau_{\gamma t}-\tau_{-a_{s-1,-t/s}}) \la q',\la j_\ep\ket_{-\kappa s/t}
	\Big|\n
	&\leq &
	|\kappa|\, ||\la j_\ep|| \,||\la j'_\ep|| \int_1^u \dd s\,  s|\gamma|
	\n
	&\to& 0 \quad(n\to \infty).
	\label{tyru}
\eeqa
In the first step we used \eqref{babounded}, in the third step we changed variables $s\mapsto s-1$ then $t\mapsto t/s$, in the fourth step we took the complex conjugate of the inner product and used unitarity of time evolution, and in the fifth step we used Lemma \ref{lemeuler}. For every $n> 0$, the second integral on the left-hand side of the first step of \eqref{tyru} is a finite sum of $n$ terms. Hence on this, the Banach limit can be applied to the integrand, by linearity. Thus,
\beqa
	\balim_{t\to\infty}
	\int_0^{u-1} \dd s\,\bra  \tau_{t + t \lfloor s\rfloor_{\eta} } \la j_\ep,\la q'\ket_{\kappa/t}
	&=& \int_0^{u-1} \dd s\,
	\balim_{t\to\infty}
	\bra  \tau_{t+t \lfloor s\rfloor_{\eta} } \la j_\ep,\la q'\ket_{\kappa/t} \n
	&=& \int_0^{u-1} \dd s\, S_{\la j_\ep,\la q'}(\kappa (1+\lfloor s\rfloor_\eta)) \n
	&\to & \int_1^{u} \dd s\, S_{\la j_\ep,\la q'}(\kappa s)\quad (n\to\infty)\label{ghtjy}
\eeqa
where in the last step, we used continuity from Lemma \ref{lemmaLconteuler}. Therefore, combining \eqref{tyru} with \eqref{ghtjy},
\beq
	\balim_{t\to\infty}
	\int_0^{u-1} \dd s\,\bra  \tau_{t + \lfloor t s\rfloor_\ep} \la j_\ep,\la q'\ket_{\kappa/t} =
	\int_1^u \dd s\,  
	S_{\la j_\ep,\la q'} (\kappa s).
\eeq

With \eqref{interm2}, we then find
\beq
	S_{\la q,\la q'}(u \kappa ) - S_{\la q,\la q'}(\kappa)
	= \ri\kappa \int_1^u \dd s\,  
	S_{\la j_\ep,\la q'} (\kappa s).
\eeq
For every $\kappa\neq0$, we change variable to obtain
\beq\label{Sqqp}
	S_{\la q,\la q'}(\kappa_2) - S_{\la q,\la q'}(\kappa_1)
	= \ri \int_{\kappa_1}^{\kappa_2} \dd s\,  
	S_{\la j_\ep,\la q'} (s)
\eeq
where $\kappa_2=u\kappa$ and $\kappa_1=\kappa$. As  $u>0$, this holds whenever $\kappa_1\kappa_2>0$. By Lemma \ref{lemmaLconteuler} (continuity at 0), we may take the limit $\kappa_1\to0$ from above or below, and the same formula holds. Therefore, \eqref{Sqqp} holds for all $\kappa_1,\kappa_2\in\R$.

By Lemma \ref{lemmaLconteuler}, the integrand in \eqref{Sqqp} is continuous. This shows continuous differentiability, and Eq.~\eqref{dSdk} for the first choice of $\la k$ in \eqref{kcur}. By Lemma \ref{lemjtj} and Theorem \ref{theocont}, the second choice of $\la j$ can be taken if $\tau$ is differentiably clustering. By Corollary \ref{cormain} of Theorem \ref{main}, if Property \ref{propspacelike} holds then we have Eq.~\eqref{dSdkbasis}. If in addition Property \ref{ass3} holds, then in \eqref{Sqqp} we may take any $\hi_0$ convergent series on $\la q'$, and we obtain Eq.~\eqref{Sqqp} for arbitrary $\la q'\in\hicons_0$; in particular on the right-hand side the equicontinuity statement of Theorem \ref{theocont} allows us to use the bounded convergence theorem. We obtain \eqref{Sqb} by using the hydrodynamic projection formula \eqref{maineq} from Theorem \ref{main}. If the basis is finite dimensional, the solution to \eqref{dSdkbasis} is \eqref{matAtheo}, using $S_{\la q_i,\la b}(0) = \bra \la q_i,\la b\ket_0$.
\eproof

\section{Proofs for quantum spin chains}\label{sectchains}

In this section we prove the main Theorems \ref{theorelaxchain}, \ref{theochain} and \ref{theoeulerchain} for quantum spin chains with finite-range interactions. For this purpose, we now show that all requirements of Subsections \ref{ssectspacetime}, \ref{ssectclus}, \ref{ssectstrong} and \ref{ssectspacelike} are satisfied in the $C^*$-algebra formulation  (Subsections \ref{ssectalgebraicformulation} and \ref{ssectcompletions}). That is, we show space-like $p_{\rm c}$-clustering (Definition \ref{deflr}), and differentiable clustering (Definition \ref{deficlusgroup}), including strong continuity of the time evolution group on $\hi_0$ with Eq.~\eqref{deltaass}. We show that this holds both if we identify the set of local observables $\lo$ with the set of all finitely-supported operators as in Subsection \ref{ssectcompletions}, and also if we identify $\lo$ with all finitely-supported operators along with all their time-evolutes (that is, with the choice $\lo=\h\lo$).

Below we fix $b,d,v_{\rm LR}$ as per Theorem \ref{theolr}.

\subsection{Quantum chain systems and sizeable clustering}\label{ssectchainsystem}

Consider the sesquilinear form \eqref{omegainner} and the construction of $\lo$ and $\hi$ from the state $\omega$ and the algebra of local operators $\mathfrak V$ in Subsection \ref{ssectcompletions}. When confusion may arise, for $\la a\in\mathfrak V$ we will denote $[\la a]\in\lo$ the associated equivalence class. Note that in quantum spin chains, $\tau$ is strongly continuous; Eq.~\eqref{derivativetaut} holds with $\delta$ induced from $\delta^{\mathfrak U}$ by \eqref{omegainner}; Eq.~\eqref{deltaass} holds; $\bra \tau_t\la a,\la b\ket$ is analytic in $t$ in a neighbourhood of 0 for every $\la a,\la b\in\lo$; and $\bra \tau_t\la a,\la b\ket$ is measurable as a function of $t$ (see Subsection \ref{ssectalgebraicformulation}). The construction in Subsection \ref{ssectcompletions} was done for $\omega=\omega_\beta$ a KMS state, but it is easy to see that it is valid for more general states. Thus, we have the following general statements.
\begin{lemma}\label{lemmachainsystem} Let $\omega$ be a space-time translation invariant state, $\omega\circ\iota_x^{\mathfrak U} = \omega\circ\tau_t^{\mathfrak U}=\omega$ for all $x\in\Z,\,t\in\R$. Then the construction of $\hi,\lo$ in Subsection \ref{ssectcompletions} makes this a dynamical system (definition in Section \ref{sectcharges}) where $\tau$ is strongly continuous and Eq.~\eqref{deltaass} holds; we call this a quantum chain system.
\end{lemma}
\proof $\lo\subset \hi$ is dense, and by the discussion above, all requirements expressed around \eqref{homo} - \eqref{homostat} hold. \eproof
\begin{corol} For every KMS state $\omega_\beta$, $\beta\geq 0$, $\hi,\lo$ is a dynamical system.
\end{corol}
\proof This follows by using Theorem \ref{golodet}. \eproof

Going beyond, we need clustering properties. Theorem \ref{golodet} guarantees uniform exponential clustering for KMS states. A more general set of states which have sufficient properties for our purposes are those which are  {\em sizeably clustering}, and whose clustering bound is a power law; these include the KMS states, as the exponential bound is stronger. The concept of sizeably clustering was introduced in \cite{Doyon2017}, and allows for a controlled dependence on the size of the supports of the operators in the clustering bound. Therefore, instead of restricting ourselves to uniformly exponentially clustering states, for generality we consider the following.
\begin{defi}\label{defisizeably} {\em \cite{Doyon2017}.}
A state $\omega$ is sizeably clustering for some $p>0$, $r\geq 0$ if there exists $u>0$ such that for every $\la a,\la b\in\mathfrak V$, we have
\beq\label{sizeably}
	|\bra \la a , \la b\ket| \leq u \,|\la a|^r\,|\la b|^r\, ||\la a||_{\mathfrak U} \,||\la b||_{\mathfrak U}\,({\rm dist}(\la a,\la b)+1)^{-p}.
\eeq
\end{defi}
\begin{theorem}  \label{theoknownresults} Every KMS state $\omega_\beta$, $\beta\geq 0$ is sizeably clustering for every $p>0$, $r\geq 0$ (Definition \ref{defisizeably}).
\end{theorem}
\proof We use Theorem \ref{golodet}, fixing $q$ as per \eqref{expnential}. For every $p\geq 0$, there exists $u'>0$ such that $\exp[-qz]\leq u'(z+1)^{-p}$ for all $z\geq 0$. Therefore $|\bra \la a,\la b\ket| \leq cu'\,||\la a||_{\mathfrak U} \,||\la b||_{\mathfrak U}\,
({\rm dist}(\la a,\la b)+1)^{-p}$, and noting that $|\la a|,\,|\la b|\geq 1$, the theorem follows.
\eproof

\subsection{Properties of sizeably clustering states}

The most nontrivial aspects of the notions introduced in the general framework are those related to clustering. We now show that the strongest clustering notions are all satisfied for quantum chain systems, whenever the state is appropriately sizeably clustering. This includes all KMS states. Below we assume $\omega$ to be space-time translation invariant,
\[
	\omega\circ\iota_x^{\mathfrak U} = \omega\circ\tau_t^{\mathfrak U}=\omega\qquad (x\in\Z,\,t\in\R).
\]

We express three theorems, whose proofs are given below. First, with the Lieb-Robinson bound (Theorem \ref{theolr}), clustering can be shown to hold not just in space, but in space-time uniformly outside light-cones; that is, in space-like regions. This is in fact a generally known result, and we express it for sizeably clustering states.
\begin{theorem}\label{theochainlightcone} Let $\omega$ be sizeably clustering for some $p> r\geq0$. Set $r' = {\rm max}\{r,1\}$. For every $v>v_{\rm LR}$ (see Theorem \ref{theolr}), there exists $u'>0$ such that, for all $\la a,\la b\in\mathfrak V$ and $t\in v^{-1}[-{\rm dist}(\la a,\la b),{\rm dist}(\la a,\la b)]$,
\beq
	|\bra \tau_t^{\mathfrak U}\la a,\la b\ket| \leq u'\,{\rm min}\{|\la a|^{r'}\,|\la b|^r,|\la a|^r\,|\la b|^{r'}\}\,||\la a||_{\mathfrak U}\,||\la b||_{\mathfrak U}\,({\rm dist}(\la a,\la b)+1)^{-(p-r)}.
\eeq
\end{theorem}
\begin{corol} \label{corochainlightcone} Let $\omega=\omega_\beta$ be a KMS state. For every $v>v_{\rm LR}$ and $p>0$, there exists $u'>0$ such that, for all $\la a,\la b\in\mathfrak V$ and $t\in v^{-1}[-{\rm dist}(\la a,\la b),{\rm dist}(\la a,\la b)]$,
\beq
	|\bra \tau_t^{\mathfrak U}\la a,\la b\ket| \leq u'\,{\rm min}\{|\la a|,|\la b|\}\,||\la a||_{\mathfrak U}\,||\la b||_{\mathfrak U}\,({\rm dist}(\la a,\la b)+1)^{-p}.
\eeq
\end{corol}
Second, it turns out that clustering holds as well on small complex time neighbourhoods.
\begin{theorem}\label{theoclusteringchainanalytic} Let $\omega$ be sizeably clustering for some $p>0$, $r\geq 0$, and let $\ell>0$. There exists $\ep>0$, which only depends on $\omega$ and $\ell$, such that for all $T\subset \R$ compact and every $0<q<p$, there exists $u''>0$ such that for all $t\in T$,  $s\in\C$ with $|s|<\ep$, and $\la a,\la b\in\mathfrak V$ with $|\la a|,\,|\la b|<\ell$, the quantity $\bra \tau_{t+s}^{\mathfrak U}\la a,\la b\ket$ is analytic in $s$ and
\beq
	|\bra \tau_{t+s}^{\mathfrak U}\la a,\la b\ket| \leq u'' \,||\la a||_{\mathfrak U}\,||\la b||_{\mathfrak U}\,({\rm dist}(\la a,\la b)+1)^{-q}.
\eeq
\end{theorem}
\begin{corol} \label{coroclusteringchainanalytic} Let $\omega=\omega_\beta$ be a KMS state and $\ell>0$. There exists $\ep>0$ such that for all $T\subset \R$ compact and $p>0$, there exists $u''>0$ such that for all $t\in T$,  $s\in\C$ with $|s|<\ep$, and $\la a,\la b\in\mathfrak V$ with $|\la a|,\,|\la b|<\ell$, the quantity $\bra \tau_{t+s}^{\mathfrak U}\la a,\la b\ket$ is analytic in $s$ and
\beq
	|\bra \tau_{t+s}^{\mathfrak U}\la a,\la b\ket| \leq u'' \,||\la a||_{\mathfrak U}\,||\la b||_{\mathfrak U}\,({\rm dist}(\la a,\la b)+1)^{-p}.
\eeq
\end{corol}
Finally, the main theorem establishes all necessary properties for the quantum chain system.
\begin{theorem} \label{theobasicchain} Let $\omega$ be sizeably clustering for some $p>r\geq 0$. Then for every $0\leq p_{\rm c}<p-r$, both the quantum chain system $\hi,\lo$, and its extension to all time evolutes $\hi,\h\lo$ (see Remark \ref{remstable}), are space-like $p_{\rm c}$-clustering (Definition \ref{deflr}), and the time evolution group $\tau$ is differentiably clustering (Definition \ref{deficlusgroup}). If instead of $p>r\geq 0$ we only impose $p>0$, $r\geq 0$, then the same statement holds with space-like $p_{\rm c}$-clustering replaced by $p_{\rm c}$-clustering (Definition \ref{defclus}).
\end{theorem}

\begin{corol} \label{corobasicchain}For every KMS state $\omega_\beta$, $\beta\geq 0$, the quantum chain system $\hi,\lo$, and its extension to all time evolutes $\hi,\h\lo$, are space-like $\infty$-clustering (Definition \ref{deflr}), and the time evolution group $\tau$ is differentiably clustering.
\end{corol}

Corollaries \ref{corobasicchain}, \ref{corochainlightcone} and \ref{coroclusteringchainanalytic} follow by using Theorem \ref{theoknownresults}. Corollaries \ref{corochainlightcone} and \ref{coroclusteringchainanalytic} can be strengthened to exponential bounds, but this fact is not required here.

The proof of Theorems \ref{theochainlightcone}, \ref{theoclusteringchainanalytic} and \ref{theobasicchain} will use the following lemma.  For every $\la a\in\mathfrak V$ and $n\in\N$, we define
\beq
	X_n^{\la a}=\{x+y:x\in {\rm supp}(\la a),\, y \in [-n,n]\cap \Z\}.
\eeq
\begin{lemma}\label{lemchainclus}
Let $\la a\in\mathfrak U$. Assume that there exists a function $\mathfrak V\ni \la b \mapsto {\rm d}(\la b) = \t{{\rm d}}({\rm supp}(\la b))\geq 1$, which satisfies the properties
\beqa\label{lemchainclusdist}
	\t{{\rm d}}(X) &\geq& \t{{\rm d}}(Y)\quad \mbox{if}\ X\subset Y\n
	\t{{\rm d}}(X_n^{\la b}) &\geq& {\rm max}\{{\rm d}(\la b)-n,1\},
\eeqa
and that there exist $w>0$, $r\geq0$ and $p>0$, such that, for all  $\la b\in {\mathfrak V}$, the following holds:
\beq\label{lemprlr}
	|\bra\la a, \la b\ket| \leq w\, |\la b|^r\,||\la b||_{\mathfrak U}\,{\rm d}(\la b)^{-p}.
\eeq
Then for every $0<q<p$, $\la b\in {\mathfrak V}$, $n\in\N$ and $t\in \R$,
\beq\label{lemchainclusresult}
	|\bra  \la a,\mathbb P_{X_n^{\la b}}\la \tau_t^{\mathfrak U}\la b\ket| \leq w'\, |\la b|^{r'}\,||\la b||_{\mathfrak U}\,{\rm d}(\la b)^{-q}
\eeq
where $r' = {\rm max}\{r,1\}$ and
\beqa
	w' &=& kw + 2b\re^{d(1+v_{\rm LR}|t|)}\,||\la a||\\
	k &=& {\rm sup}\Big\{\frc{(1+2(p/d) \log z)^r\,z^{q}}{
	({\rm max}\{z-(p/d) \log z,1\})^{p}}:z\in[1,\infty)\Big\}<\infty
\eeqa
(recall that $b,d,v_{\rm LR}$ are from Theorem \ref{theolr}).
\end{lemma}
\proof This is a slight extension of the statements established in the proof of \cite[Thm 6.3]{Doyon2017}. We note that, using in particular ${\rm supp}(\mathbb P_{X_n^{\la b}}\la \tau_t^{\mathfrak U}\la b)\subset X_n^{\la b}$,
\beq\label{leminequalities}
	|\mathbb P_{X_n^{\la b}}\la \tau_t^{\mathfrak U}\la b|\leq |X_n^{\la b}| \leq |\la b|(1+2n), \quad {\rm d}(\mathbb P_{X_n^{\la b}}\la \tau_t^{\mathfrak U}\la b) \geq {\rm max}\{{\rm d}(\la b)-n,1\},\quad ||\mathbb P_{X_n^{\la b}}\la \tau_t^{\mathfrak U}\la b||_{\mathfrak U} \leq ||\la b||_{\mathfrak U}.
\eeq
The first inequality is obtained by considering the ``worst case scenario", where ${\rm supp}(\la b)$ is composed of points in $\Z$ separated by distances greater than $n$, and by using $|\la b|\geq 1$. Below we denote $z = {\rm d}(\la b)$.

For every $n \leq \frc{p}d \log z$, from \eqref{lemprlr} follows
\beqa
	|\bra \la a,\mathbb P_{X_n^{\la b}}\la \tau_t^{\mathfrak U}\la b\ket| &\leq & w\, |\la b|^r(1+2n)^r\,||\la b||_{\mathfrak U}\,({\rm max}\{z-n,1\})^{-p}\n
	&\leq & w\, |\la b|^r(1+2(p/d) \log z)^r\,||\la b||_{\mathfrak U}\,({\rm max}\{z-(p/d) \log z,1\})^{-p}.\no
\eeqa
For every $0<q<p$, the function
\[
	[1,\infty)\ni z\mapsto \frc{(1+2(p/d) \log z)^r({\rm max}\{z-(p/d) \log z,1\})^{-p}}{
	z^{-q}}
\]
is bounded from above, because it is bounded on every compact subsets and it converges to 0 as $z\to\infty$. Let us denote the supremum by $k>0$ (it only depends on $p,q,d$). Then, for every $0<q<p$, there exists $k>0$ such that for all $\la b\in {\mathfrak V}$, $n \leq \frc{p}d \log z$,  and $t\in\R$, 
\beq\label{bounb1}
	|\bra \la a,\mathbb P_{X_n^{\la b}}\la \tau_t^{\mathfrak U}\la b\ket| \leq kw\, |\la b|^r\,||\la b||_{\mathfrak U}\,z^{-q}.
\eeq

Using \eqref{LR}, the triangle inequality, and the fact that ${\rm dist}(\la b , \Z\setminus X_n^{\la b}) = n$ for all $n\in\N$, we find that for all $t\in \R$, $m,n\in \N$,  $\la b\in  \mathfrak V$ and $x\in\Z$,
\beqa\label{lemprlr2}
	|\bra \la a,\mathbb P_{X_n^{\la b}}\la \tau_t^{\mathfrak U}\la b\ket| &\leq&
	|\bra \la a,\mathbb P_{X_m^{\la b}}\la \tau_t^{\mathfrak U}\la b\ket|+\\ && +\,
	b\,|\la b|\,||\la a||\,||\la b||_{\mathfrak U}\Big(
	\exp\Big[-d\,\big(m - v_{\rm LR}|t|\,\big)\Big] +  \exp\Big[-d\,\big(n - v_{\rm LR}|t|\,\big)\Big]\Big).\no
\eeqa
For $|\bra \la a,\mathbb P_{X_n^{\la b}}\la \tau_t^{\mathfrak U}\la b\ket|$ let us use the bound \eqref{bounb1} if $ n < \lfloor \frc{p}d \log z\rfloor$, and, otherwise, the bound \eqref{lemprlr2} with \eqref{bounb1} for $|\bra \la a,\mathbb P_{X_m^{\la b}}\la \tau_t^{\mathfrak U}\la b\ket|$ and with
\beq\label{mchoice}
	m = \lfloor \frc{p}d \log z\rfloor >\frc{p}d \log z - 1.
\eeq
Therefore in this latter case, we find that for every $0<q<p$, there exists $k>0$ such that for all $\la b\in {\mathfrak V}$, $n\geq \frc{p}d \log z$ and $t\in \R$,
\beqa
	|\bra \la a,\mathbb P_{X_n^{\la b}}\la \tau_t^{\mathfrak U}\la b\ket| &\leq&
	|\bra\la a,\mathbb P_{X_m^{\la b}}\la \tau_t^{\mathfrak U}\la b\ket|+
	2b\,|\la b|\,||\la a||\,||\la b||_{\mathfrak U}
	\exp\Big[-d\,\big(m - v_{\rm LR}|t|\,\big)\Big] \n
	&\leq& kw\, |\la b|^r\,||\la b||_{\mathfrak U}\,z^{-q}+
	2b\re^d\,|\la b|\,||\la a||\,||\la b||_{\mathfrak U}\,\re^{dv_{\rm LR}|t|} \,z^{-p}.\label{bounb2}
\eeqa

With $r' = {\rm max}\{r,1\}$, we conclude the bound \eqref{lemchainclusresult} from \eqref{bounb1} and \eqref{bounb2}.
\eproof

\subsection{Proofs of Theorems \ref{theochainlightcone}, \ref{theoclusteringchainanalytic} and \ref{theobasicchain}.}

We assume that $\omega$ is sizeably clustering for some $p>0$, $r\geq 0$, and fix the values of $p$ and $r$. We denote $r' = {\rm max}\{r,1\}$. We prove in turn five statements.

\medskip
\noindent{\bf I.} The dynamical system $\hi,\lo$ is $p_{\rm c}$-clustering for every $0\leq p_{\rm c}<p$  (Definition \ref{defclus}).

This is part of space-like clustering, and contains a clustering statement for local observables, and uniform clustering statement for families approximating the time evolution in terms of local observables.

Clustering holds on all local pairs of elements $(\la a,\la b)\in \lo\times \lo$, for any $p>0$. Indeed, we relate the translation length $x$ in Definition \ref{defclus0} to the distance between the observables as follows. For nonzero elements $[\la a],[\la b]\in\lo$, choose representatives $\la a,\la b\in\mathfrak V$. By a simple geometrical analysis, it is clear that ${\rm dist}(\iota_x^{\mathfrak U} \la a,\la b)\geq \max\{|x|-{\rm diam}(\la a,\la b),0\}$ (recall ${\rm diam}(\la a,\la b) = {\rm diam}({\rm supp}(\la a)\cup {\rm supp}(\la b))$). Hence,
\beqa\label{distiota}
	({\rm dist}(\iota_x^{\mathfrak U} \la a,\la b)+1)^{-p} &\leq&
	\big({\rm max}\{|x|+1-{\rm diam}(\la a,\la b),1\}\big)^{-p}\n
	&\leq & (|x|+1)^{-p}({\rm diam}(\la a,\la b)+1)^p.
\eeqa
Therefore, thanks to sizeable clustering Definition \ref{defisizeably}, for every $\la a,\la b\in\lo$, clustering holds as in Definition \ref{defclus0} with
\beq
	c= u \,|\la a|^r\,|\la b|^r \, ||\la a||_{\mathfrak U} \,||\la b||_{\mathfrak U}\,
	({\rm diam}(\la a,\la b)+1)^p.
\eeq

For the uniformity statement of Definition \ref{defclus}, we use the Lieb-Robinson bound as expressed in \eqref{LR}, and apply Lemma \ref{lemchainclus} to $\la a\in\mathfrak V$. We choose ${\rm d}(\la b) = {\rm dist}(\la a,\la b)+1$ and clearly the requirements \eqref{lemchainclusdist} and \eqref{lemprlr} are satisfied thanks to \eqref{sizeably}, with $w=u\,|\la a|^r\,||\la a||_{\mathfrak U}$. The explicit result of the lemma, considered for every such $\la a$ and using $||\la a||\leq ||\la a||_{\mathfrak U}$, implies that for every $0<q<p$ and compact $T\subset\R$, there exists $w>0$ such that for all $\la a,\la b\in{\mathfrak V}$, $n\in\N$ and $t\in T$,
\beq\label{clusapx}
	|\bra  \la a,\mathbb P_{X_n^{\la b}}\la \tau_t^{\mathfrak U}\la b\ket| \leq w\,|\la a|^{r}\, |\la b|^{r'}\,||\la a||_{\mathfrak U}\,||\la b||_{\mathfrak U}\,({\rm dist}(\la a,\la b)+1)^{-q}.
\eeq
We now use this in order to apply again the lemma with the replacement $\la a \to \mathbb P_{X_m^{\la a}} \tau_s^{\mathfrak U}\la a$ for $\la a\in\mathfrak V$. We conclude, using in particular the last inequality in \eqref{leminequalities}, that for every $0<q<p$ and compact $T\in\R$, there exists $w>0$ such that for all $\la a,\la b\in \mathfrak V$, all $m,n\in\N$ and all $s,t\in T$,
\beq\label{concluLR}
	|\bra  \mathbb P_{X_m^{\la a}} \tau_s^{\mathfrak U}\la a,\mathbb P_{X_n^{\la b}}\la \tau_t^{\mathfrak U}\la b\ket| \leq w\,|\la a|^{r'}\, |\la b|^{r'}\,||\la a||_{\mathfrak U}\,||\la b||_{\mathfrak U}\,({\rm dist}(\la a,\la b)+1)^{-q}.
\eeq
Specialising this to $s=t$, passing to the space $\lo$, we can therefore define $\sigma_n$ (see Definition \ref{defclus}) as follows:
\beq\label{seqchain}
	\mbox{For each $[\la a]\in\lo$, select $\la a \in [\la a]\subset \mathfrak V$ and set}\ 
	\sigma_n{\tau_t [\la a]} =  [\mathbb P_{X_n^{\la a}}\tau_t^{\frak U}\la a].
\eeq
Thanks to \eqref{LR} we have $\lim_n\mathbb P_{X_n^{\la a}}\tau_t^{\frak U}\la a = \tau_t^{\mathfrak U}\la a$ in $\mathfrak U$. Hence (by boundedness of the state) $\lim_n [\mathbb P_{X_n^{\la a}}\tau_t^{\frak U}\la a]$ converges in $\hi$, and it is simple to see that it converges to $\tau_t [\la a]$, thus $\lim_n \sigma_n{\tau_t \la a} = \tau_t \la a$ on $\hi\ni \la a$. Using \eqref{distiota},  we have shown the uniformity statement in Definition \ref{defclus}. Hence, the dynamical system $\hi,\lo$ is $p_{\rm c}$-clustering for every $0\leq p_{\rm c}<p$.

\medskip

\noindent{\bf II.} Both dynamical systems $\hi,\lo$ and $\hi,\h\lo$, are $p_{\rm c}$-clustering for every $0\leq p_{\rm c}<p$, and with respect to both, $\tau$ is continuously clustering  (Definition \ref{deficlusgroup}).

Taking the limit on $n$ and $m$, we find from \eqref{concluLR},
\beq
	|\bra \tau_s^{\mathfrak U}\la a,\la \tau_t^{\mathfrak U}\la b\ket| \leq w\,|\la a|^{r'}\, |\la b|^{r'}\,||\la a||_{\mathfrak U}\,||\la b||_{\mathfrak U}\,({\rm dist}(\la a,\la b)+1)^{-q}.
\eeq
Therefore, for every $\ep>0$ a clustering bound is obtained uniformly for all $s,t\in\R,\,|s|,|t|<\ep$, and continuous clustering holds, Definition \ref{deficlusgroup}, for all $0\leq p_{\rm c}<p$. In particular, this also implies that the dynamical system $\hi,\h\lo$, where the local observables include all time-evolutes, is also $p_{\rm c}$-clustering and continuously clustering.

\medskip
\noindent{\bf III.} With respect to $\hi,\lo$, the time evolution $\tau$ is differentiably clustering (Definition \ref{deficlusgroup}).

We recall that $\{\tau_t:t\in\R\}$ form a strongly continuous one-parameter unitary group on $\hi$; that Eq.~\eqref{derivativetaut} holds with $\delta$ induced from $\delta^{\mathfrak U}$ by \eqref{omegainner}; that Eq.~\eqref{deltaass} holds; and that $\bra \tau_t\la a,\la b\ket$ is analytic in $t$ in a neighbourhood of 0 for every $\la a,\la b\in\lo$. Differentiable clustering with respect to $\hi,\lo$ is then established by showing the analytic clustering statement in Lemma \ref{lemmaanalytic}.

Consider the convergent Taylor series representation of $\tau_t^{\frak U} \la a$,
\beq\label{tautuabseries}
	\bra \tau_t^{\frak U} \la a,\la b\ket = \sum_{n=0}^\infty
	\frc{t^n}{n!} \bra (\delta^{\frak U})^n\la a,\la b\ket\quad
	(\la a,\la b\in\mathfrak V,\,t\in\C\ \mbox{small enough}).
\eeq
With sizeable clustering, for every $\la a,\la b\in\mathfrak V$ and $t\in\C$ near enough to 0, we have
\beq\label{tautuab}
	\Big|\frc{t^n}{n!} \bra (\delta^{\frak U})^n\la a,\la b\ket\Big|
	\leq u\,|\la b|^r\,||\la b||_{\mathfrak U}
	\frc{|t|^n}{n!}\, |(\delta^{\frak U})^n \la a|^r\, ||(\delta^{\frak U})^n \la a||_{\mathfrak U}\,({\rm dist}((\delta^{\frak U})^n\la a,\la b)+1)^{-p}.
\eeq
By considering each term in the explicit commutator \eqref{deltaU}, it is easy to see that $(\delta^{\frak U})^n \la a$ has a support of size at most $|\la a|+2n|\la h|$, and thus
\beq\label{bounsizedeltau}
	|(\delta^{\frak U})^n \la a| \leq |\la a|+2n|\la h|.
\eeq
Further, we have (see e.g.~\cite[App B]{Doyon2017})
\beq\label{boundeltau}
	||(\delta^{\frak U})^n \la a||_{\mathfrak U}\leq n! \,2^n |\la h|^n(|\la a|+|\la h|)^n\, ||\la h||_{\mathfrak U}^n\, ||\la a||_{\mathfrak U}.
\eeq
Also,
\beqa\label{boundistdeltau}
	({\rm dist}((\delta^{\frak U})^n\la a,\la b)+1)^{-p} &\leq&
	\big({\rm max}\{{\rm dist}(\la a,\la b)+1-n|\la h|,1\}\big)^{-p}\n
	&\leq & ({\rm dist}(\la a,\la b)+1)^{-p}(1+n|h|)^p.
\eeqa
Putting these bounds together,
\beq\label{analclus}
	\Big|\frc{t^n}{n!} \bra (\delta^{\frak U})^n\la a,\la b\ket\Big| \leq u\, |\la b|^r\,||\la a||_{\mathfrak U}\,||\la b||_{\mathfrak U}\,
	({\rm dist}(\la a,\la b)+1)^{-p}
	\big(2|\la h|(|\la a|+|\la h|)||\la h||_{\mathfrak U}\,|t|\big)^n
	(|\la a| +2n|\la h|)^r (1+n|\la h|)^p.
\eeq
Therefore the series on the right-hand side of \eqref{tautuabseries} is absolutely convergent whenever
\beq\label{radiust}
	|t| < t_{\la a} := \big(2|\la h|(|\la a|+|\la h|)||\la h||_{\mathfrak U}\big)^{-1}.
\eeq
This implies that $\bra \tau_t^{\frak U} \la a,\la b\ket$ can be analytically continued in $t$ to this region. Also, in this region,
\beq\label{resultanlyticw}
	|\bra \tau_t^{\frak U} \la a,\la b\ket|\leq
	c\, 
	({\rm dist}(\la a,\la b)+1)^{-p}
\eeq
where
\beq\label{analyticw}
	c = u\,|\la b|^r\,||\la a||_{\mathfrak U}\,||\la b||_{\mathfrak U}\,\sum_{n=0}^\infty |t/t_{\la a}|^n
	(|\la a| +2n|\la h|)^r (1+n|\la h|)^p <\infty.
\eeq
Replacing $\la a \to \iota_x^{\frak U} \la a$, using \eqref{distiota}, and passing to the space $\lo$, we find that for every $\la a,\la b\in\lo$, there is $\ep>0$ such that clustering holds uniformly for the family of pairs $\{(\tau_t\la a,\la b):t\in\C,\;|t|<\ep\}$. Thus Lemma \ref{lemmaanalytic} applies, and the dynamical system $\hi,\lo$ is differentiably clustering.

\medskip
\noindent{\bf IV.} With respect to $\hi,\h\lo$, the time evolution $
\tau$ is differentiably clustering (Definition \ref{deficlusgroup}). Proof of Theorem \ref{theoclusteringchainanalytic}.

Let
\beq
	\la a_n(s) = \frc{s^n}{n!} (\delta^{\frak U})^n\la a\in\mathfrak V\quad (\la a\in\mathfrak V,\,n\in\N,\,s\in\C).
\eeq
We apply the bound Eq.~\eqref{clusapx} (with $n\to\infty$ in this equation), and thus for every $0<q<p$ and compact $T\subset \R$, there exists $w>0$ such that for all $\la a,\la b\in\mathfrak V$ and $t\in T$,
\beq
	| \bra \la a_n(s),\tau_t^{\mathfrak U}\la b\ket|
	\leq w\,|\la a_n(s)|^r\,|\la b|^{r'}\,||\la a_n(s)||_{\mathfrak U}\,||\la b||_{\mathfrak U}\,
	\,({\rm dist}(\la a_n(s),\la b)+1)^{-q}.
\eeq
From \eqref{bounsizedeltau}, \eqref{boundeltau} and \eqref{boundistdeltau}, we get a bound as in \eqref{analclus}:
\beq
	| \bra \la a_n(s),\tau_t^{\mathfrak U}\la b\ket|
	\leq w\, |\la b|^{r'}\,||\la a||_{\mathfrak U}\,||\la b||_{\mathfrak U}\,
	({\rm dist}(\la a,\la b)+1)^{-q}
	|s/t_{\la a}|^n
	(|\la a| +2n|\la h|)^r (1+n|\la h|)^q.
\eeq
Therefore, for every $|s|<t_{\la a}$, the series
\beq
	\bra \tau_{t+s}^{\mathfrak U}\la a,\la b\ket
	= \sum_{n=0}^{\infty}\bra \la a_n(s),\tau_{-t}^{\mathfrak U}\la b\ket
\eeq
is absolutely convergent, showing analyticity in $s$. As $t_a$ (Eq.~\eqref{radiust}) only depends on $|\la a|$ and not on other properties of $\la a$ or  on $\la b$, the analyticity statement of Theorem \ref{theoclusteringchainanalytic} follows. Further, we also conclude that for every $0<q<p$ and compact $T\subset \R$, we have for all $\la a,\la b\in\mathfrak V$, $t\in T$ and $|s|\leq \ep<t_{\la a}$,
\beq
	|\bra \tau_{t+s}^{\mathfrak U}\la a,\la b\ket|\leq
	c \,({\rm dist}(\la a,\la b)+1)^{-q}
\eeq
where
\beq
	c = w\, |\la b|^{r'}\,||\la a||_{\mathfrak U}\,||\la b||_{\mathfrak U}\,
	\sum_{n=0}^\infty (\ep/t_{\la a})^n
	(|\la a| +2n|\la h|)^r (1+n|\la h|)^q <\infty.
\eeq
This shows the bound in Theorem \ref{theoclusteringchainanalytic}, thus completing the proof of the theorem.

Replacing $\la a \to \iota_x^{\frak U} \la a$, using \eqref{distiota}, and passing to the space $\lo$, we find that for every $\la a,\la b\in\lo$ and $t\in\R$, there is $\ep>0$ such that clustering holds uniformly for the family of pairs $\{(\tau_{t+s}\la a,\la b):s\in\C,\,|s|<\ep\}$. By the group properties of $\tau_t$, Lemma \ref{lemmaanalytic} implies that the the time evolution $\tau$ is differentiably clustering with respect to the dynamical system $\hi,\h\lo$.

\medskip
\noindent{\bf V.} If $p>r$, then both the dynamical systems $\hi,\lo$ and $\hi,\h\lo$ are space-like $q$-clustering with velocity $v_{\rm c}$, for every $0<q<p-r$ and $v_{\rm c}>v_{\rm LR}$ (Definition \ref{deflr}). Proof of Theorem \ref{theochainlightcone}.

By Remark \ref{remaspacelike}, it is sufficient to establish this for the dynamical system $\hi,\lo$. For this purpose, we need the assumption that $p>r$.t

Recall the argument for Eq.~\eqref{lemprlr2}: using the expression \eqref{LR} of the Lieb-Robinson bound, the triangle inequality, and the fact that ${\rm dist}(\la b , \Z\setminus X_n^{\la b}) = n$ for all $n\in\N$, it follows that for all $t\in \R$, all $m,n\in \N$ and all $\la a,\la b\in {\mathfrak V}$,
\beqa\label{lemprlr2b}
	|\bra \la a,\mathbb P_{X_n^{\la b}}\la \tau_t^{\mathfrak U}\la b\ket| &\leq&
	|\bra \la a,\mathbb P_{X_m^{\la b}}\la \tau_t^{\mathfrak U}\la b\ket|+\\ && +\,
	b\,|\la b|\,||\la a||_{\mathfrak U}\,||\la b||_{\mathfrak U}\Big(
	\exp\Big[-d\,\big(m - v_{\rm LR}|t|\,\big)\Big] +  \exp\Big[-d\,\big(n - v_{\rm LR}|t|\,\big)\Big]\Big).\no
\eeqa
Set $z = {\rm dist}(\la a,\la b)+1$, choose $0<\ep'<\ep<1$, and set the compact $T_z=v_{\rm LR}^{-1} \ep' \,[-z, z]$.

With the inequalities \eqref{leminequalities} and sizeable clustering, we obtain for all $\N\ni n\leq \ep z$
\beq
	|\bra \la a,\mathbb P_{X_n^{\la b}}\la \tau_t^{\mathfrak U}\la b\ket| \leq  u\, |\la a|^r\,|\la b|^r(1+2\ep z)^r\,||\la a||_{\mathfrak U}\,||\la b||_{\mathfrak U}\,({\rm max}\{z-\ep z,1\})^{-p}.
\eeq
The function
\beq
	[1,\infty) \ni z\mapsto \frc{(1+2\ep z)^r({\rm max}\{z-\ep z,1\})^{-p}}{z^{-(p-r)}}
\eeq
has a finite upper bound. Therefore, there exists $k>0$ such that for all $t\in \R$, all $\N\ni n\leq \ep z$ and all $\la a,\la b\in {\mathfrak V}$,
\beq\label{lemspacelikebound}
	|\bra \la a,\mathbb P_{X_n^{\la b}}\la \tau_t^{\mathfrak U}\la b\ket| \leq  ku\, |\la a|^r\,|\la b|^r\,||\la a||_{\mathfrak U}\,||\la b||_{\mathfrak U}\,z^{-(p-r)}.
\eeq

Let us consider $|\bra \la a,\mathbb P_{X_n^{\la b}}\la \tau_t^{\mathfrak U}\la b\ket|$. We use the bound \eqref{lemspacelikebound} if $ n < \lfloor \ep z\rfloor$, and, otherwise, the bound \eqref{lemprlr2b} with $m = \lfloor \ep z
\rfloor$, which becomes
\beq
	|\bra \la a,\mathbb P_{X_n^{\la b}}\la \tau_t^{\mathfrak U}\la b\ket| \leq
	|\bra \la a,\mathbb P_{X_m^{\la b}}\la \tau_t^{\mathfrak U}\la b\ket|+
	2b\,|\la b|\,||\la a||_{\mathfrak U}\,||\la b||_{\mathfrak U}
	\exp\Big[-d\,\big(m - v_{\rm LR}|t|\,\big)\Big].
\eeq
In this case ($n\geq \lfloor\ep z\rfloor$), we use \eqref{lemspacelikebound} (at $n=m$) for $|\bra \la a,\mathbb P_{X_m^{\la b}}\la \tau_t^{\mathfrak U}\la b\ket|$, and we further have, for all $t\in T_z$ and using $m>\ep z-1$,
\beq
	\exp\Big[-d\,\big(m - v_{\rm LR}|t|\,\big)\Big]
	\leq \exp\Big[-d\,(\ep-\ep')z+d\Big].
\eeq
Note that the power law decay $z^{-(p-r)}$ dominates the exponential decay, and we may use $|\la b|^r\leq |\la b|^{r'}$ and $|\la b|\leq |\la a|^r|\la b|^{r'}$.

We conclude that for all $v_{\rm c}>v_{\rm LR}$ (taking $v_{\rm c} = v_{\rm LR}/\ep'$), there exists $u'>0$ such that for all $n\in \N$, $\la a,\la b\in{\mathfrak V}$ and $t\in v_{\rm c}^{-1}[-{\rm dist}(\la a,\la b),{\rm dist}(\la a,\la b)]$,
\beq
	|\bra \la a,\mathbb P_{X_n^{\la b}} \tau_t^{\mathfrak U}\la b\ket| \leq u'\,|\la a|^r\,|\la b|^{r'}\,||\la a||_{\mathfrak U}\,||\la b||_{\mathfrak U}\,({\rm dist}(\la a,\la b)+1)^{-(p-r)}.
\eeq
Taking the limit on $n$, we have $\lim_n \mathbb P_{X_n^{\la b}}\la \tau_t^{\mathfrak U}\la b = \tau_t^{\mathfrak U}\la b$, and symmetrising on $\la a,\,\la b$, this shows Theorem \ref{theochainlightcone}. Using \eqref{distiota}, we may pass to the space $\lo$. Let $v_{\rm c}>v_{\rm LR}$. Then for every $\la a,\la b\in\lo$, there exists $c>0$ and $0<v<v_{\rm c}$ such that, for all $x\in\Z$ and $t\in v^{-1}[-|x|,|x|]$
\beq
	|\bra \iota_x \tau_t \la a,\la b\ket| \leq c\,(|x|+1)^{-(p-r)}.
\eeq
Thus we have space-like $p_{\rm c}$-clustering for every $p_{\rm c}<p-r$ and velocity $v_{\rm c} > v_{\rm LR}$. \eproof

\subsection{Ergodicity, hydrodynamic projections and Euler equations}\label{ssectproofschain}

We finally provide, with the above results, the proofs of the quantum spin chain theorems expressed in Section \ref{sectresultschain}.

\medskip

\proofof{Theorem \ref{theorelaxchain}} By Corollary \ref{corobasicchain}, the dynamical system $\hi,\lo$ is space-like $\infty$-clustering, hence by Lemma \ref{lemmaspacelike} it satisfies Property \ref{propspacelike}a. As for any $\la a\in\mathfrak U$ there corresponds an element of $\hi$, the result follows using \eqref{omegainner}.
\eproof

\medskip

\proofof{Theorem \ref{theochain}} As the dynamical system $\hi,\lo$ is $\infty$-clustering (Corollary \ref{corobasicchain}), hence 1-clustering, Point I follows from Theorem \ref{theodrude}, and the first part of Point II from Theorem \ref{theocont}. As the dynamical system is in fact space-like clustering, it satisfies Property \ref{propspacelike} (Lemma \ref{lemmaspacelike}), hence the last part of Point II follows from Theorem \ref{main}.
\eproof

\medskip

\proofof{Theorem \ref{theoeulerchain}} Point I: By Corollary \ref{corobasicchain}, the dynamical system $\hi,\h\lo$ is space-like $\infty$-clustering, and with respect to it, $\tau$ is differentiably clustering. Thus by Theorems \ref{theocomplete} and \ref{theocurrent} it can be extended so that points (1)-(3) hold. By Theorem \ref{theocomplete}, the dynamical system $\hi,\h\lo^\#$ is still space-like $\infty$-clustering, and with respect to it, $\tau$ is still differentiably clustering. Point II then follows from Theorem \ref{theoeuler}, as space-like clustering guarantees that Property \ref{propspacelike} holds (Lemma \ref{lemmaspacelike}).
\eproof

\section{Discussion}\label{sectcon}

We have developed a general framework, based on Hilbert space structures, for the long-time, small-wavenumber behaviours of strongly interacting many-body systems. We have shown an almost-everywhere ergodicity theorem, as well as a hydrodynamic projection formula and a linearised Euler equation. We have shown in particular that all properties of the general framework hold in quantum spin chains with finite-range interactions, thus proving in all generality the above resuls in this family of models.

We make a number of remarks about the results.

\medskip
\noindent {\em1. There is time-like ergodicity, thus decay of operator strength almost everywhere within the light cone.} In our proof, projection is seen to arise thanks to almost-everywhere ergodicity, which is thus a sufficient relaxation property. In fact -- see the last discussion point below -- large-time vanishing holds not only for zero-frequency time-averages, but also at any frequency. Physically, almost-everywhere ergodicity can be pictured as follows. By the Lieb-Robinson bound, the time evolute of a local observable is an observable supported on a growing region, lying within a ``light cone" in space-time \cite{LiebRobinson}. Intuitively, the strength of the observable, or at least its time-average, should nevertheless decay over time, within this light-cone, as the observable carries a finite quantity of information, energy, etc. The theorem says that, for any given frequency, the observable may in fact divide up into non-decaying parts at, say, a countable number of velocities within this light-cone. However, it must indeed decay at almost every velocity. This theorem is discussed in more depth and generalised in \cite{ampelogiannis2021almost,ampelogiannis2021ergo}.

\medskip
\noindent
{\em2.  The space of ballistic waves is universally defined as the space of extensive (or pseudolocal) conserved charges.} The idea that projections onto slow modes gives rise to hydrodynamics has been discussed widely in the past, especially starting with the works of Zwanzig and Mori \cite{zwanzig1961lectures,mori1965transport}. We have defined unambiguously the degrees of freedom $\hicons_0$ onto which projection must occur at the Euler scale: the space of thermodynamically extensive conserved charges. It is important to note that these are properties not only of the dynamics, but also of the state $\omega$: extensivity is a notion that is state-dependent. Further, they are properties of statistical models on infinite space; there is no such notion on finite lengths. Physically, the projection onto the correct, reduced space of large-scale degrees of freedom occurs, because the infinite length of the system allows the asymptotic regions to absorb the infinite amount of information lost in going to the Euler scale.

\medskip
\noindent
{\em3. The space of ballistic waves is infinite-dimensional in integrable models.} In integrable models, the space $\hicons_0$ can be shown to be infinite-dimensional, using various standard methods \cite{faddeev1996algebraic,grabowski1995structure}. The above is then expected to reproduce the results of generalised hydrodynamics for correlation functions, obtained in \cite{SciPostPhys.3.6.039,doyoncorrelations}. However,  establishing this precise link would require establishing rigorously the connection between the space $\hicons_0$ and the space of root densities of the thermodynamic Bethe ansatz (see the discussions in \cite{IlievskiInteracPart,pozsgay2017generalized,1742-5468-2016-6-064002}), which is still missing. Another open problem, in the case where $\hicons_0$ is infinite-dimensional, is that of establishing the existence and uniqueness of solutions to the linearised Euler equation \eqref{dSdkintro}.

\medskip
\noindent
{\em4. The problem of finite-dimensionality is nontrivial in non-integrable models.} Perhaps the most important open problem, a seemingly very difficult one, is that of determining finite-dimensionality of $\hicons_0$ in non-integrable systems. The recent progress \cite{shiraishi2019proof} in proving the absence local conserved charges in certain non-integrable quantum spin chains is an interesting first step. One may in fact conjecture that finite-dimensionality is intimately related to non-integrability. Thus, in particular, integrability, or the lack thereof, must be defined with respect to a choice of both a dynamics and a state. It would be particularly interesting to apply the framework to quantum and classical gases, which, even without integrability, possess a nontrivial Euler hydrodynamic limit, thanks to a conserved momentum.

\medskip
\noindent
{\em5. The Drude weights are a simple special case.} As mentioned, the projection result \eqref{hpf} at $\kappa=0$, when applied to current observables $\la j_i$, is a result for the Drude weights. In fact, for general observables, at $\kappa=0$, the limit in \eqref{defS} exists and the result \eqref{hpf} holds by a simple application of von Neumann's mean ergodic theorem for unitary operators \cite{RudinFunctional}. The question of what conserved charge is involved in saturating the Mazur bound (the Suzuki equality) has been discussed recently \cite{dhar2020revisiting}; Eq.~\eqref{hpf} gives a rigorous answer in models on infinite space.

\medskip
\noindent
{\em6.  The algebraic structure of conserved charges does not influence the Euler scale.} We find that all conserved charges -- be their algebra abelian or not -- contribute emergent ballistic degrees of freedom. In particular, no requirement is made for the charges to commute with the density matrix used to define the Gibbs state. This is despite the fact that the heuristic linear-response argument  requires this, or otherwise involves the Kubo-Mori-Bogoliubov inner product \cite{kubo1954magnetic,mori1965transport} (but we expect the same results to hold using this inner product). The question of how non-abelian charges affect hydrodynamics has been discussed in the literature, see the recent viewpoints given in \cite{krajnik2020undular,glorioso2020hydro}, and in particular their connection with super-diffusion \cite{ilievski2020superuni}.

\medskip
\noindent
{\em7. Only homogeneous conserved charges contribute.} Further, we show that the charges must be homogeneous (translation invariant), a point which does not seem to have been emphasised yet; non-homogeneous conserved charges, such as the ``strictly local conservation laws" discussed in \cite{buca2020quantum}, do not enter the linearised Euler scale with respect to homogeneous states.

\medskip
\noindent
{\em8. The linearised Euler scale near to arbitrary frequencies and wavelengths can be accessed by the same theory.} Finally, as our theorems are based on a very general abstract framework, there is room for many possible extensions. In particular, different definitions of the notion of space- and time-translations may be used, with respect to which homogeneity and stationarity may be imposed. This suggests that the general Euler-scale structure holds not just at large wavelengths and low frequencies, but rather as an expansion around given wavelengths and frequencies: if $\tau_t$ and $\iota_x$ are the usual, homogeneous time and space translation isomorphisms, then $\re^{\ri\omega t}\tau_t$ and $\re^{\ri kx}\iota_x$ also are appropriate time and space translation isomorphisms, satisfying all required properties. These give access to hydrodynamic quantities near to frequency $\omega$ and wavenumber $k$. This is relevant in respect of recent works \cite{buca2019nonstationary,medenjak2020rigo,buca2020quantum}. Our results therefore establish that the linearised Euler-scale structure is extremely universal. For our ergodic result, this extension is performed explicitly in \cite{ampelogiannis2021almost,ampelogiannis2021ergo}.

\medskip
\noindent
{\em9. Many extensions are possible.} We assumed the statistical model to lie on a the discrete space $\Z$. The framework can nevertheless be applied to models lying on the line $\R$, by for instance restricting to observables lying on any subset $\cong\Z$, or by gathering degrees of freedom (such as field configurations) into cells. We also assumed the space of local observables to be countable dimensional. If the space is larger, then our proof of almost-everywhere ergodicity fails. Under the assumption of almost-everywhere ergodicity, however, it is simple to see that the general hydrodynamic projection result still holds. We assumed time translations to form a group isomorphic to $\R$. We expect that a large part of the theory can be adapted to models with a discrete group of time translations isomorphic to $\Z$. This would be of interest in current research on the dynamics of cellular automata \cite{prosen2016cellular}. A large part of the framework is immediately generalisable to higher dimensions, and is applicable to a wider family of quantum chains and states than those considered, as algebraic decay of correlations is sufficient for many of the results. It is also possible that, combining with the construction of hydrodynamic spaces in \cite{DoyonDiffusion2019}, similar arguments as those presented here can lead to diffusive-scale hydrodynamic equations.

{\bf Acknowledgments.} This work originated from discussions with J. M. Maillet and K. Gawedski, while visiting \'Ecole Normale Sup\'erieure de Lyon in March 2019 as an invited professor. I am grateful for the support from \'ENS Lyon. The second version of the work has greatly benefited from discussions with Dimitrios Ampelogiannis, Amirali Hannani and Stefano Olla. The work and its context has also benefited from discussions with D. Bernard, B. Bu\c{c}a, O. A. Castro-Alvaredo, W. De Roeck, D. Karevski, R. Kuehn, A. Lucas, M. Medenjak, T. Prosen, A. Pushnitski, T. Sasamoto and H. Spohn. I acknowledge funding from the Royal Society under a Leverhulme Trust Senior Research Fellowship, ``Emergent hydrodynamics in integrable systems: non-equilibrium theory", ref.~SRF\textbackslash R1\textbackslash 180103, under which part of this work was done. I am grateful to the Tokyo Institute of Technology for support and hospitality during an invited professorship (October 2019), where part of this research was performed. This research was also supported in part by the International Centre for Theoretical Sciences (ICTS) during a visit for the program Thermalization, Many body localization and Hydrodynamics (code: ICTS/hydrodynamics2019/11).

\appendix

\section{Banach limits}\label{appbanach}

We are first looking for a Banach limit $\phi$ on the Banach space of bounded functions $f: \R_+\to\C$, with norm $||f|| =\limsup_{t\to\infty} |f(t)|$, with the properties of positivity, if $f(t)\geq0$ for all $t>0$ then $\phi(f)\geq0$, invariance under affine transformations, for all $s(t)=at+b$ ($a>0$ and $b\in\R$) we have $\phi(f\circ s) = \phi(f)$, and compatibility with the limit, $\phi(f) = \lim_{t\to\infty} f(t)$ if the limit exists. Note that it is also possible to define Banach limits with the restriction of re-parametrisation invariance, $s(t)$ with $\lim_{t\to\infty} s(t) = \infty$; however it is not useful to do so here. See \cite{bookconway} for a discussion of Banach limits.

On the linear space of bounded functions $f:\R_+\to \C$, we consider the equivalence under the relation $f\equiv g$ if, for $f,g$, there exists, at large enough $t$, a re-parametrisation $s(t)=at+b$ such that $f(t) = g(s(t))$:
\beq
	f \equiv g \quad \Leftrightarrow \quad
	\exists\; t_0\in\R,\,a>0,\,b\in\R\ \big|\ 
	f(t) = g(s(t)) \ \forall\ t\geq t_0.
\eeq
The linear space $S$ of equivalence classes $\hat f$ under this equivalence relation is normed by $||\hat f|| = ||f|| = \limsup_{t\to\infty}|f(t)|$ (for any representative $f$ of $\hat f$). The limit operation is a linear functional on the subspace $\{\hat f\}$ formed of equivalence classes of converging functions $f$, and the result of the limit operation on an equivalence class $\hat f$ that has a limit is bounded by the norm $||\hat f||$. By \cite[Thm 3.3]{RudinFunctional}, there exists a linear functional $\h\phi$ on $S$ that extends the limit operation, whose result is also bounded by the norm, hence $||\h\phi||=1$. In order to show the property of positivity, let $f(t)\geq 0\;\forall t\in \R_+$. Then $||\hat{e}-\hat f_1 ||\leq 1$ where $e(t)=1$ and $f_1(t) = f(t)/||f||$ has norm 1. But if $\h\phi(\h f)<0$, then $\h\phi(\h e-\h f_1) = 1-\h\phi(\h f_1) = 1-\h\phi(\h f)/||f||>1$, which is impossible as $||\h\phi||=1$. Finally, we just have to set $\phi(f) = \h\phi(\h f)$.

We then consider the space of bounded functions $f:\R_+\to\C$ with the additional property that $f$ be measurable. On this, we define the limit $\balim_{t\to\infty}$ by composition of some given Banach limit $\phi$ as above, with the Ces\`aro limit:
\beq
	\balim_{t\to\infty} f(t) = \phi(F),\quad
	F(t) = \frc1t \int_0^t\dd s\,f(s).
\eeq
Clearly, this preserves positivity, affine invariance and compatibility with the limit. Further, we have the bound
\beq
	\big|\balim_{t\to\infty} f(t)\big| \leq
	\limsup_{t\to\infty} \frc1t \Big| \int_0^t \dd s\,f(s)\Big|
	\leq
	\limsup_{t\to\infty} |f(t)|.
\eeq
Hence $\balim_{t\to\infty}$ is a Banach limit.

\def\doi#1{\\ doi: \href{https://www.doi.org/#1}{#1}}

\end{document}